\documentclass[a4paper,11pt]{article}
\pdfoutput=1
\usepackage{jheppub} 
\usepackage{epsfig}
\usepackage[caption=false]{subfig}
\usepackage{amsmath}
\usepackage{amsfonts}
\usepackage{amssymb}
\usepackage{float}
\usepackage{color}
\usepackage{graphicx}
\usepackage{graphics}
\usepackage{hyperref}
\usepackage{adjustbox,array}
\hypersetup{}
\usepackage[utf8x]{inputenc} 
\usepackage{epstopdf}
\usepackage{multirow}
\usepackage{slashed}
\usepackage{booktabs}
\usepackage{gensymb}
\usepackage{comment}
\usepackage{diagbox}
\usepackage{afterpage}

\usepackage{colortbl}
\usepackage{array}

\definecolor{Gray}{gray}{0.85}
\newcolumntype{?}{!{\vrule width 1pt}}

\hypersetup{
 colorlinks=true,
 linkcolor=blue,
 urlcolor=blue,
 citecolor=red
}

\preprint{CTPU-PTC-23-08}

\title{The Strong Force meets the Dark Sector: a robust estimate of QCD uncertainties for anti-matter dark matter searches}

\author[a]{Adil Jueid,}
\author[b]{Jochem Kip,}
\author[c]{Roberto Ruiz de Austri}
\author[d]{and Peter Skands}

\affiliation[a]{Particle Theory  and Cosmology Group, Center for Theoretical Physics of the Universe, Institute for Basic Science (IBS), Daejeon, 34126, Republic of Korea}
\affiliation[b]{Institute for Mathematics, Astrophysics and Particle Physics, Radboud University, \\
Nijmegen, Heyendaalseweg 135, Nijmegen, the Netherlands}
\affiliation[c]{Instituto de F\'{\i}sica Corpuscular, CSIC-Universitat de Val\`encia, \\ E-46980 Paterna, Valencia, Spain}
 \affiliation[d]{School of Physics and Astronomy, Monash University, \\ Wellington Rd, Clayton VIC-3800, Australia}

\emailAdd{adiljueid@ibs.re.kr}
\emailAdd{jochem.kip@ru.nl}
\emailAdd{rruiz@ific.uv.es}
\emailAdd{peter.skands@monash.edu}

\begin{document}

\abstract{In dark-matter annihilation channels to hadronic final states, stable particles --- such as positrons, photons,  antiprotons, and antineutrinos --- are produced via complex sequences of phenomena including QED/QCD radiation, hadronisation, and hadron decays. These processes are normally modelled by Monte Carlo event generators whose limited accuracy imply intrinsic QCD uncertainties on the predictions for indirect-detection experiments like \textsc{Fermi-LAT}, \textsc{Pamela}, \textsc{IceCube} or \textsc{Ams}--02. In this article, we perform a complete analysis of QCD uncertainties in antimatter spectra from dark-matter annihilation, based on parametric variations of the \textsc{Pythia}~8 event generator. After performing several retunings of light-quark fragmentation functions, we define a set of variations that span a conservative estimate of the QCD uncertainties. We estimate the effects on antimatter spectra for various annihilation channels and final-state particle species, and discuss their impact on fitted values for the dark-matter mass and thermally-averaged annihilation cross section. We find dramatic impacts which can go up to $\mathcal{O}(40)$ GeV for uncertainties on the dark-matter mass and up to $\mathcal{O}(10\%)$ for the annihilation cross section. We provide the spectra in tabulated form including QCD uncertainties and code snippets to perform fast dark-matter fits, in this \href{https://github.com/ajueid/qcd-dm.github.io.git}{github} repository.}


\maketitle

\section{Introduction}
\label{sec:intro}

Various theories beyond the Standard Model (SM) predict the annihilation of weakly interacting massive particles (WIMPs) into SM particles that produce stable final-state particles such as photons, anti-protons, neutrinos, or positrons as a result of complex sequences of processes (for a review see {\it e.g.}, \cite{Jungman:1995df,Bergstrom:2000pn,Bertone:2004pz,Feng:2010gw}). Contributions of dark-matter (DM) annihilation products  to the cosmic-ray (CR) fluxes may leave footprints in various experiments like Fermi Large Area Telescope (\textsc{Fermi--LAT}), or the Alpha Magnetic Spectrometer~(AMS). 
Secondary CR antiprotons and positrons are important tools to study the nature and the properties of various sources in the galactic region and beyond. The idea of using antiprotons in DM indirect detection is not recent\footnote{In fact, an excess over the astrophysical predictions was reported right after the first evidence for the existence of antiprotons in 1979 \cite{Golden:1979bw} (further confirmation was done in 1981 \cite{Buffington:1981zz}). An attempt to explain this excess was performed shortly after this discovery wherein a massive photino DM with mass of $m_{\tilde{\gamma}} \sim 3~{\rm GeV}$ can reproduce both the correct relic density and the total antiproton flux \cite{Silk:1984zy, Stecker:1985jc}.}. Recently, a measurement of the antiproton flux and the $\bar{p}/p$ flux ratio has been performed by the AMS--02 collaboration at the International Space Station over the rigidity range $1$--$450$~GV  \cite{AMS:2016oqu}. With data collected between 2011 and 2015, about $3.49\times 10^{5}$ antiproton events have been reported which render the statistical uncertainties a very subleading contribution to the total errors in most rigidity regions. Interestingly an excess over the expected background has been reported by several analyses in the rigidity range of $10$--$20$ GV \cite{Cuoco:2016eej,Cui:2016ppb,Cuoco:2017rxb,Reinert:2017aga,Cui:2018klo,Cuoco:2019kuu,Cholis:2019ejx,Lin:2019ljc,Abdughani:2021pdc,Hernandez-Arellano:2021bpt,Biekotter:2021ovi}. It was pointed out that DM with mass of about $m_X \sim 50$--$100$ GeV annihilating predominantly into hadronic final states can explain this excess, with most of the analyses preferring a DM mass of roughly $m_X \sim 60$~GeV. Moreover, similar DM properties (mass range, and thermal annihilation cross sections) have been considered in the so-called gamma ray Galactic Center Excess (GCE) found in data reported by the \textsc{Fermi}--LAT collaboration \cite{Goodenough:2009gk, Vitale:2009hr, Hooper:2010mq, Gordon:2013vta, Hooper:2011ti, Daylan:2014rsa, Calore:2014xka, Abazajian:2014fta,Zhou:2014lva, Caron:2015wda, vanBeekveld:2016hbo, Butter:2016tjc, Karwin:2016tsw, Achterberg:2017emt}. 

While the statistical uncertainties on the antiproton flux are now very small, a proper treatment of systematic uncertainties and their correlations can be very important in DM analyses \cite{diMauro:2014zea,Kappl:2014hha,Kachelriess:2015wpa,Winkler:2017xor,Korsmeier:2018gcy}. It was pointed out that proper treatment of systematic errors may not only reduce the antiproton excess but can even completely exclude it \cite{Heisig:2020nse}. On the other hand, the AMS--02 collaboration has released measurements of both the $e^+/e^-$ ratio and the positron flux that both pointed toward an excess with respect to the astrophysical backgrounds in the region $\simeq 10$--$300$ GeV \cite{AMS:2019rhg} which is consistent with the previous measurements but with smaller error bars. The fact that the observed spectra are not expected in astrophysics has prompted several explanations including DM (see e.g. \cite{Bergstrom:2008gr,Cirelli:2008pk,Bergstrom:2009fa} for details about the different DM possibilities).\\

For DM masses above a few GeV, QCD jet fragmentation can be the leading source of antimatter production in DM annihilation. A large number of antineutrinos, positrons and antiprotons can be produced from complex sequences of physical processes that can include resonance decays (if the DM annihilate to intermediate resonances like $W/Z/H$ bosons or the top quark), QED and QCD bremsstrahlung, hadronisation, and hadron decays. The modeling of hadronisation in particle production from DM annihilation is usually done using multi-purpose Monte Carlo (MC) event generators \cite{Buckley:2011ms} which are either based on the string \cite{Artru:1974hr,Andersson:1983ia} or the cluster \cite{Webber:1983if,Winter:2003tt} models. This is due to the fact that hadronisation cannot be solved from first principles in QCD but only using phenomenological models typically involving several free parameters. Proper estimates of QCD uncertainties that stem from hadronisation are seldom rigorously addressed in DM literature. We note that  comparisons between different MC event generators such as \textsc{Herwig} and \textsc{Pythia} have been done in \cite{Cirelli:2010xx,Cembranos:2013cfa}. Using gamma rays as an example, it was found that the differences in particle spectra predicted in different MC event generators can be observed in the tails of the spectra while there is a high level of agreement between them in the bulk of the spectra \cite{Cembranos:2013cfa}. This finding has been confirmed in a previous study where we have used the most recent and widely used MC event generators \cite{Amoroso:2018qga}. This level of agreement is due to the fact that the default parameter sets for the MC models are being optimised to essentially provide ``central'' fits to roughly the same set of constraining data, comprised mostly of LEP measurements \cite{Buckley:2009bj,Buckley:2010ar,Skands:2010ak,Platzer:2011bc,Karneyeu:2013aha,Skands:2014pea,Fischer:2014bja,Fischer:2016vfv,Reichelt:2017hts,Kile:2017ryy}. Therefore, the envelope of different MC event generators does not reliably span the theory uncertainty allowed by data in the bulk of the spectra while it can overestimate the uncertainty in both the high- and low-energy tails of the spectra. To address this issue, we have provided for the {\it first time} a conservative estimate of QCD uncertainties within \textsc{Pythia}~8 on gamma-ray \cite{Amoroso:2018qga} and antiproton \cite{Jueid:2022qjg} production from DM annihilation  (see also \cite{Amoroso:2020mjm, Jueid:2021dlz,Jueid:2022pzm} for short summaries of these studies). In the present work, we aim of to complete these studies by performing a comprehensive analysis of the QCD uncertainties on antimatter production from DM annihilation. 

We first revisit the constraints from LEP measurements on the parameters of the Lund fragmentation function while this time thoroughly discussing the differences between the various measurements of baryon spectra at the $Z$--pole. We then perform several (re)tunings based on the baseline \textsc{Monash} tune \cite{Skands:2014pea} of \textsc{Pythia}~8.244 event generator \cite{Sjostrand:2014zea}. In this paper, we perform for the first time a Bayesian analysis of the fragmentation-function parameters, finding very good agreement with the results of the frequentist fit. We then estimate the various QCD uncertainties both connected to parton-shower modeling and hadronisation. This paper provides also a first unified model for the production of gamma rays, positrons, antineutrinos and antiprotons in DM annihilation within \textsc{Pythia}~8. The uncertainties are found to be important and range from a few percent to about $50\%$ depending on  particle species,  DM mass,  annihilation channel, and particle energy. Therefore, we recommend  DM groups to use the results of this work in their analyses\footnote{The new data tables and code snippets to perform fast DM fits can be found in \href{https://github.com/ajueid/qcd-dm.github.io.git}{github}.}. \\

The remainder of this paper is organised as follows. In section \ref{sec:physics}, we discuss the physics modeling of antimatter production in a generic DM annihilation process. The last part of section \ref{sec:physics} is essential to determine the relevant constraining data at LEP. In section \ref{sec:measurements}, we discuss the relevant experimental measurements of baryon spectra at LEP  and the consistency between the theory predictions and experimental measurements using three state-of-art MC event generators.  The fitting procedure of this work is discussed in section \ref{sec:setup}. In section \ref{sec:tunes} we discuss the results of the various tunings. A comprehensive discussion of the different types of uncertainties and their estimates is done in section \ref{sec:uncertainty} where we also study quantitatively the impact on the energy spectra for a few selected DM masses, annihilation channels and particle species. Section \ref{sec:DMfit} is devoted to the impact of QCD uncertainties on DM indirect detection experiments for the spectra of antiprotons, electron antineutrinos and photons. We draw our conclusions in section \ref{sec:conclusions}.

\section{Antimatter from dark-matter annihilation}
\label{sec:physics}

\subsection{General features of stable antiparticle production in \textsc{Pythia}~8}

To properly assess the QCD uncertainties on antimatter spectra in DM indirect searches, we first study their production in a generic annihilation or decay process of a dark-matter candidate with mass in the GeV--TeV range. For this, let us consider the following generic annihilation process\footnote{This discussion applies to the case of decaying dark-matter as well. For instance, $\chi \to {\rm SM}~{\rm SM}$ is theoretically possible in models breaking the $Z_2$ symmetry through {\it e.g.} nonminimal interaction of dark matter with gravity (see {\it e.g.} \cite{Cata:2016dsg, Azri:2020bzl, Shaposhnikov:2020aen} for more details).}
\begin{equation}
    \chi \chi \to X_1 X_2 \ldots X_N \to \bigg(\displaystyle\prod_{i=1}^{a_1} Y_{1i} \bigg) \bigg(\displaystyle\prod_{j=1}^{a_2} Y_{2j} \bigg) \ldots \bigg(\displaystyle\prod_{z=1}^{a_N} Y_{Nz} \bigg).
    \label{eq:annihilation}
\end{equation}
We have assumed the narrow-width approximation which enables us to factorise the process into a production part $\chi \chi \to \prod_{i=1}^{N} X_i$ and a decay part $X_i \to \prod_{k=1}^{a_i} Y_{ik}$. In equation \ref{eq:annihilation}, $X_i$ refers to any parton-level SM particle which could be a resonance such as the Higgs boson, $W/Z$-bosons, or the top quark or a non-resonant state like gluons or light quarks. The $X_i$ states are assumed to produce $a_i$ states, through $X_i \to \prod_{k=1}^{a_i} Y_{ik}$, after a complex sequence of processes such as QED bremsstrahlung, QCD parton showering, hadronisation and hadron decays. The produced antiparticles are, therefore, part of this final state and can be detected in indirect detection experiments. Unlike the case of gamma rays, both the total rate as well as the shape of the antiparticle spectra are slightly affected by QED bremsstrahlung. This process occurs if $X_i$ or $Y_i$ contains electrically charged particles or photons and will lead to production of additional photons and/or charged fermions. Besides QED bremsstrahlung, coloured fermions produced in DM annihilation, either promptly or through the decay of heavy resonances, will undergo QCD bremsstrahlung wherein additional coloured particles are produced. This phenomenon is characterised by an enhancement of probabilities for emissions of soft and/or collinear gluons, with the latter being suppressed if the produced fermions are heavy. Furthermore, the rates of $g\to q\bar{q}$ are enhanced at low gluon virtualities: $Q^2 = (p_q + p_{\bar{q}})^2 \to 0$. The rate of QCD radiation processes is mainly controlled by the effective value of the strong coupling constant $\alpha_S$ which is evaluated at a scale proportional to the shower evolution variable\footnote{In \textsc{Pythia}~8, the shower evolution scale is the transverse momentum of the branching parton.}. We note that the value of $\alpha_S(M_Z)$ in \textsc{Pythia}~8 is larger than $\alpha_S(M_Z)^{\overline{{\rm MS}}}$ 
for two reasons: {\it (i)} the so-called Catani-Webber-Marchesini (CMW) scheme involves a set of universal corrections in the soft limit \cite{Catani:1990rr} which is equivalent to a net $10\%$ increase in the value of $\alpha_S(M_Z)$ and {\it (ii)} an agreement between data and theory in experimental measurements of $3$-jets observables in $e^+ e^-$ collisions at LEP energies is reached if $\alpha_S(M_Z)$ is increased by a further $\sim10\%$ \cite{Skands:2010ak,Skands:2014pea}.  

Finally, at a scale $Q_{\rm IR} \simeq \mathcal{O}(1)~{\rm GeV}$, any coloured particle must hadronise to produce a set of colourless hadrons. This process, called fragmentation is modeled within \textsc{Pythia}~8 with the Lund string model \cite{Andersson:1983ia, Sjostrand:1982fn, Sjostrand:1984ic}. The longitudinal part of the description of the hadronisation process is given by the left-right symmetric {\it fragmentation function}, $f(z)$, which gives the probability for a hadron to take a fraction $z \in [0, 1]$ of the remaining energy at each step of the (iterative) string fragmentation process. The general form can be written as 
\begin{equation}
        f(z,m_{\perp h}) \propto N \frac{(1-z)^a}{z}\exp\left(\frac{-b m_{\perp h}^2}{z}\right)~,
\label{eq:fz}
\end{equation}
where $N$ is a normalisation constant that guarantees the distribution to be normalised to unit integral, and $m_{\perp h}= \sqrt{m_h^2 + p_{\perp h}^2}$ is called the ``transverse mass'', with $m_h$ the mass of the produced hadron and $p_{\perp h}$ its momentum transverse to the string direction, and $a$ and $b$ are tunable parameters which are denoted in \textsc{Pythia}~8 by \texttt{StringZ:aLund} and \texttt{StringZ:bLund} respectively. As was pointed in a previous work \cite{Amoroso:2018qga}, the $a$ and $b$ parameters are highly correlated (in the context of fits to measurements) since the former controls the high-$z$ tail, while the latter mainly controls the low-$z$ one, while most of the data is sensitive to the average $z$ which is given by a combination of the two which does not have a simple analytical expression. A new reparametrisation of the fragmentation function exists for which the $b$ parameter is replaced by $\langle z_\rho \rangle$ which represents the average longitudinal momentum fraction taken by mainly the $\rho$ mesons, {\it i.e.} 
\begin{equation}
\left<z_\rho\right> = \int_0^1 \mathrm{d}z \ z f(z,\left<m_{\perp\rho}\right>)~.
\label{eq:zrho}
\end{equation}

This equation is solved numerically for $b$ at the initialisation stage (which requires that the option \texttt{StringZ:deriveBLund = on}) where the following parameters:
\begin{eqnarray}
\left<m_{\perp\rho}\right>^2 & = & 
m^2_\rho + 2( \mbox{\texttt{StringPT:sigma}})^2~,
\\
\left<z_\rho\right> & = &\mbox{\texttt{StringZ:avgZLund}}~,
\end{eqnarray}
are used. 

 In the string-fragmentation picture, baryons are produced similarly to mesons, by allowing the breaking of strings by the production of diquarks-antidiquark pairs; these can be thought as bound states of two quarks (in an antiduqark) or two antiquarks (in a diquark). This basic picture entails a very strong (anti)correlation of the produced baryons in both flavour and phase space, due to the fact that a baryon originating from a diquark produced in a string breaking is associated with an anti-baryon as the new end-point of the residual string. Experimental measurements of $\Lambda^0 \bar{\Lambda}^0$ correlations by the \textsc{Opal} collaboration \cite{Abbiendi:1998ux} do not find such strong correlations. To address this, the so-called \emph{pop-corn} mechanism was suggested \cite{Andersson:1984af, Eden:1996xi}, in which baryons are produced such that the string breaking occurs with the production of one or more $q\bar{q}$ pairs ``in between'' the diquark-antidiquark pair. This picture enables the production of one or more mesons between two baryons and therefore decrease the correlations between them. Note that in the context of DM indirect searches, the correlation between baryons is not relevant as the produced protons travel for long distances before they reach the detector. In \textsc{Pythia8}, baryon production is controlled by an additional parameter denoted by $a_{\rm Diquark}$ such that the $a$ parameter in $f(z)$ is modified as $a \to a + a_{\rm Diquark}$. The extra parameter relevant for baryon production is denoted in \textsc{Pythia}~8 by \verb|StringZ:aExtraDiquark|.

\subsection{The origin of positrons, antineutrinos and antiprotons in a generic dark-matter annihilation process}
After discussing the general features of antiparticle production in a generic dark-matter annihilation process in \textsc{Pythia}~8, we turn into a detailed investigation of the origin of these particle species. For this task, we consider a generic dark matter with mass of $1000$ GeV and focus on four annihilation channels: $q\bar{q}:~q=u,d,s,b$, $t\bar{t}$, $VV:~V=W,Z$ and $HH$. The reason is that at this mass value, the universality of the fragmentation function implies that all these annihilation channels have approximately the same features. In  the following we assume that $\sigma(\chi\chi \to WW) = \sigma(\chi\chi \to ZZ)$ without any loss of generality.

\begin{figure}[!t]
\centering
\includegraphics[width=0.9\linewidth]{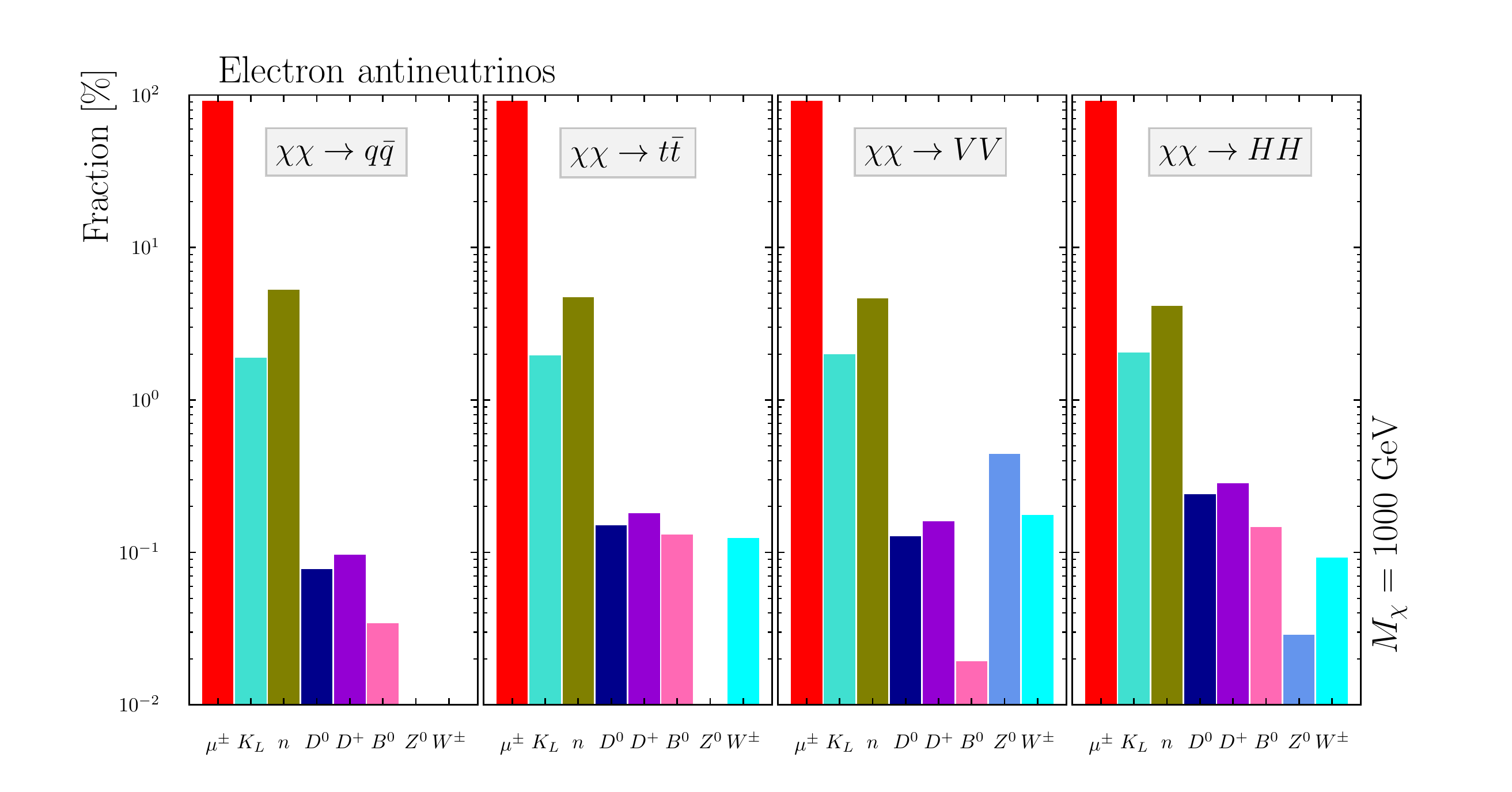}
\caption{The mean contribution to $\bar{\nu}_e$ production in DM annihilation with $M_{\chi} = 10000~{\rm GeV}$. From the left to the right we show the $q\bar{q}$, $t\bar{t}$, $VV$ and $HH$ channels. One shows $\bar{\nu}_e$ produced from the decay of muons (red), $K_L$ (turquoise), $n, \bar{n}$ (olive), $D^0$ (dark blue), $D^+$ (violet), $B^0$ (hot pink), $Z^0$ (light blue) and $W^\pm$ (cyan).}
\label{fig:origin:nue} 
\end{figure}

\begin{figure}[!t]
\centering
\includegraphics[width=0.9\linewidth]{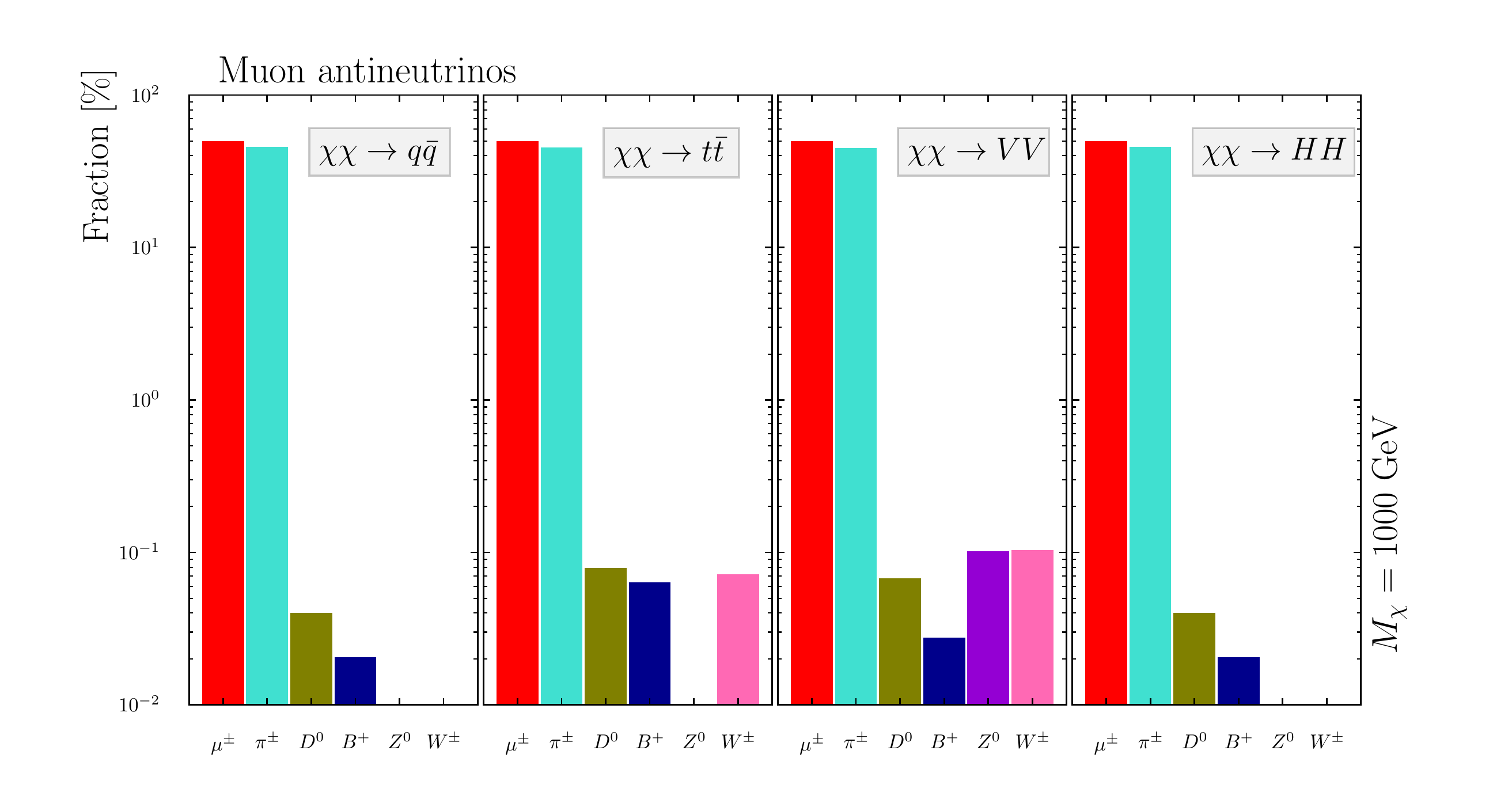}
\caption{Same as figure \ref{fig:origin:nue} but for $\bar{\nu}_\mu$. Here, $\bar{\nu}_\mu$ are produced from the decay of muons (red), $\pi^\pm$ (turquoise), $D^0$ (olive), $B^+$ (dark blue), $Z^0$ (purple) and $W^\pm$ (hot pink).}
\label{fig:origin:num}
\end{figure}

\begin{figure}[!h]
\centering
\includegraphics[width=0.9\linewidth]{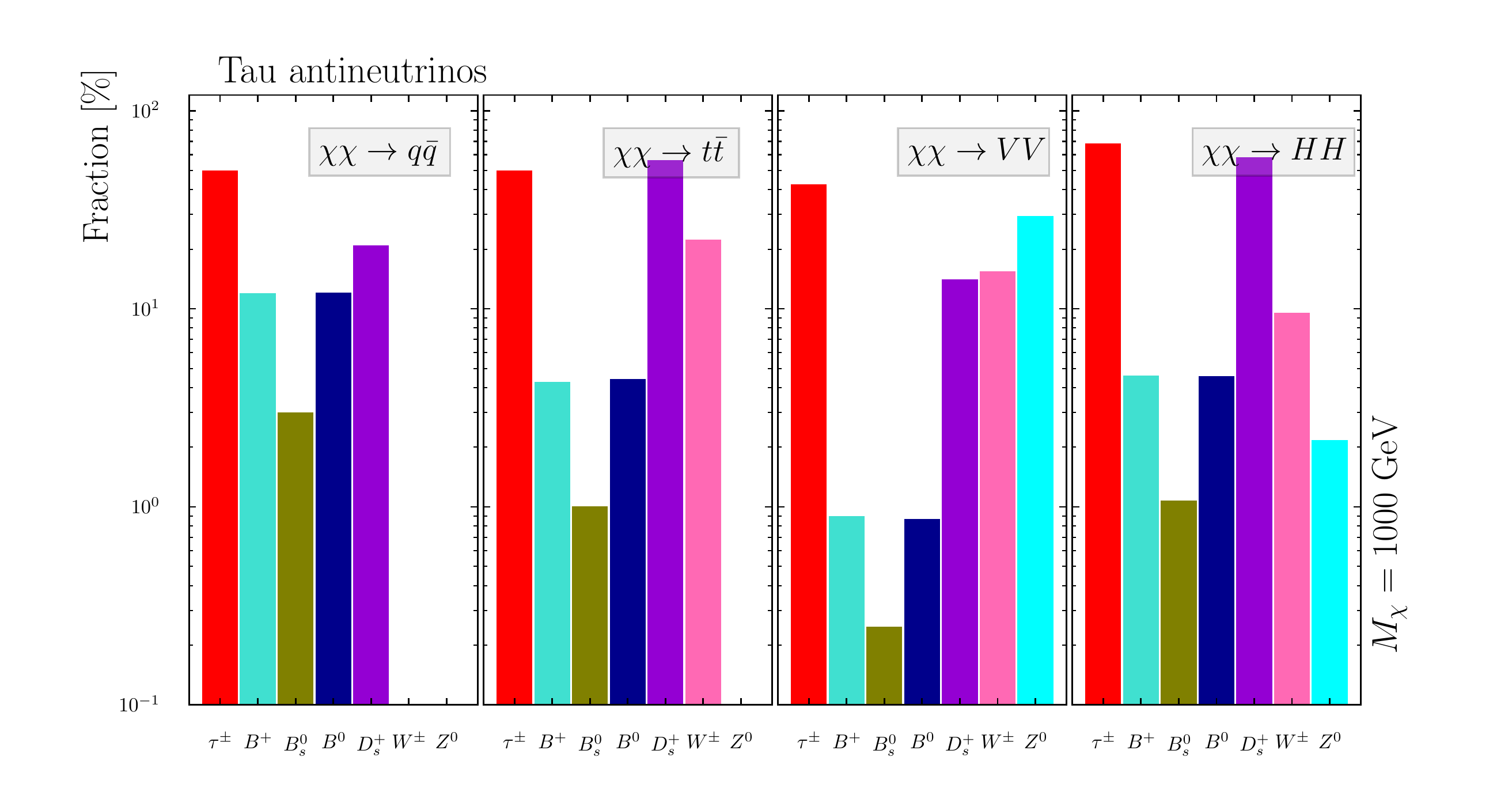}
\caption{Same as figure \ref{fig:origin:nue} but for $\bar{\nu}_\tau$. The $\bar{\nu}_\tau$ produced from the decay of $\tau^\pm$, $B^+$, $B_s^0$, $B^0$, $D_s^+$, $W^\pm$ and $Z^0$ are shown in red, turquoise, olive, dark blue, purple, hot pink, and cyan respectively.}
\label{fig:origin:nuta}
\end{figure}

\subsubsection{Antineutrinos}

We start with the spectra of antineutrinos and we split them into electron antineutrinos ($\bar{\nu}_e$), muon antineutrinos ($\bar{\nu}_\mu$) and tau antineutrinos ($\bar{\nu}_\tau$) and they are shown in figures \ref{fig:origin:nue}--\ref{fig:origin:nuta}. 

First, most of the $\bar{\nu}_e$ are coming from the decay of $\mu^\pm$ in $\mu^- \to e^- \bar{\nu}_e \nu_\mu$. The contribution of the muons to the rate of $\bar{\nu}_e$ is $90\%$ irrespective of the annihilation channel. This is followed by the contribution of (anti)-neutrons through $n \to p e^- \bar{\nu}_e$ which is about $4\%$--$5\%$ depending on the annihilation channel. The other contributions are small (below $\simeq 2\%$); one note among others $K_L$, $D^0$, $D^+$, $B^0$, $W^- \to e^- \bar{\nu}_e$ and $Z^0 \to \nu_e \bar{\nu}_e$. The last two contributions ($Z\to \nu_e$ and $W^+ \to \nu_e$) are possible in the $t\bar{t}$, $VV$ and $HH$ channels. We note that most of the muons that leads to the $\bar{\nu}_e$ are coming from the decays of charged pions, {\it i.e.}, $\pi^- \to \mu^- \bar{\nu}_\mu$. Charged pions are produced in abundance through fragmentation of quarks/gluons and/or decay of heavier hadrons \cite{Amoroso:2018qga}. Therefore, one can find a direct connection between the modeling of gamma rays and electron antineutrinos. 

The situation is slightly different for $\bar{\nu}_\mu$ where can see that both $\mu^+$ and $\pi^+$ contribute to about $47$--$50\%$ of the total rate while the other sources give negligible contributions (see figure \ref{fig:origin:num}). Similar to $\bar{\nu}_e$, most of the muons that give rise to $\bar{\nu}_\mu$ are coming from the decay of $\pi^+$. Finally, one note that $\bar{\nu}_\tau$ have extremely different type of sources -- the total rate of $\bar{\nu}_\tau$ is negligibly small as compared to $\bar{\nu}_e$ and $\bar{\nu}_\mu$ --. $\bar{\nu}_\tau$ are produced from $\tau^+$, $D_s^0$, $B_s^0$, $B^+$, $B^0$, $W^\pm$ and $Z^0$ with average contributions of about $1\%$--$50\%$ depending on the annihilation channel (see figure \ref{fig:origin:nuta}).

\begin{figure}[!t]
\centering
\includegraphics[width=0.95\linewidth]{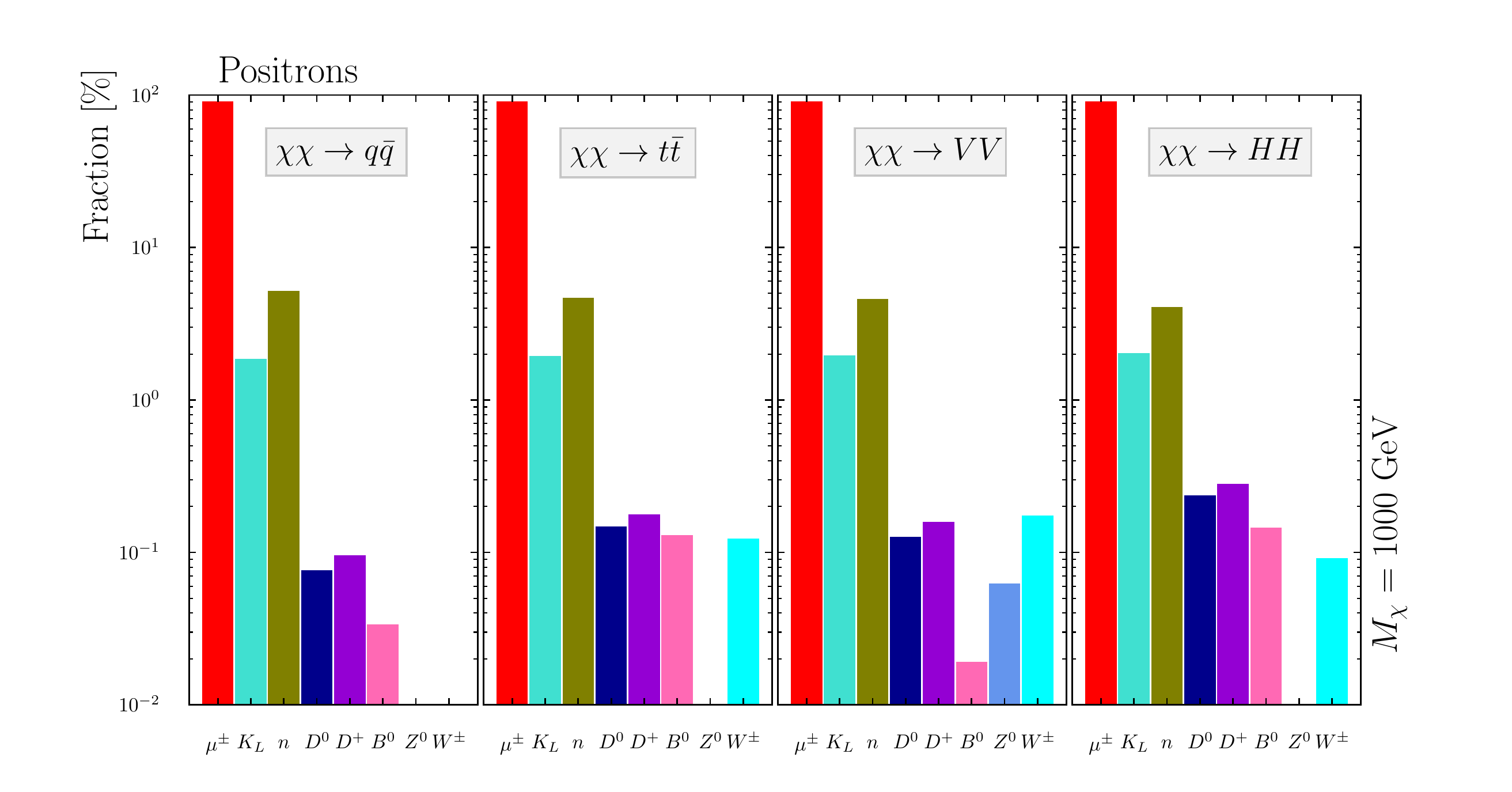}
\caption{Same as figure \ref{fig:origin:nue} but for $e^+$. Here, $e^+$ are produced from the decay of muons (red), $K_L$ (turquoise), $n$ (olive), $D^0$ (dark blue), $D^+$ (purple), $B^0$ (hot pink), $Z^0$ (light blue) and $W^\pm$ (cyan).}
\label{fig:origin:e}
\end{figure}

\subsubsection{Positrons}

In figure \ref{fig:origin:e}, we show the mean contribution of different particles to the rate of $e^+$ in generic DM annihilation for DM mass of $1000$~GeV. We can see that, similarly to $\bar{\nu}_e$ (see figure \ref{fig:origin:nue}), most of $e^+$ are produced from the decay of $\mu^+$ independently of the annihilation channel. The other contributions are similar to the case of $\bar{\nu}_e$ production. 

\subsubsection{Antiprotons}

\begin{figure}[!t]
    \centering
    \includegraphics[width=0.95\linewidth]{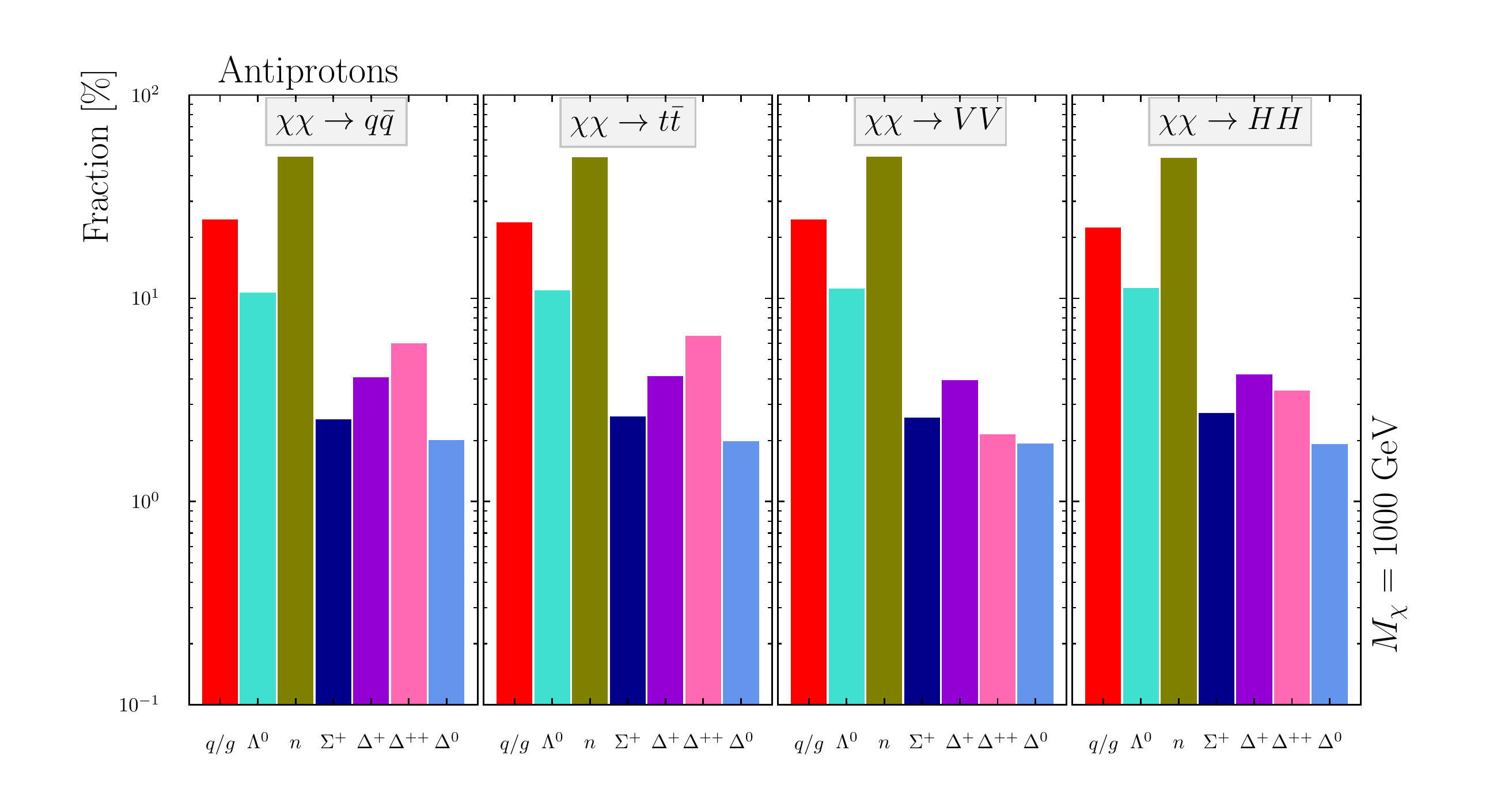}
    \caption{Same as in figure \ref{fig:origin:nue} but for $\bar{p}$. Here, one shows $p/\bar{p}$ produced from QCD fragmentation (red), decay of $\Lambda^0$ (turquoise), $n$ (olive), $\Sigma^+$ (dark blue), $\Delta^+$ (purple), $\Delta^{++}$ (hot pink) and $\Delta^0$ (light blue).}
    \label{fig:origin:p}
\end{figure}

In general (anti-)protons can be split into two categories: \emph{(i)} primary (anti-)protons produced directly from the string fragmentation of quarks and gluons and \emph{(ii)} secondary (anti-)protons produced from the decay of heavier baryons. In figure \ref{fig:origin:p}, we display the mean contributions to the (anti-)proton yields in DM annihilation for $M_\chi = 1000$ GeV. We can see that about $22\%$ of the produced (anti)protons are emerging from $q/g$-fragmentation. On the other hand, the dominant large fraction ($\simeq 50\%$) of (anti)-protons is originated from the decay of (anti-)neutrons in decays $\bar{n} \to \bar{p} e^+ \nu_e$ as can be clearly seen in figure \ref{fig:origin:p}. This is followed by the contribution of five baryons which mainly contribute to secondary protons: $\Lambda^0$, $\Sigma^+$ and $\Delta(1232)$ -- summing contributions from $\Delta^+, \Delta^{0}$ and $\Delta^{++}$ -- contribute to about $2\%$--$7\%$ of the produced (anti)-protons respectively. \\

We conclude that the spectra of antiparticles are correlated to each other with the dominant hadrons to be modeled properly are charged pions, protons and $\Lambda^0$ baryons. We note that in ref. \cite{Amoroso:2018qga}, we have discussed in detail the modeling of pion spectra at LEP and therefore we will not discuss it anymore here (although we do include these measurements in our fits). In the next section, we discuss in detail the measurements of baryon spectra at LEP and assess the agreement between the corresponding theory predictions and experimental measurements for three multi-purpose MC event generators.  
\section{Experimental measurements and generator predictions}
\label{sec:measurements}
\subsection{Introduction}
From the discussion in section \ref{sec:physics}, it is clear that the modeling of the spectra of anti-protons will be improved if one includes all the relevant measurements of proton spectrum performed at LEP. Besides the measurements of the proton spectrum itself, one may expect some improvements from measurements of the spectra of the following baryons:
\begin{itemize}
    \item {\it{$\Lambda^0$}}: $\Lambda^0$ is the dominant source of secondary protons at LEP (about $22\%$ of the total protons are coming from $\Lambda^0$ baryons). The mean multiplicity of $\Lambda^0$ was measured by several collaborations at LEP: $\langle n_{\Lambda^0} \rangle = 0.357\pm 0.017$ \cite{Abreu:1993mm}, $\langle n_{\Lambda^0} \rangle = 0.348\pm 0.013$ \cite{Abreu:1996na} performed in $1993$ and $1996$ respectively. We note that $\Lambda^0$ baryons decay with $63.8\%$ branching ratios into $p\pi^-$ \cite{ParticleDataGroup:2020ssz} and therefore we expect a strong correlation between the scaled momentum of $\Lambda^0$ and of $p/\bar{p}$ since most of $\Lambda^0$ baryons are reconstructed using tracks identified with charged pions and protons. 
    
    \item {\it{$\Delta^{++}$}}: About $11\%$ of protons at the $Z$-pole are produced from the decay of the $\Delta^{++}$ which decays with $100\%$ branching ratio into $p\pi$. Given that $\Delta^{++}$ and $p$ are members of the same multiplet, we may expect important constraints on $p/\bar{p}$ due to isospin. However, there are only two measurements of $\Delta^{++}$ performed by \textsc{Delphi} \cite{DELPHI:1995ysj} and \textsc{Opal} \cite{OPAL:1995otk}. These measurements suffer from large uncertainties due to the difficulty in isolating the $\Delta^{++}$ signal from the overwhelming backgrounds. We, therefore, do not include these measurements in the fits.

    \item {\it{$\Sigma^\pm$}}: About $5\%$ of protons at LEP are coming from the decays of $\Sigma^+$. The corresponding branching ratio is ${\rm BR}(\Sigma^+ \to p\pi^0) = 51.57\%$ \cite{ParticleDataGroup:2020ssz}. There are two measurements of $\Sigma^+$ scaled momentum at LEP by \textsc{Delphi} \cite{DELPHI:2000oqt} and \textsc{Opal} \cite{OPAL:1996dbo} collaborations. These measurements will not be used in the fit for the same arguments we used to exclude $\Delta^{++}$ measurements.
\end{itemize}

Therefore, the main constraining observables in this study will consist of a set of measurements of $\Lambda$ and $p/\bar{p}$ energy--momentum distributions. To guarantee a good agreement with the results of the previous study \cite{Amoroso:2018qga}, we also include measurements of meson spectra, event shapes and particle multiplicities. Before going into a discussion of the setup used in this study, we discuss briefly the various measurements performed by LEP at the $Z$--pole using data collected between 1992 and 1999\footnote{The measurements reported on by the experimental collaborations at LEP of the $\Lambda$ or $p/\bar{p}$ spectra correspond the measured variables being either $x_p = |p_{\rm hadron}|/|p_{\rm beam}|$ (\textsc{Aleph} \cite{Barate:1996fi} and \textsc{Delphi} \cite{Abreu:1993mm, Abreu:1995cu, Abreu:1998vq}), $x_E = E_{\rm hadron}/|p_{\rm beam}|$ (\textsc{Opal} \cite{Alexander:1996qj}), $\xi = \log(1/x_p)$ (\textsc{Aleph} \cite{Barate:1999gb}) or $|p_{\rm hadron}|$ (\textsc{Delphi} \cite{Abreu:1995cu, Abreu:1998vq} and \textsc{Opal} \cite{Akers:1994ez}).}. 


\subsection{Note about the relevant measurements}

\paragraph{{\bf ALEPH~(1996--2000).}} The \textsc{Aleph} collaboration has reported on three measurements of $\Lambda$ scaled momentum -- $x_p$ in \cite{Barate:1996fi} and $\xi$ in \cite{Barate:1999gb} -- and one measurement of $p/\bar{p}$ scaled momentum \cite{Barate:1996fi}. The first measurements rely on the initial data which contain 520,000 inclusive hadronic events \cite{Barate:1996fi}. In a second paper published in 2000, the \textsc{Aleph} collaboration used a more complete dataset consisting of 3.7 million hadronic events \cite{Barate:1999gb}. The measurement of identified particle spectra by \textsc{Aleph} can be made by a simultaneous measurement of the hadronic momentum and ionisation energy loss ${\rm d}E/{\rm d}x$ in the Time Projection Chamber (TPC). There are holes in the reported measurement of $p/\bar{p}$ spectrum in the regions $x_p \in [1.8, 2.4] \times 10^{-2} \cup [2.8, 7] \times 10^{-2}$ due to the strong overlap between the bands of $p/\bar{p}$ and of the other hadrons ($\pi^\pm, K^\pm$). Most of the systematic uncertainties can be roughly categorised into three components:  track reconstruction efficiencies, efficiencies in the $\Lambda^0$ reconstruction and the background calibration. These errors have been corrected for by using the predictions of \textsc{JetSet} \cite{Sjostrand:1993yb}. We note that these correction factors can be large in some kinematics regions.

\begin{figure}[!t]
    \centering
    \includegraphics[width=0.49\linewidth]{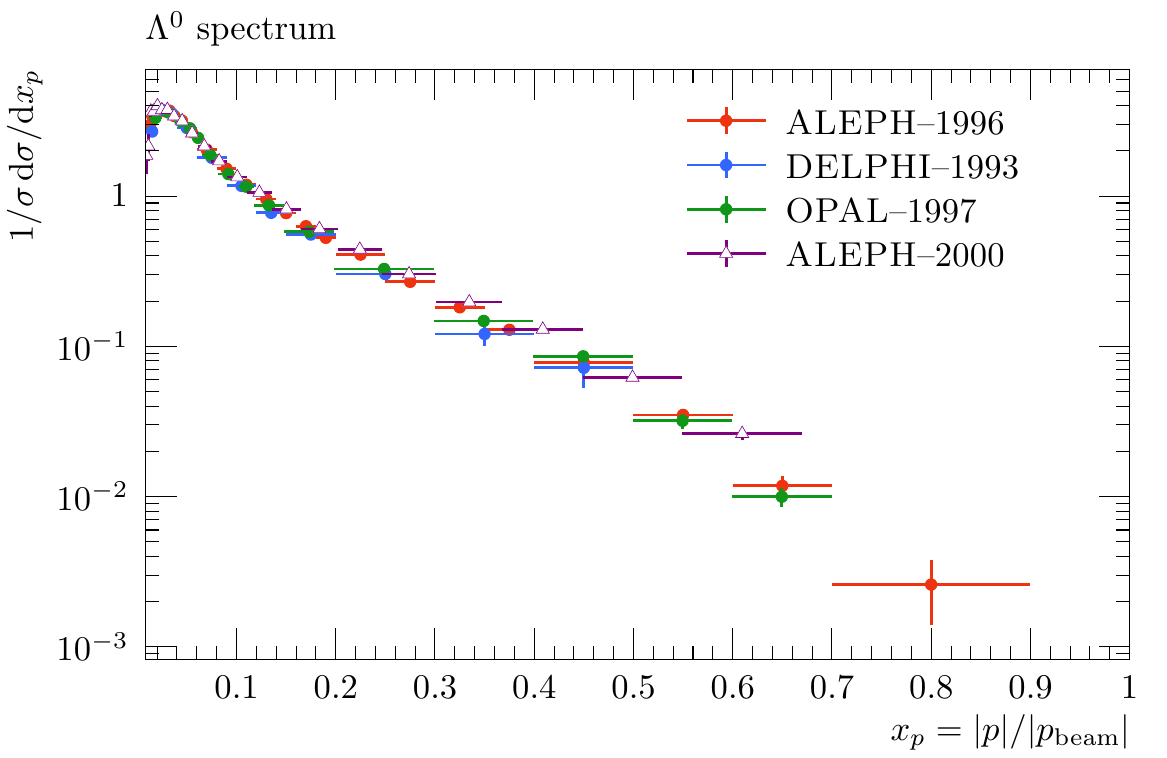} \hfill
    \includegraphics[width=0.49\linewidth]{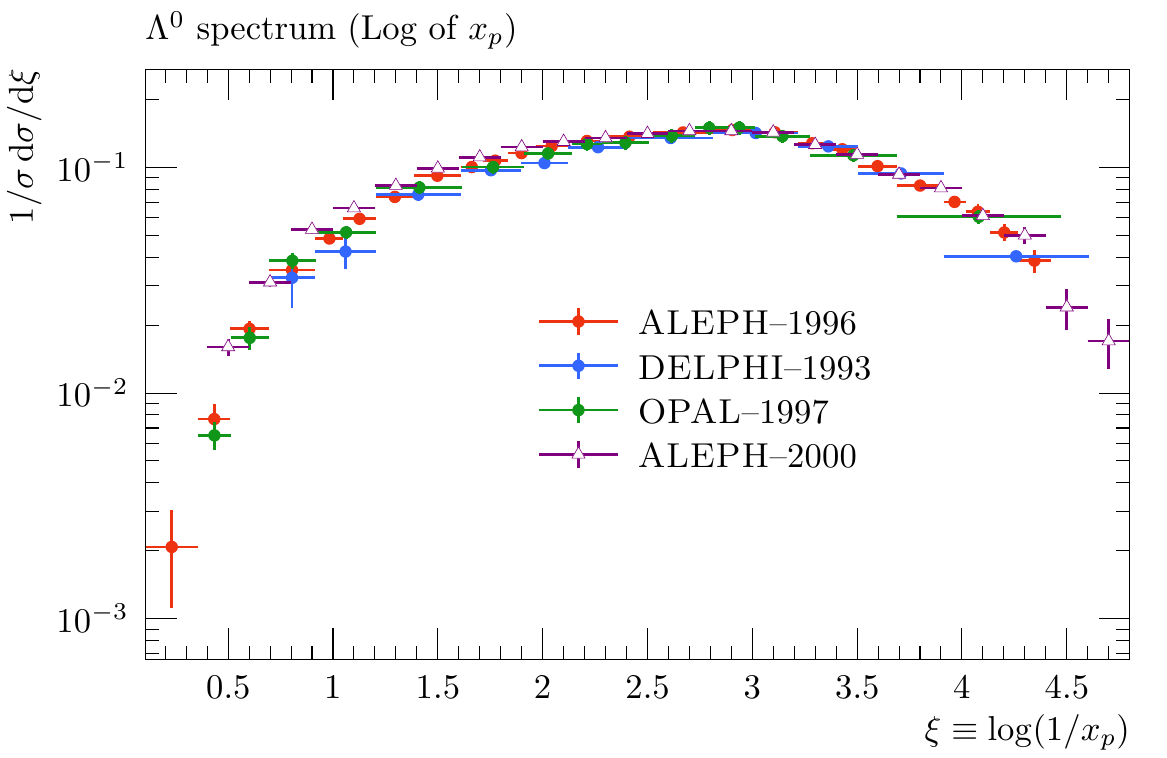}
    \vfill
    \includegraphics[width=0.49\linewidth]{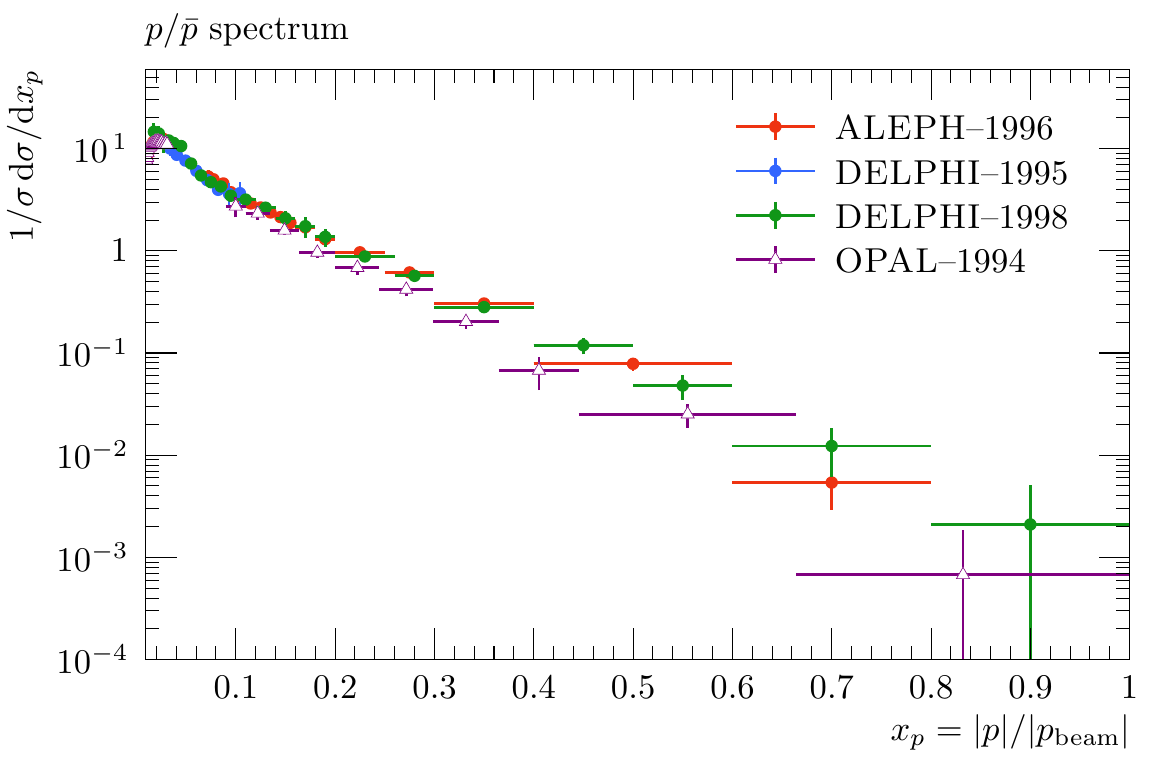}
    \hfill
    \includegraphics[width=0.49\linewidth]{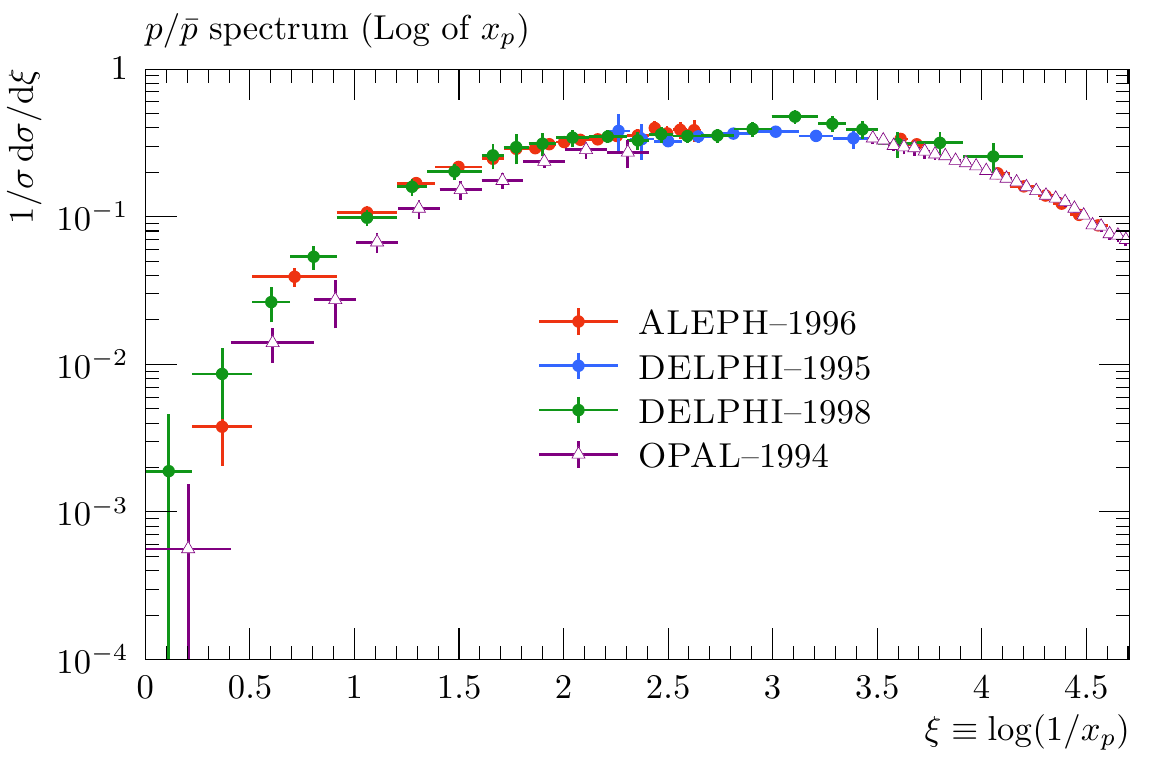}
    \caption{{\it Upper panels:} Comparison between the different measurements of the $\Lambda^0$ scaled momentum $x_p$~({\it left}) and $\log(1/x_p)$~({\it right}). Here, we show the results from \textsc{Aleph} (red and violet), \textsc{Delphi} (blue) and  \textsc{Opal} (green). {\it Bottom panels}: Same as in the upper panels but for the $p/\bar{p}$ scaled momentum ({\it left}) and its logarithm ({\it right}). For the proton case, we show the results from \textsc{Aleph} (red), \textsc{Delphi} (blue and green) and \textsc{Opal} (violet). In both panels, the errors correspond to statistical and systematic uncertainties summed up in quadrature. Data is taken from \cite{Abreu:1993mm, Barate:1996fi, Akers:1994ez, Abreu:1995cu, Abreu:1998vq, Abe:1998zs, Barate:1999gb, Abe:2003iy}.}
    \label{fig:comparisons}
\end{figure}  

\paragraph{{\bf DELPHI~(1993--1998).}} The \textsc{Delphi} collaboration has performed a detailed analysis of the scaled momentum of $\Lambda^0$ and $p/\bar{p}$ using data collected in the period of 1991--1998 \cite{Abreu:1993mm, Abreu:1995cu, Abreu:1998vq}. In a first measurement of $\Lambda^0$ spectrum and $\Lambda^0$--$\bar{\Lambda}^0$ correlations, 993,000 hadronic events were used \cite{Abreu:1993mm}. In \cite{Abreu:1995cu}, a measurement of proton momentum for $x_p \in [0.03, 0.1]$ has been carried using 17000 hadronic events. This measurement has been superseded by a more recent one which relied on a larger statistical sample consisting of 1,400,000 hadronic events and covering a wider range of proton momenta, {\it i.e.} $x_p \in [1.53 \times 10^{-2}, 1]$ \cite{Abreu:1998vq}. Contrarily to \textsc{Aleph}, the reconstruction of the particle momenta in \textsc{Delphi} is based on the measurement of the ionization angle in the Ring Imaging CHerenkov (RICH) detector\footnote{Note that the Cherenkov angle has a dependence on the particle mass and the refractive index of the radiator -- two radiators have been used in this analysis --. The probability density of observing the measured Cherenkov angle $\theta_C^i$ for a track "$i$" depends on various parameters and was fitted taking into account three particle species $\pi$ (which also includes electrons and muons since they cannot be distinguished from pions with this method), $K^\pm$ and $p/\bar{p}$. Finally, the Likelihood function includes an additional constant term which depends on the noise.} . A number of selection cuts have been applied to improve the quality of particle identification in RICH (see section 3 of Ref.~\cite{Abreu:1995cu} for more details). The fraction of $p/\bar{p}$ particles were determined from a fit to the Cherenkov angle distribution in specific momentum ranges. The systematic errors mainly arise from the parametrization of the backgrounds. To account for these errors, correction factors ranging from $1\%$ to $10\%$ have been applied using the MC simulations of \textsc{JetSet} event generator. 

 
\paragraph{{\bf OPAL~(1994--1997).}} \textsc{Opal} has measured the spectra of charged hadrons using $\mathcal{L} =24.9~{\rm pb}^{-1}$  of data collected in 1992 \cite{Akers:1994ez} and of strange baryons using approximately 3.65 million events collected between 1990--1994 \cite{Alexander:1996qj}. The determination of the hadron yields has been done from the simultaneous measurement of the track momentum and differential energy loss. Since the identification of charged hadrons cannot be done unambiguously, the \textsc{Opal} collaboration has used a statistical method to fit the number of particles measured in the data. Correction factors of order $20$--$30\%$ have been applied to account for effects of geometrical and kinematical acceptance, nuclear corrections and decay in flight \cite{Akers:1994ez}. The $\Lambda^0$ baryons have been reconstructed from the tracks associated to their decay products ($p\pi$) using two methods optimised to either have a good mass and momentum resolution or optimised to give a higher efficiency over a broader $\Lambda^0$ momentum range. The total systematic uncertainty is about $2.7\%~ (3.3\%)$ for the first (second) method while the statistical uncertainties are subleading.

\subsection{Conclusions about the included measurements}

In Fig.~\ref{fig:comparisons}, we show the comparisons between the different experimental measurements of $\Lambda^0$~spectrum and $p/\bar{p}$ spectrum. As was pointed out previously, the measurements of baryon spectra were presented for different variables. In order to be able to easily compare between the different measurements, we scale all the normalized distributions to be either in $x_p$ or in $\log(1/x_p)$ in case they depend on a different variable, {\it i.e.} under the change of variable $x_E \to x_p$, the differential normalized cross section changes as $1/\sigma {\rm d}\sigma/{\rm d}x_E \to |J|^{-1} 1/\sigma {\rm d}\sigma/{\rm d}x_p$ with $J = \partial x_E/\partial x_p$ being the Jacobian of the transformation.  The normalized cross sections in $x_p$ are shown in the left panels of Fig. \ref{fig:comparisons} to identify differences in the tails of the scaled momentum. On the other hand, the normalized cross sections in $\log(1/x_p)$ are very useful to display the differences in the bulk and the peak regions  (right panels of Fig. \ref{fig:comparisons}). We can see the following differences between the different measurements:
\begin{itemize}
    \item There are some tensions between the measurement of the scaled momentum of $p/\bar{p}$ performed by \textsc{Opal} and the other experiments for $x_p > 0.1$ (the \textsc{Opal} result is below all the others).
    \item The old \textsc{Delphi} measurement (blue) of $p/\bar{p}$ momentum is inconsistent with the new one (green) for few bins of $\xi \simeq 3$--$3.2$. Note that both these \textsc{Delphi}~measurements cover the hole left by \textsc{Aleph}--1996. Furthermore, the trend of the data seems to be more consistent with \textsc{Delphi}--1998 rather than \textsc{Delphi}--1995 (very large corrections have been applied to the proton momentum in the \textsc{Delphi}--1995 measurement). 
    \item The \textsc{Delphi}--1993 measurement of $\Lambda^0$ scaled momentum seems to be inconsistent with the others for for $\xi < 1.1$ (the discrepancy is mild as compared to the proton case).
\end{itemize}

\begin{figure}[!t]
\centering
\includegraphics[width=0.49\linewidth]{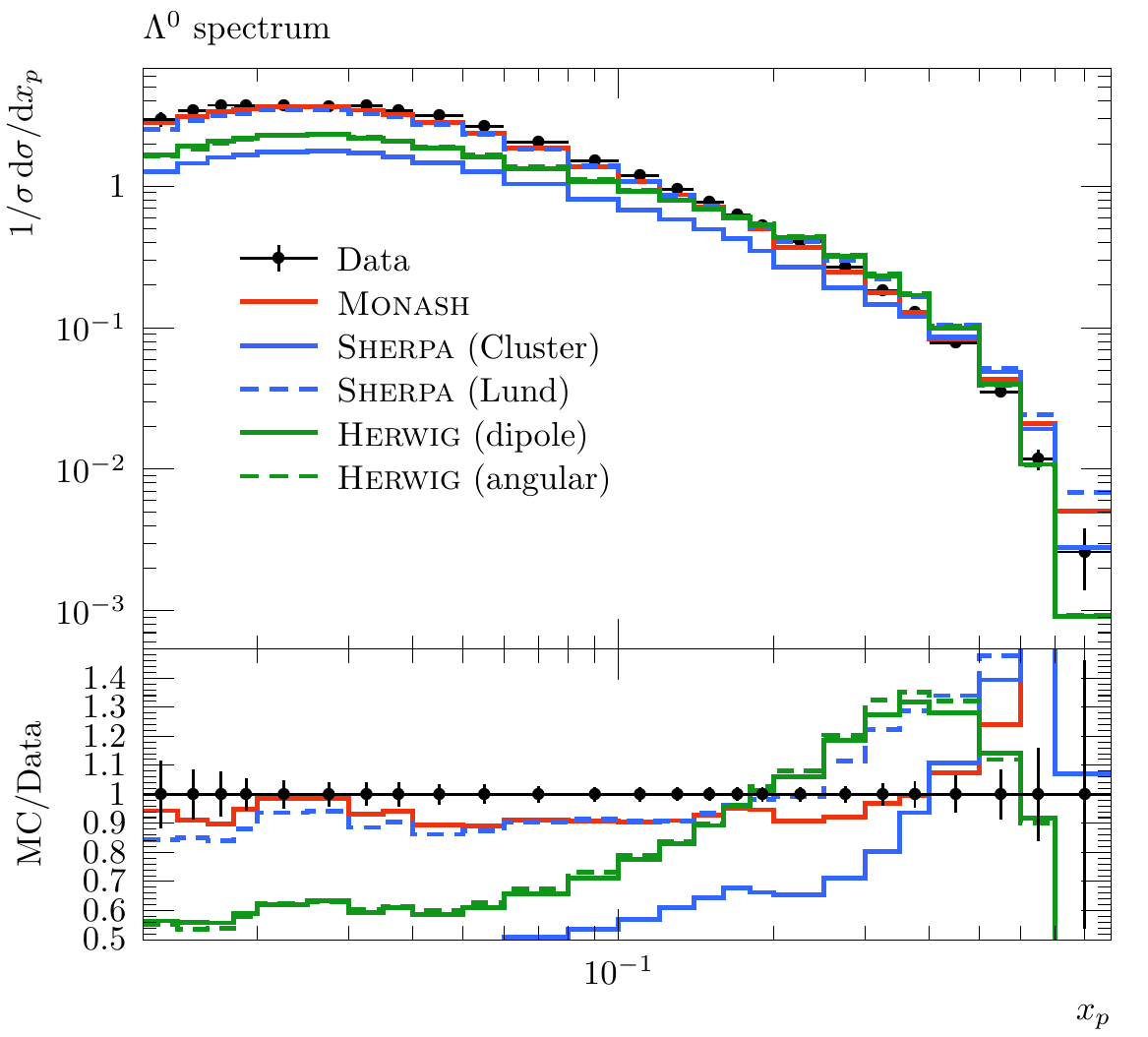}
\hfill
\includegraphics[width=0.49\linewidth]{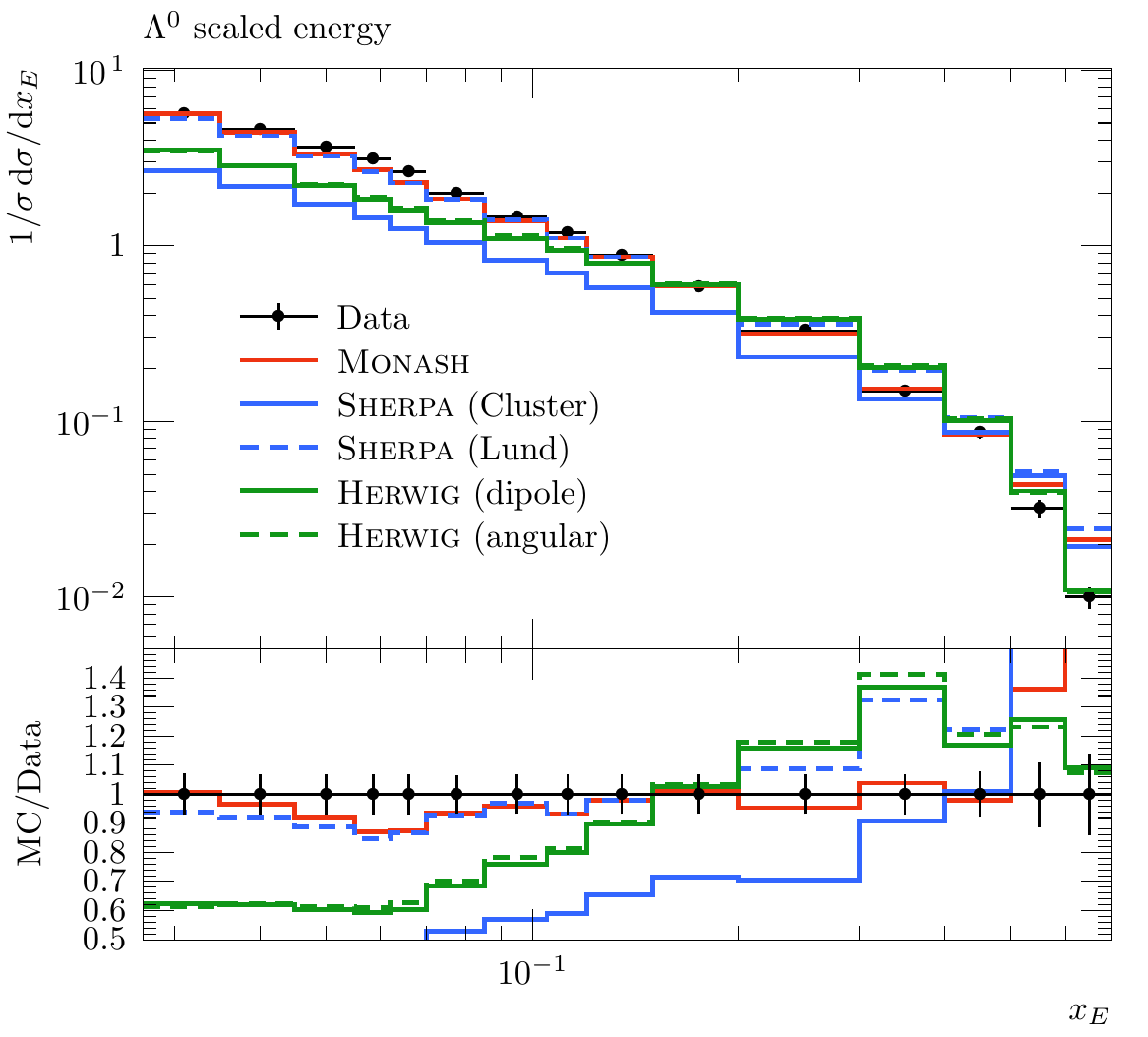}
\vfill 
\includegraphics[width=0.49\linewidth]{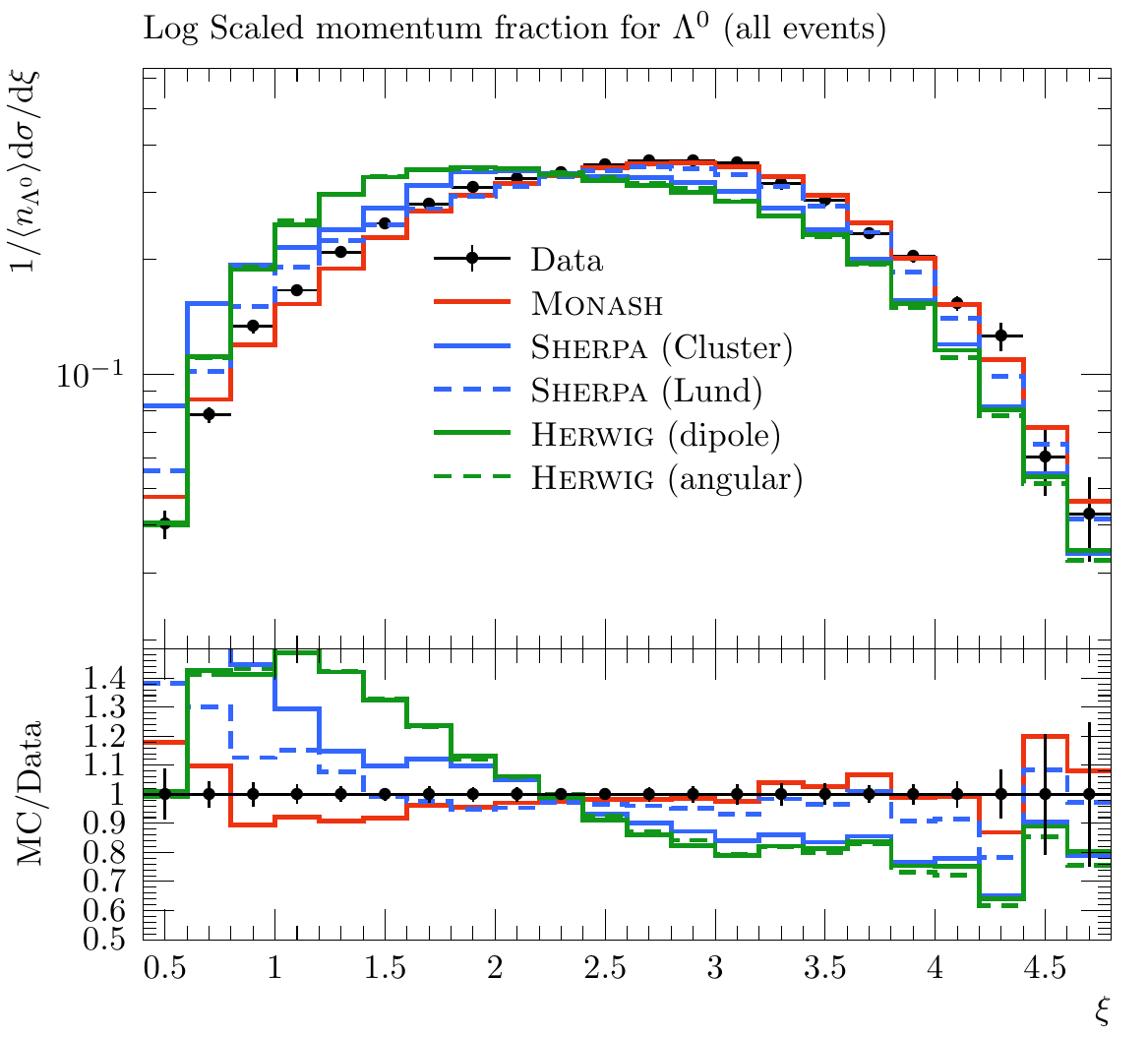}
\hfill
\includegraphics[width=0.49\linewidth]{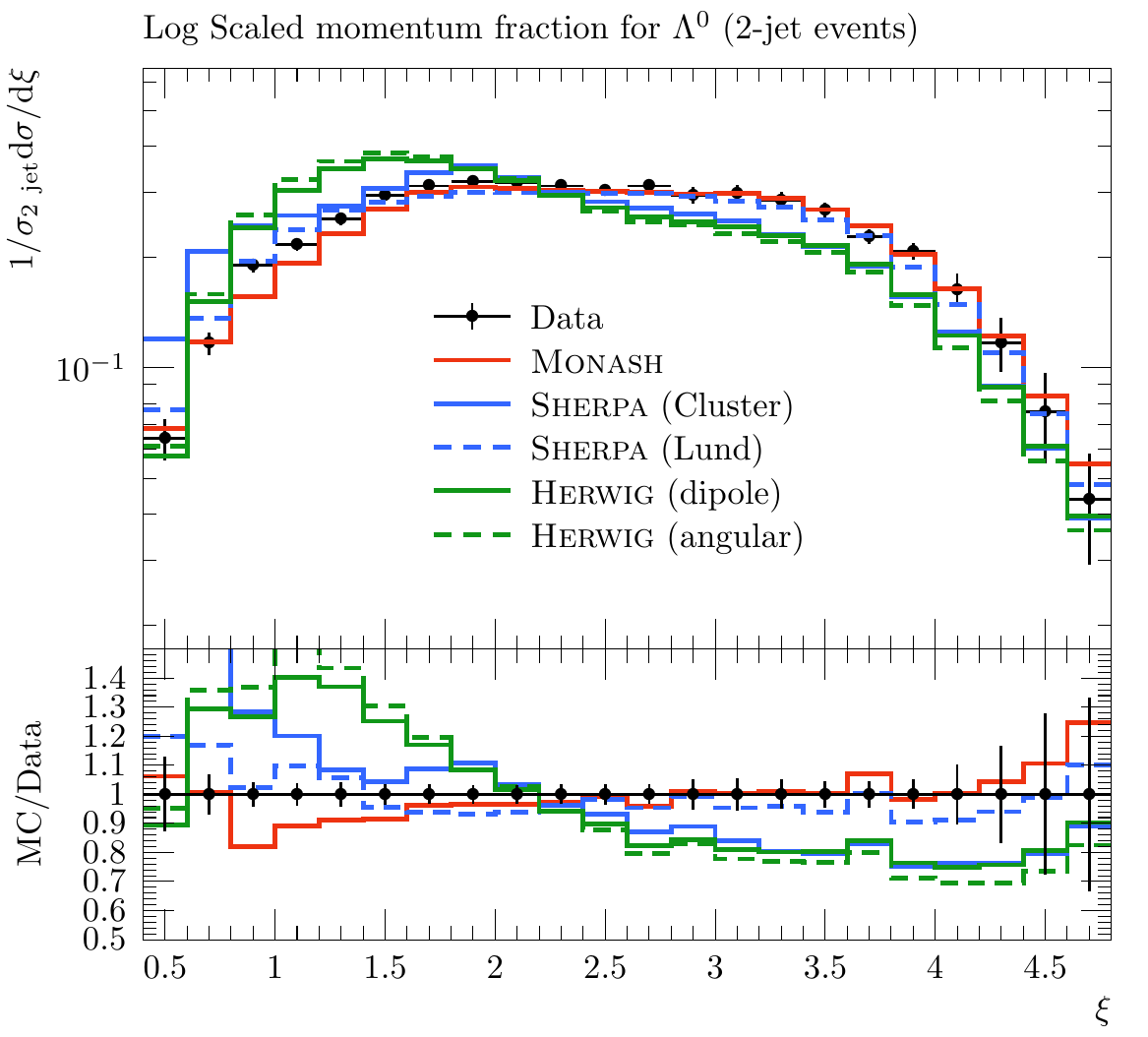}
\caption{Comparison between theory predictions and experimental measurements of $\Lambda^0$ (scaled)-momentum distribution at LEP. Here we show the $\Lambda^0$ spectrum (left upper panel) \cite{Barate:1996fi}, $\Lambda^0$ scaled energy (right upper panel) \cite{Alexander:1996qj}, Log of scaled momentum for $\Lambda^0$ in all events (left lower panel) and in $2$--jet event (right lower panel) \cite{Barate:1999gb}. The \textsc{Pythia}~8 prediction is shown in red, the \textsc{Sherpa}~2 predictions are shown in blue for the cluster (solid) and Lund (dashed) models while the \textsc{Herwig}~7 is shown in green for dipole (solid) and angular-based (dashed) shower algorithms.}
\label{fig:generators:comparison:Lambda}
\end{figure}

Based on these observations, the tunes of the fragmentation function are not expected to find a robust best-fit point for $\texttt{StringZ:aExtraDiquark}$ due to the mutual tensions between some of the measurements unless some of them are discarded in our fits. We, first, tune individually to various measurements and display the results as they are. In the more general tune that includes also the measurements of the meson scaled momenta, event shapes and mean multiplicities, we do not include the measurements of $p/\bar{p}$ and $\Lambda^0$ scaled momenta performed by \textsc{Aleph}--1996 \cite{Barate:1996fi} since the former has a hole in the peak region while the latter is superseded by \textsc{Aleph}--2000 measurement \cite{Barate:1999gb}. On the other hand, we do not include \textsc{Opal}--1994 measurement as it is inconsistent with all the other measurements for $x_p > 0.1$ (\textsc{Pythia}~8 cannot find a good agreement for this region for any choice of fragmentation function parameters). Finally, \textsc{Delphi}--1995 measurement of proton spectrum is not included in the four-dimensional parameter space tune as it was superseded by the \textsc{Delphi}--1998 measurement.

\subsection{How good are the current theory predictions?}
\label{sec:MCgenerators}
We discuss in this section the level of agreement between the theory predictions of the three commonly used multi-purpose Monte Carlo event generators and the experimental measurements of the baryon spectra at LEP (proton and $\Lambda^0$). For this task, we use the \textsc{Pythia}~8 version 307 with the baseline Monash tune \cite{Skands:2014pea}, \textsc{Sherpa} version 2.2.12 \cite{Gleisberg:2008ta} with a shower model based on the Catani-Seymour subtraction (CSS) method \cite{Schumann:2007mg} and two hadronisation models: the cluster model which is provided by \textsc{Ahadic}++ \cite{Winter:2003tt} (the default in \textsc{Sherpa}) and the Lund string model based on \textsc{Pythia}~6 \cite{Sjostrand:2006za} and \textsc{Herwig} version 7.2.3 \cite{Bellm:2015jjp} with two radiation models: the angular-based parton-shower algorithms \cite{Gieseke:2003rz} and the dipole-based algorithm \cite{Platzer:2009jq, Platzer:2011bc}. It is found that the different parton-shower algorithms in the three MC event generators yield similar predictions in several collider observables within the theory uncertainties (see chapter V.I. of ref. \cite{Andersen:2016qtm}). In ref. \cite{Amoroso:2018qga}, we have found that the three MC event generators agree pretty well in various experimental measurements of event shapes, pion and photon spectra.  \\ 

\begin{figure}[!t]
    \centering
    \includegraphics[width=0.49\linewidth]{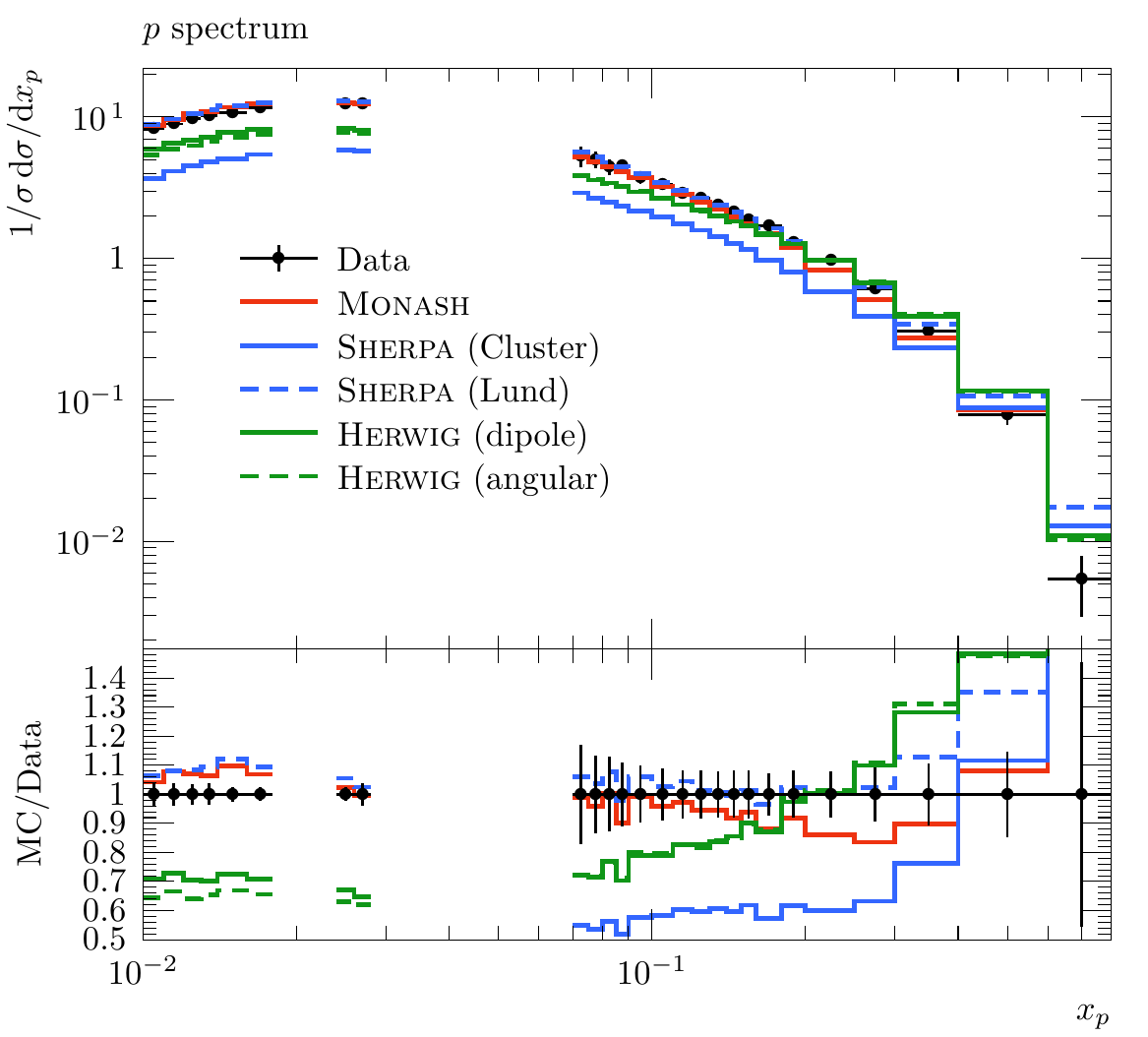}
    \hfill
    \includegraphics[width=0.49\linewidth]{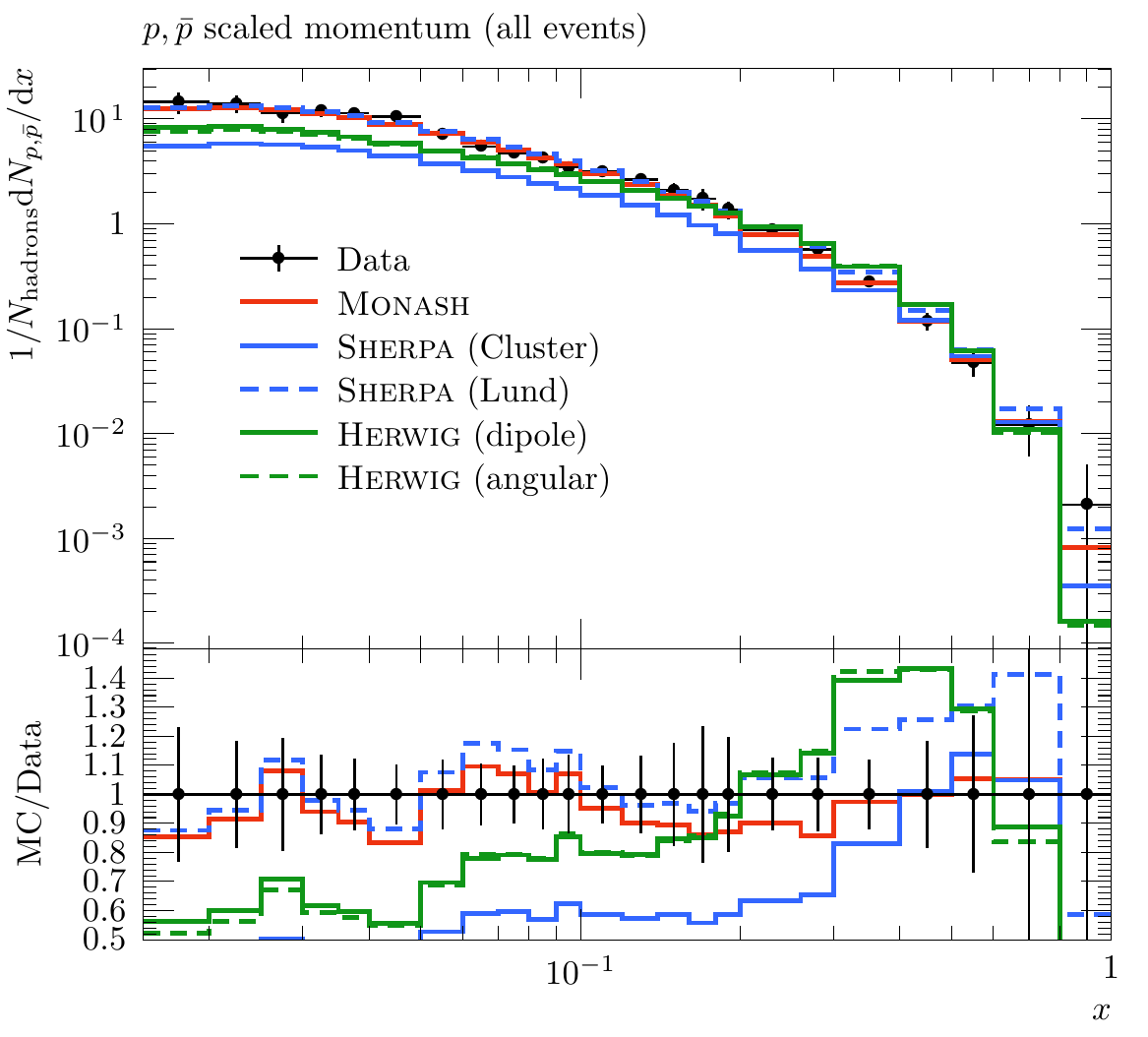}
    \caption{Same as figure \ref{fig:generators:comparison:Lambda} but for $p$ spectrum. Here we show the $p$ spectrum (left panel) \cite{Barate:1996fi} and $p,\bar{p}$ scaled momentum (right panel) \cite{Abreu:1998vq}.}
    \label{fig:generators:comparison:p}
\end{figure}

In figures \ref{fig:generators:comparison:Lambda} and \ref{fig:generators:comparison:p}, we show the comparison between the aforementioned generators for a set of selected measurements of proton and $\Lambda^0$ spectra at LEP. We can see that the generators based on the cluster model {\it i.e.} \textsc{Herwig}~7 and \textsc{Sherpa}~2 do not agree with data and have discrepancy of more than $40\%$ in some regions with respect to the experimental measurement. On the other hand, \textsc{Sherpa}~2 with the Lund model agrees quite well with \textsc{Pythia}~8 as well as with data.

\section{Fitting procedure}
\label{sec:setup}

\begin{table}[t!]
\setlength\tabcolsep{8pt}
  \begin{center}
    \begin{tabular}{llcll}
      \toprule
      parameter & \textsc{Pythia8} setting & Variation range & \textsc{Monash} & Tune-3D \cite{Amoroso:2018qga} \\
      \midrule
      $\sigma_{\perp}$~[GeV] & \verb|StringPT:Sigma|   & 0.0 -- 1.0 & 0.335 & 0.3174 \\
      $a$              & \verb|StringZ:aLund|    & 0.0 -- 2.0 & 0.68 & 0.5999 \\
      $b$              & \verb|StringZ:bLund|    & 0.2 -- 2.0 & 0.98 & -- \\
      $\left<z_\rho\right>$       & \verb|StringZ:avgZLund| & 0.3 -- 0.7 & (0.55) & 0.5278 \\
      $a_{\rm DiQuark}$ & \verb|StringZ:aExtraDiquark| & 0.0 -- 2.0 & 0.97 & 0.97 \\
      \bottomrule
    \end{tabular}
  \end{center}
  \caption{\label{tab:ranges} Parameter ranges used for the \textsc{Pythia} 8 tuning,
    and their corresponding values in the \textsc{Monash} tune \cite{Skands:2014pea} and in a tune performed in \cite{Amoroso:2018qga}.}
\end{table}

In this study, we use \textsc{Pythia8} version 8.244 \cite{Sjostrand:2014zea} to generate Monte Carlo samples for different values of the string fragmentation function parameters assuming the \textsc{Monash} tune \cite{Skands:2014pea} as our baseline. The different measurements used in our tunings are implemented in the validation package \textsc{Rivet} version 3.1.3 \cite{Buckley:2010ar, Bierlich:2019rhm}. Frequentist-type tunings are performed using the optimisation tool \textsc{Professor} version 2.3.3 \cite{Buckley:2009bj}.  Analytical expressions for the physical dependence of the observables on the different parameters are derived by fitting the Monte Carlo predictions to a set of points in the parameter space (called anchor points). The best-fit points for the parameters are determined by a standard $\chi^2$-minimisation method -- \textsc{Minuit} \cite{James:1975dr} --  implemented in \textsc{Professor} and which uses the analytical polynomial 
interpolations.

The parameters of interest are four which are $a$, and $\langle z_\rho\rangle$ which controls the longitudinal momentum taking by the hadrons inside the QCD jets, $a_{\rm Diquark}$ which is mainly connected to production of baryons in QCD jets. Finally, the $\sigma$ parameter is connected to the mean transverse momentum taken away by a hadron in the fragmentation process (see Table \ref{tab:ranges} for their default values and their allowed range in \textsc{Pythia8}). Two baseline retunings are performed throughout this study: 
\begin{itemize}
    \item In the first tune, we fix the parameters $a, \langle z_\rho \rangle$ and $\sigma$ to the values derived in a previous study \cite{Amoroso:2018qga}, \emph{i.e.}
    \begin{eqnarray}
    \verb|StringZ:aLund| &=& 0.5999^{+0.2000}_{-0.2000}, \nonumber \\
    \verb|StringZ:avgZLund| &=& 0.5278^{+0.0270}_{-0.0230}, \\
    \verb|StringPT:sigma| &=& 0.3174^{+0.0420}_{-0.0370}, \nonumber
    \label{eq:tune2018}
    \end{eqnarray}
    and tune $a_{\rm Diquark}$ to a set of constraining measurements compromising of proton, and $\Lambda^0$ spectra. This tuning is called a one-dimensional tuning. The set of the measurements used in this optimisation part is summarised in Table \ref{tab:measurements:protons}. 
    
    \item In a second tune, we fit the four-dimensional parameter space using the measurements listed in Tables \ref{tab:measurements:protons}-\ref{tab:measurements:eventshapes}. The additional measurements used in the tunes include meson scaled momenta, event shapes, jet rates, and charged multiplicities, and identified particle multiplicities.
\end{itemize}

For comparison and as an extra check of the results of the frequensit fit, we further perform a Bayesian fit using \textsc{MultiNest} \cite{Feroz:2008xx}. To increase the precision of the posteriors we generate one thousand live points with a tolerance of $10^{-3}$. Since the probability distribution functions (PDFs) associated to data are unimodal Gaussian PDFs, we expect that the results of frequentist and Bayesian fits to have a perfect agreement. 

\begin{figure}[!t]
    \centering
    \includegraphics[width=0.49\textwidth]{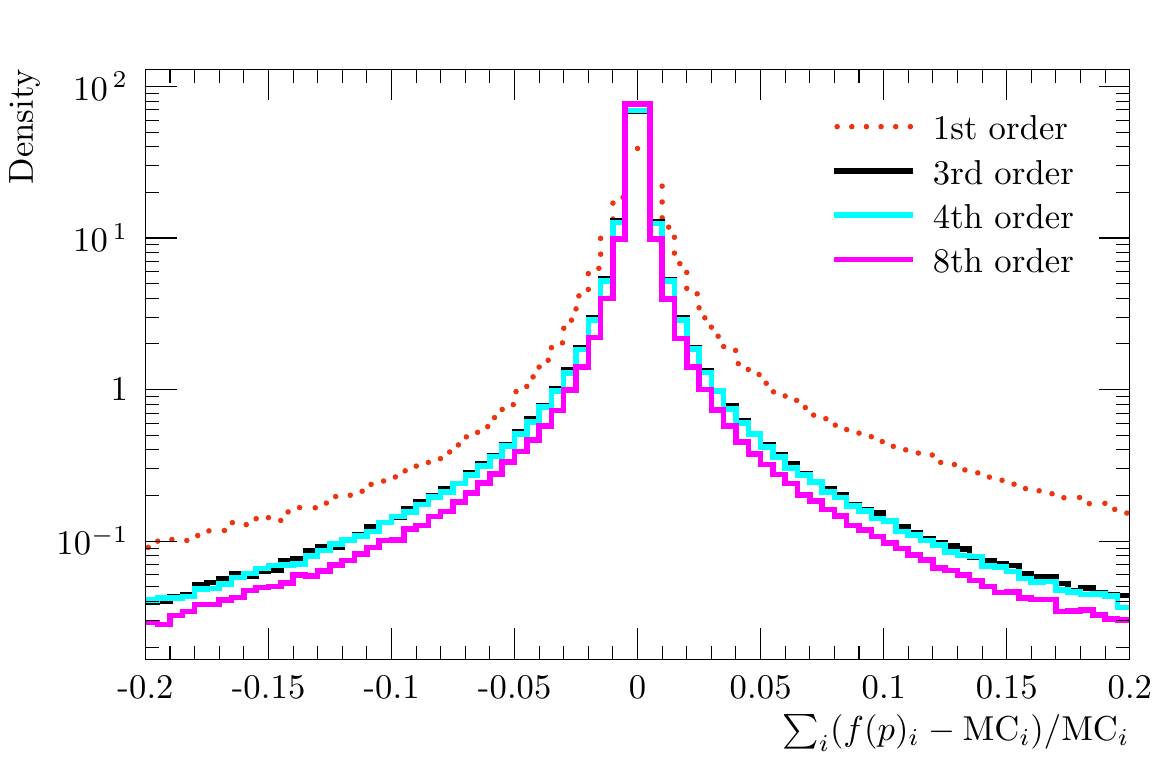}
    \hfill
    \includegraphics[width=0.49\textwidth]{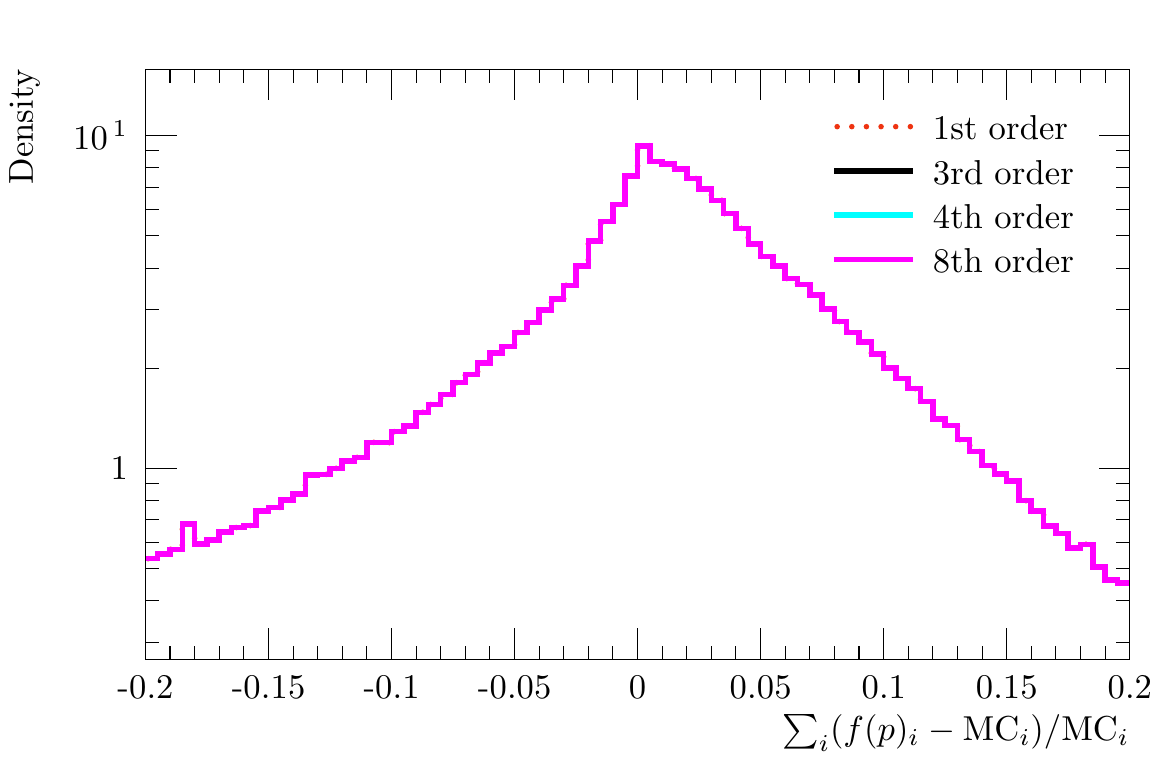}
    \caption{The distributions for the interpolated observable values ({\it left}) and their errors ({\it right}). These distributions are evaluated over all the 1000 MC runs and sum over all the observables used in the fit (defined in Tables \ref{tab:measurements:protons}-\ref{tab:measurements:eventshapes}). Here, we show the first order (red dotted), third order (blue), fourth order (cyan) and eighth order (magenta) interpolation polynomial. The residual for the central values of the observables show clearly that the eighth order polynomial performs better than the other interpolations. The residuals for the errors are independent of the order we used in the interpolations.}
    \label{fig:residuals}
\end{figure}

The goodness-of-fit is defined as 
\begin{equation}
 \chi^2 = \sum_{\mathcal{O}} \sum_{b\in \mathcal{O}} \bigg(\frac{f_{(b)}(\{p_i\}) - \mathcal{R}_b}{\Delta_b}\bigg)^2,
\label{eq:GoF}
\end{equation}
where $\mathcal{R}_b$ is the central value for the experimental measurement $\mathcal{O}$ at a bin $b$, $f_{(b)}(\{p_i\})$ is the analytical expression of the response function which is a polynomial of the parameters, and $\Delta_b$ is the total error. The number of degrees-of-freedom ($N_{\rm df}$) is the number of degrees-of-freedom which is defined as the number of measurements minus the number of independent parameters, {\it i.e.}
\begin{eqnarray}
N_{\rm df} = \sum_{\mathcal{O}} |b \in \mathcal{O}| - N_{\rm parameters}.
\label{eq:NDF}
\end{eqnarray}
We turn now to a brief discussion of the treatment of the errors ($\Delta_b$) used in the goodness-of-fit definition. Experimental errors are the quadratic sum of statistical and systematic uncertainties. Besides, we do not assume any correlations between the different measurements as this information is not provided by the experimental collaborations. The Monte Carlo uncertainties connected to the size of the samples used in our interpolations are summed in quadrature with the experimental errors. Finally, we add a flat $5\%$ uncertainty on each bin and for each observable which is used as a protection against overfitting effects and as sanity limit for the accuracy in both the perturbative (high order corrections,...) and non-perturbative (high twist terms, ...) regimes. We note that the introduction of this flat uncertainty will also reduce the value of $\chi^2/N_{\rm df}$ to be consistent with unity. The resulting variations around the best-fit points can reasonably defines conservative estimates of the QCD uncertainties on the predictions. The total error per bin $b$ is defined by
\begin{eqnarray}
\Delta_b = \sqrt{\sigma_{b, \rm exp}^2 + \sigma_{b, {\rm MC}}^2 + \sigma_{b, {\rm th}}^2},
\end{eqnarray}
with $\sigma_{b, {\rm th}} = 0.05 \times f_{(b)}(\{p_i\})$. The polynomial dependence of the true Monte Carlo response is parametrised as follows
\begin{eqnarray}
    f_{(b)}(\{p_i\}) &=& \alpha_0^{(b)} + \sum_{i=1}^4 \beta_i^{(b)} p_i + \sum_{i,j=1}^4 \gamma_{ij}^{(b)} p_i p_j + \sum_{i,j,k = 1}^4 \delta_{ijk}^{(b)} p_i p_j p_k + \sum_{i,j,k,\ell=1}^4 \epsilon_{ijk\ell}^{(b)} p_i p_j p_k p_\ell \nonumber \\
    &+& \sum_{i,j,k,\ell,m=1}^4 \zeta_{ijk \ell m}^{(b)} p_i p_j p_k p_\ell p_m + \sum_{i,j,k,\ell,m,n=1}^4 \theta_{ijk \ell m n}^{(b)} p_i p_j p_k p_\ell p_m p_n \\
    &+& \sum_{i,j,k,\ell,m,n,r=1}^4 \omega_{ijk \ell m n r}^{(b)} p_i p_j p_k p_\ell p_m p_n p_r + \sum_{i,j,k,\ell,m,n,r,s=1}^4 \rho_{ijk \ell m n r s}^{(b)} p_i p_j p_k p_\ell p_m p_n p_r p_s, \nonumber
    \label{eq:interp}
\end{eqnarray}
with $\alpha, \beta, \gamma, \delta, \epsilon, \zeta, \theta, \omega,~{\rm and}~\rho$ are the polynomial coefficients determined in the fit and $\{p_i\} = \{a, \langle z_\rho \rangle, \sigma_\perp, a_{\rm Diquark} \}$ are the parameters of the  Lund fragmentation function. The order of the polynomial function plays a crucial role in both the quality of the fits and the consistency of the interpolated results with the true MC response at the minimum of the model parameters. To see which polynomial is most suitable in our tune, we compute the distributions of for the interpolated values (called residuals) and their errors for few polynomial functions. The results of these are shown in figure \ref{fig:residuals} where we display the residuals for $1^{\rm st}$ order (dashed red), $3^{\rm rd}$ order (navy), $4^{\rm th}$ order (cyan) and $8^{\rm th}$ order (magenta) polynomial functions. We can see that the $8^{\rm th}$ order polynomial performs better than the others as most of the density of the residuals is within $5\%$ of the true MC response. Nevertheless, the $4^{\rm th}$ order polynomial interpolation has a good performance as well (we also checked that the differences between the predictions for the polynomials of order $4$ and $8$ interpolations at the minimum and found good agreement). We must stress out that using polynomials with orders higher than $4$ for the interpolation will cause additional overfitting effects. Therefore, we will use the $4^{\rm th}$ order interpolation polynomial throughout this study and set $\zeta = \theta = \omega = \rho = 0$ in equation \ref{eq:interp}. Finally, the distributions of the residuals for the errors are showing similar behavior which may be explained by the fact that we have used a large number of events for our MC sampling ($2$ million events per parameter point).
\section{Results}
\label{sec:tunes}

\begin{figure}[!tb]
    \centering
    \vspace{-0.5cm}
    \includegraphics[width=0.90\linewidth]{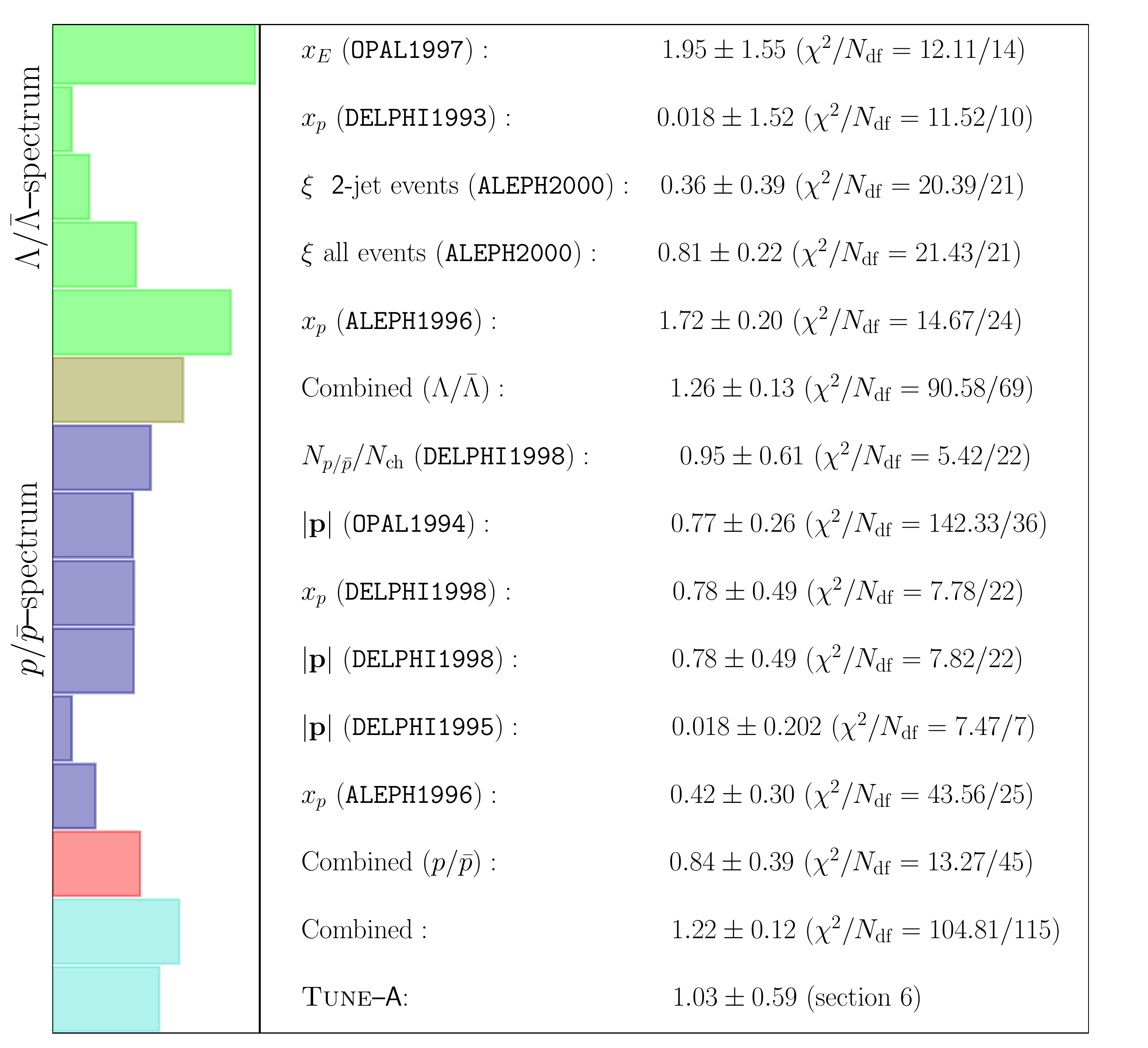}
    \caption{Best fit point for $\texttt{StringZ:aExtraDiquark}$, the corresponding $68\%$ errors and the associated $\chi^2/N_{\rm df}$ for the different observables and their combinations. Here, we show $\Lambda$ scaled momentum/energy (light green), combination of all $\Lambda$ measurements (olive), $p/\bar{p}$ momenta (purple), combination of all $p/\bar{p}$ measurements (red) and the combined measurements of $\Lambda$ and $p/\bar{p}$ momenta (turquoise). More details are can be found in the main text.}
    \label{fig:tunes1D:results}
\end{figure}

The results of the one-parameter fits are displayed in figure \ref{fig:tunes1D:results} which are shown in the form of horizontal bar plots. For the one-dimensional tunes, we show the best-fit point for tunes to ({\it i}) $\Lambda^0$ scaled momentum/energy (green bar) including data from \textsc{Aleph}--1996 \cite{Barate:1996fi}, \textsc{Aleph}--2000 \cite{Barate:1999gb}, \textsc{Delphi}--1993 \cite{Abreu:1993mm}, and \textsc{Opal}--1997 \cite{Alexander:1996qj}  and to ({\it ii}) proton (scaled)-momentum (blue bar) which includes data from \textsc{Aleph}--1996 \cite{Barate:1996fi}, \textsc{Delphi}--1995 \cite{Abreu:1995cu}, \textsc{Delphi}--1998 \cite{Abreu:1998vq} and \textsc{Opal}--1994 \cite{Akers:1994ez}. The tune of \texttt{StringZ:aExtraDiquark} resulting from a combination of $\Lambda^0$~($p/\bar{p}$) measurements is shown in olive~(red) bar while the one resulting from  a combination of all the measurements is shown in turquoise. Finally, a combination procedure described in section \ref{sec:uncertainty} is also shown with the label Tune--A. We first discuss the impact of the individual measurements on the best-fit point of \texttt{StringZ:aExtraDiquark} and we close this section by a discussion of the combinations. First, we can see that all the individual tunes but the ones using data from \textsc{Aleph}--1996, \textsc{Delphi}--1993, \textsc{Delphi}--1995 and \textsc{Opal}--1994 give results that are consistent with each other; {\it i.e.} the best-fit point floats around $0.8$--$0.93$ with uncertainties of about $0.20$--$0.61$. The tune the \textsc{Opal}--1997 measurement prefers large values of \texttt{StringZ:aExtradDiquark}, of $1.97$\footnote{This is due to the fact that the last two bins, corresponding to $x_E \geq 0.5$, of the $x_E$ distribution forces \texttt{StringZ:aExtraDiquark} to take the largest possible. Removing these two bins will reduce the best-fit point from $1.95$ to $0.91$.}. We note that the consistency between the theory and data for this measurement is almost independent of whether we include or not these two bins in the tune. Finally, we must note that the result of the fit of the proton momentum performed by \textsc{Delphi}--1995 prefers very small values of \texttt{StringZ:aExtraDiquark}. 
\begin{table}[t!]
\setlength\tabcolsep{3pt}
  \begin{center}
    \begin{tabular}{lcccccc}
      \toprule
      Tune     & \verb|aLund| & \verb|avgZLund| & \verb|sigma| & \verb|aExtraDiquark| & \verb|bLund| &  $\chi^2/N_{\rm df}$\\
      \midrule
      \textsc{Aleph}  & $0.758\substack{+0.074\\ -0.074}$ & $0.541\substack{+0.007\\ -0.007}$ & $0.297\substack{+0.005\\-0.005}$ & $1.218\substack{+0.358\\-0.358}$ & $1.040$ & $116.22/296$ \\
      \textsc{Delphi} & $0.358_{-0.054}^{+0.054}$ & $0.497_{-0.007}^{+0.007}$ & $0.287_{-0.006}^{+0.006}$ & $0.782_{-0.298}^{+0.298}$ & $0.533$
      &  $144.37/268$ \\
      \textsc{L3}     &  $0.478_{-0.063}^{+0.063}$  & $0.557_{-0.006}^{+0.006}$ &  $0.315_{-0.007}^{+0.007}$ & $1.998_{-0.049}^{+0.049}$ & $0.897$ & $84.70/140$ \\
      \textsc{Opal}   & $0.588_{-0.086}^{+0.086}$ & $0.536_{-0.005}^{+0.005}$ &     $0.300_{-0.005}^{+0.005}$    &  $1.998_{-0.204}^{+0.204}$ & $0.872$ &     $53.54/136$ \\
      \midrule
      \textsc{Combined} &       $0.601_{-0.038}^{+0.038}$ & $0.540_{-0.004}^{+0.004}$ &  $0.307_{-0.002}^{+0.002}$ & $1.671_{-0.196}^{+0.196}$ & $0.897$ &     $676.69/852$ \\
      \bottomrule
    \end{tabular}
  \end{center}
    \caption{\label{tab:experiments} Results of the tunes performed 
    separately to all the considered measurements from a given experiment.} 
\end{table}

\begin{figure}[!h]
    \centering
    \vspace{-1cm}
    \includegraphics[width=0.80\linewidth]{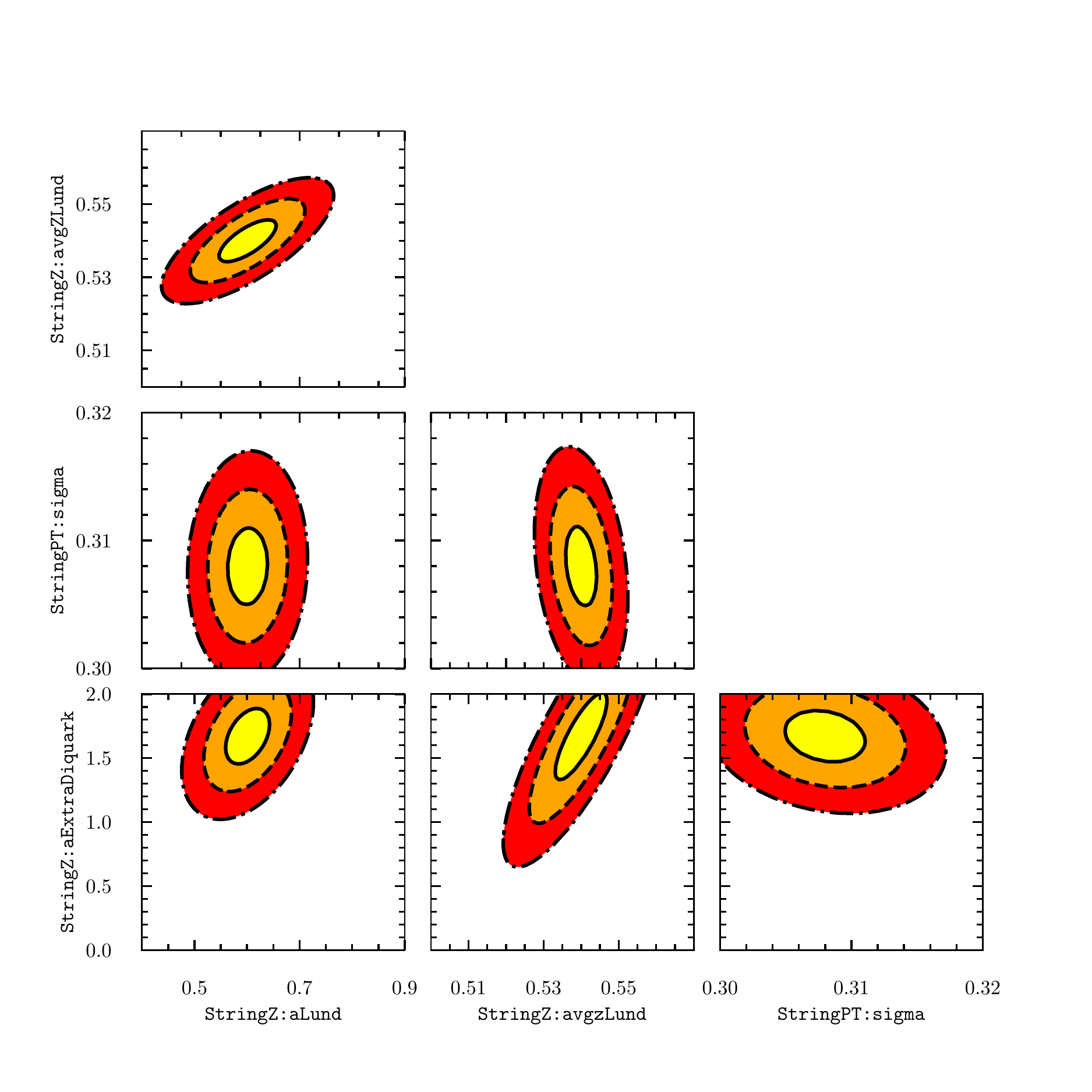}
    \caption{Results of tunes performed to all the data shown in Table  \ref{tab:measurements:protons}-\ref{tab:measurements:eventshapes}. The results are projected on different parameters we used in the four-dimensional parameter space tune. The contours corresponding to $68\%$, $95\%$ and $99.5\%$ confidence levels are shown in yellow, orange, and red respectively.}
    \label{fig:tune:results}
\end{figure}

We turn now to a discussion about the combination procedure. First, we dot not include the measurement of $p/\bar{p}$ spectrum by \textsc{Opal}--1994 since it is inconsistent with the others for $x_p > 0.1$~(see figure \ref{fig:comparisons})\footnote{We have checked that the value of $\texttt{StringZ:aExtraDiquark}$ at the minimum is almost unaffected if the OPAL\_1994\_S2927284 measurement of the proton ($p/\bar{p}$) momentum is added to the fit. One must note that the net effect of including that measurement is worsening the quality of the fit since $\chi^2/N_{\rm df}$ increases by almost a factor of one. This is unsurprising due to the fact that the OPAL measurement itself is inconsistent with the other experimental measurements of proton (scaled)-momentum.}. The measurements of $p/\bar{p}$ and $\Lambda^0$ scaled momenta performed by \textsc{Aleph}--1996 have been removed from the combination as well. There are two reasons for this choice: first, the tuning of \texttt{StringZ:aExtraDiquark} to the corresponding $p/\bar{p}$ measurement prefers small value of about $0.42$ (see figure \ref{fig:tunes1D:results}) and the agreement between theory and data at the best-fit point is not good enough ($\chi^2 \simeq 44$ even after including the $5\%$ theory uncertainty). Second, the measurement of $\Lambda^0$ scaled momentum has already been superseded by a more recent one performed by \textsc{Aleph}--2000 \cite{Barate:1999gb} which includes more data. On the other hand, the tuning of \texttt{StringZ:aExtraDiquark} to \textsc{Aleph}--1996 measurement of $\Lambda^0$ scaled momentum prefers large value of about $1.72$ as shown in figure \ref{fig:tunes1D:results}. The measurements of $p/\bar{p}$ (scaled) momentum by \textsc{Delphi}--1995 \cite{Abreu:1995cu} have been removed since these measurements have already been superseded by the more recent \textsc{Delphi}--1998 \cite{Abreu:1998vq} measurement which covers a wider range of momenta $p \in [0.5, 45]~{\rm GeV}/c^2$.

After all these considerations, two measurements of $p/\bar{p}$ and four measurements of $\Lambda$ (scaled) momenta have been used in the combinations. The result of the tune from a combination of $\Lambda$ measurement only is about $1.26\pm 0.13$ with a $\chi^2/N_{\rm df}$ of order $1.2$. The combination of the $p/\bar{p}$ scaled momentum and $N_{p/\bar{p}}/N_{\rm charged}$ measurements gives a best-fit point of $0.84\pm 0.39$ consistent with the previous combination within the quoted uncertainties. Finally, we note that the combination of all the six mentioned measurements give a best-fit point of $1.22\pm 0.12$ with a good value of $\chi^2/N_{\rm df} \sim 0.91$.


Now, we discuss the results of the four-dimensional parameter space tuning. The results of the fits are shown in table \ref{tab:experiments} for individual experiments and their combinations. We can see that the best-fit points of \texttt{StringZ:aLund}, \texttt{StringZ:avgZLund} and \texttt{StringPT:sigma} are consistent with the results of a previous study \cite{Amoroso:2018qga}. On the other hand, the best-fit point of \texttt{StringZ:aExtraDiquark} is larger than what we found in the one-dimensional parameter space tune. Note that this value is driven by the results from two experiments: \textsc{L3} and \textsc{Opal}. The correlations are found to be 
\begin{eqnarray}
C_{ij} =
\left(
\begin{array}{rrrr}
1.000 & 0.718 & 0.057 & 0.415 \\
0.718 & 1.000 & -0.270 & 0.816 \\
0.057 & -0.270 & 1.000 & -0.204 \\
0.415 & 0.816 & -0.204 & 1.000
\end{array}
\right),
\label{eq:correlation}
\end{eqnarray}

\begin{figure}[!tbp]
    \centering
    \includegraphics[width=0.80\textwidth]{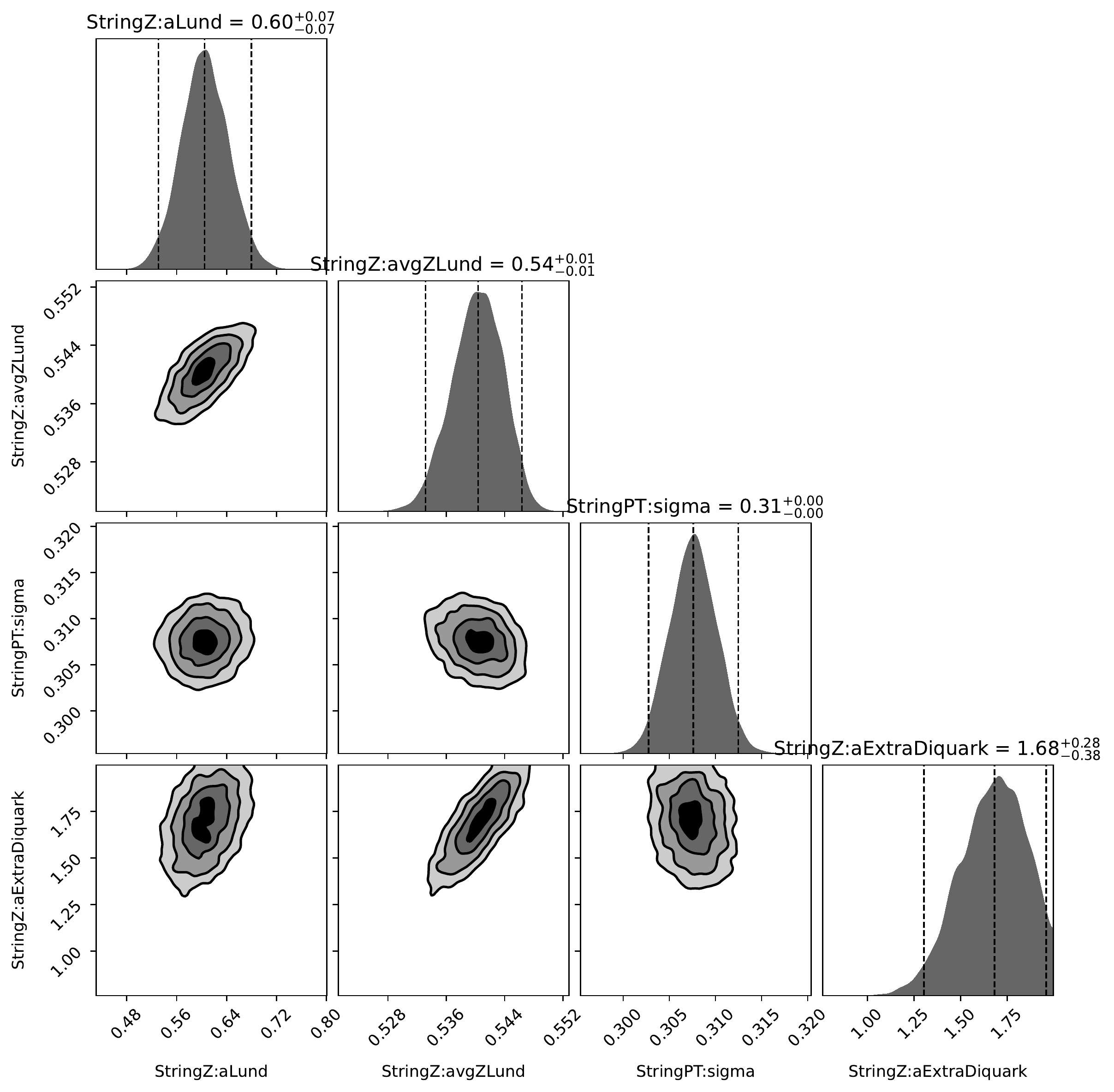}
    \caption{One- and two-dimensional marginalized posterior distributions for the uni-modal four-dimensional parameter space fit. Here, the contours show the $68\%$ and $95\%$ Bayesian credible intervals. The results include all the measurements listed in Tables \ref{tab:measurements:protons}-\ref{tab:measurements:eventshapes}.}
    \label{fig:bayesian:results:combined}
\end{figure}

where the coefficients are ordered as follow $\{a, \langle z_\rho \rangle, \sigma_\perp, a_{\rm Diquark}\}$. It is clear from equation \ref{eq:correlation} that the correlation between the fragmentation function parameters are small except that \texttt{StringZ:avgZLund} is extremely highly correlated with both \texttt{StringZ:aLund} and \texttt{StringZ:aExtraDiquark}. In figure \ref{fig:tune:results} we show the $68\%$, $95\%$ and $99.5\%$ CL intervals projected on the different fragmentation function parameters. This is a clear visualization of the results of equation \ref{eq:correlation} and Table \ref{tab:experiments}. As was pointed in the previous subsection, we have found that different measurements prefer different values of the \texttt{StringZ:aExtraDiquark} parameter. We have also checked that these results do not correspond to flat directions as the best-fit point of \texttt{StringZ:aExtraDiquark} does not significantly depend on the choice of the prior (unless the prior is too close to the best-fit point). We finally note that these tune results give fairly good agreement with data and the results are competitive with the baseline \textsc{Monash} tune.

We close this section by showing the results of the Bayesian fit which constitutes a very good cross-check of the results of the frequentist analysis. These results are shown in figure \ref{fig:bayesian:results:combined} where we can see a very good agreement with those of the frequentist fit shown in figure \ref{fig:tune:results}. The results of the Bayesian fit along with the $95\%$ credible levels are shown below:
\begin{eqnarray*}
   \verb|StringZ:aLund| &=&  0.60 \pm 0.07, \\
   \verb|StringZ:avgZLund| &=& 0.54 \pm 0.01, \\
   \verb|StringPT:sigma| &=& 0.31 \pm 0.01, \\
   \verb|StringZ:aExtraDiquark| &=& 1.68^{+0.28}_{-0.38},\\
\end{eqnarray*}
which agrees very well with the results of the frequentist fit.
\section{Uncertainty estimates}
\label{sec:uncertainty}
In this section, we discuss the different sources of QCD uncertainties that may affect the particle spectra from dark-matter annihilation. We start with a discussion of the parton-shower uncertainties and how they are estimated within \textsc{Pythia}~8. The formalism of the shower uncertainty estimates used here is based on the method developed in reference \cite{Mrenna:2016sih}. We then move to the discussion of the hadronisation uncertainties and we estimate their size. The estimate of these uncertainties is done with two different methods: Hessian variations that are widely used in the estimate of parton distribution functions (PDFs) uncertainties (see  {\it e.g.} \cite{Pumplin:2001ct} for more details about the Hessian method) and a manual method which we used in a previous analysis \cite{Amoroso:2018qga}. 
\subsection{Perturbative uncertainties}

\begin{figure}[!t]
    \centering
    \includegraphics[width=0.8\textwidth]{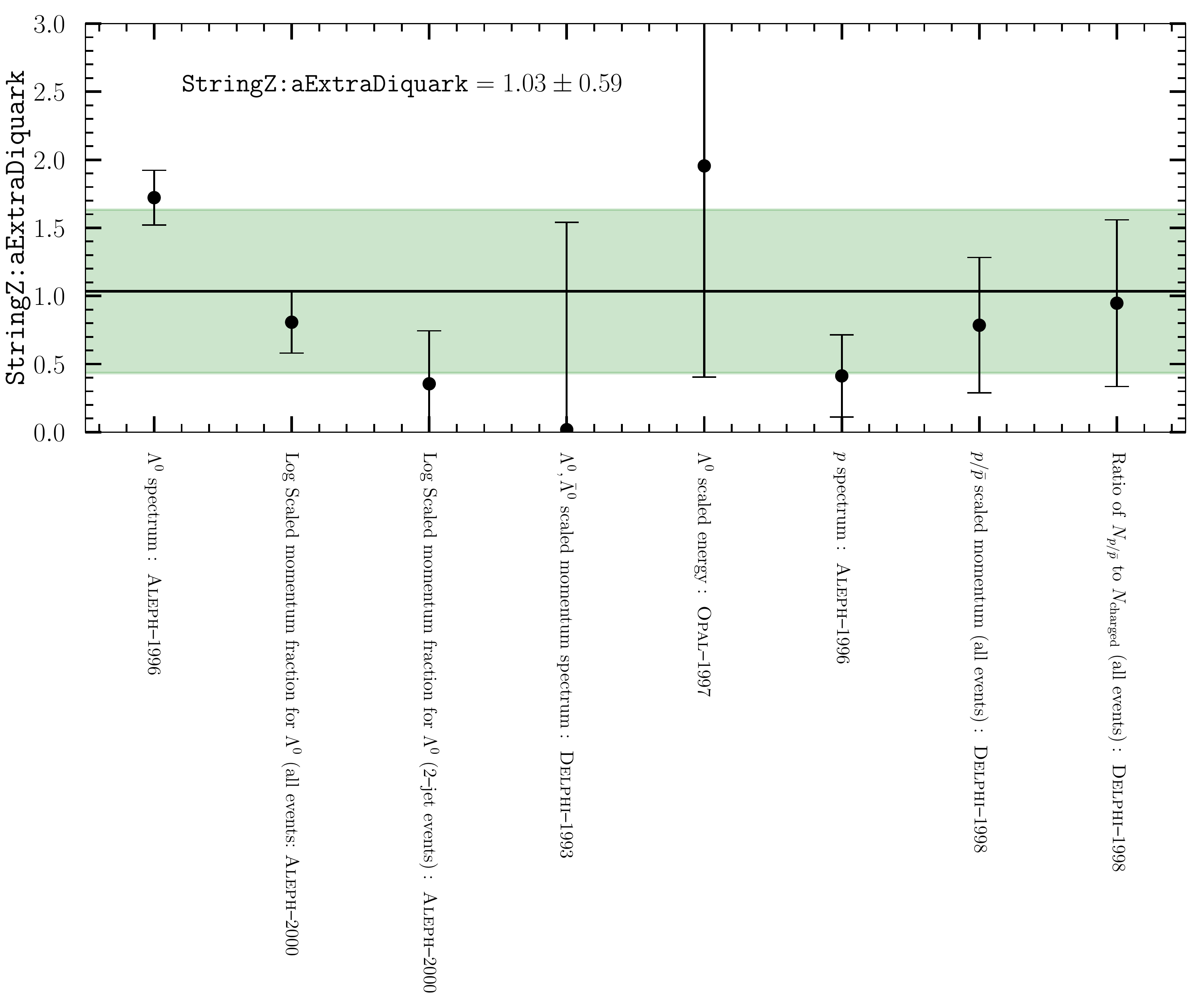}
    \caption{The tunes results of the \texttt{StringZ:aExtraDiquark} parameter performed separately to each of the eight measurements of proton or $\Lambda$ spectra. The weighted average of the tunes to the individual measurements is shown with a black line while the green shaded area corresponds to the $68\%$ CL interval on \texttt{StringZ:aExtraDiquark}.}
    \label{fig:weightedfit:result}
\end{figure}

The perturbative uncertainties are split into catergories: scale-variation uncertainties and non-singular uncertainties. Before digging into the details of the estimate of these uncertainties, we first remind the reader that the parton showers in \textsc{Pythia}~8 are based on a dipole type $p_\perp$--evolution which has been available since \textsc{Pythia}~6.3 \cite{Sjostrand:2004ef} and is used for both QED and QCD emissions. There are strong arguments that the renormalisation scale, at which the parton branchings are estimated, should be equal to the transverse momentum of the branching parton. This scale choice is accompanied by a universal factor that absorbs the leading second-order corrections in the soft-limit \cite{Amati:1980ch, Catani:1990rr}. This usually tends to increase the value of the strong coupling $\alpha_S(M_Z^2)$ by about $10\%$. Therefore, it makes complete sense to estimate uncertainties that are originated from the shower scale variations which define the most important source of perturbative uncertainties. This first class of the shower uncertainties are estimated by varying the renormalisation scale by a factor of two in each direction with respect to the nominal scale choice. Let us consider a variation defined by $\mu_R \equiv p_\perp \to k p_\perp$ with $k = 1/2, 2$. Under this variation, the gluon-emission probability
\begin{eqnarray}
P(t, z) = \frac{\alpha_S(p_\perp)}{2\pi} \frac{P(z)}{t},
\end{eqnarray}
changes into 
\begin{eqnarray}
P'(t, z) = \frac{\alpha_S(k p_\perp)}{2\pi} \bigg(1 + \frac{\alpha_S(\mu)}{2\pi} \beta_0 \ln k\bigg) \frac{P(z)}{t},
\label{eq:splitting}
\end{eqnarray}
where $t = p_\perp^2, z$ is the longitudinal momentum fraction, $P(z)$ is the Dokshitzer-Gribov-Lipatov-Altarelli-Parisi (DGLAP) splitting kernel for the $a\to bc$ branching, $\beta_0 = (11 N_c - 2 n_f)/3$ is the one-loop beta function with $N_c = 3$ being the number of colour degrees of freedom and $n_f$ is the number of active quarks with mass $m_q$ below $\mu_R = p_\perp$. To guarantee that scale variations are as  conservative as possible, a number of modifications to equation \ref{eq:splitting} are in order. First, the scale $\mu$ at which $\alpha_S$ is evaluated (the second term inside the parenthesis) is chosen to be the maximum of the dipole mass $m_{\rm dip}$ and $k p_\perp$; {\it i.e.} $\mu \equiv \mu_{\rm max} = \max({m_{\rm dip}, k p_\perp}$). Second, another factor $\zeta$ that depends on the singularity of the splitting is introduced.  In this case, the second term inside the parenthesis of equation \ref{eq:splitting} is multiplied by $(1 - \zeta)$ so that the correction factor vanishes linearly outside the soft limit. We note that in \textsc{Pythia}~8, a limit on the allowed value of $\alpha_S$ that changes under the scale variation is imposed {\it i.e.} $|\Delta \alpha_S| \leq 0.2$ in order to prevent branchings near the cut-off scale from generating important changes to the event weights. The variations of the non-universal hard components of the DGLAP kernels are also possible with the new formalism. Under these variations, the shower splitting kernels transform as 
\begin{eqnarray}
P(z) \frac{{\rm d}Q^2}{Q^2} \to \bigg(P(z) + \frac{c_{\rm NS} Q^2}{m_{\rm dip}^2}\bigg) \frac{{\rm d}t}{t},
\end{eqnarray}
where $Q^2$ is the virtuality of the parent branching parton, and $c_{\rm NS}$ is a factor that corresponds to the variations -- by default $c_{\rm NS} = \pm 2$ is used but the user is free to change it. We close this discussion by noting that Matrix-Element Corrections (MECs), switched on by default in \textsc{Pythia}~8, lead to very small variations of the non-singular terms of the DGLAP splittings. It was found that switching off these corrections would lead to comprehensively larger envelopes \cite{Mrenna:2016sih}. We have checked that this the case but the error band from non-singular variations without MECs strongly overlaps with those from scale variations. Therefore, if the total perturbative uncertainty is taken as the envelope of the variations and not their sum in quadrature then there is no major difference between switching MECs on or off.

\subsection{Fragmentation function uncertainties}
\subsubsection{Weighted fit} 
This method consists of taking the values of the best-fit points of the $a_{\rm Diquark}$ parameter from eight different measurements -- shown in figure \ref{fig:tunes1D:results} and estimate a new combined best-fit point (weighted best-fit point) and the associated error. In what follows, we do not take the results from the fits to the measurement of the (anti-)proton momentum by \textsc{Delphi}--1995 \cite{Abreu:1995cu} and \textsc{Opal}--1994 \cite{Akers:1994ez}. Let us denote the best-fit point of \texttt{StringZ:aExtraDiquark} from a measurement $i$~(here $i=1,\cdots,8$) by $x_i \pm \sigma_i$ (with $\sigma_i$ being the \texttt{MIGRAD} error on $x_i$). If one assumes a Gaussian probability distribution function for $x_i$ and no correlation between the different measurements, we define a global $\chi^2$ of the variable $x$ as 
\begin{eqnarray}
\chi^2 = \sum_{i=1}^8 \bigg(\frac{x - x_i}{\sigma_i}\bigg)^2.
\label{eq:chi2:combo}
\end{eqnarray}
The combined best-fit point value of \texttt{StringZ:aExtraDiquark} is obtained by minimizing the $\chi^2$ measure defined in equation \ref{eq:chi2:combo}:
\begin{eqnarray}
\frac{\partial \chi^2}{\partial x}\bigg|_{x = \hat{x}} = 2 \sum_{i=1}^8 \bigg(\frac{\hat{x} - x_i}{\sigma_i^2}\bigg) = 0.
\end{eqnarray}
Solving this equation will give
\begin{eqnarray}
\hat{x} = \frac{\sum_i x_i \sigma^{-2}_i}{\sum_i \sigma_i^{-2}} = 1.03,
\end{eqnarray}
where the numerical value is obtained by using the results shown in figure \ref{fig:tunes1D:results}. The error on $\hat{x}$ is obtained by differentiating $\chi^2$ two times with respect to $\hat{x}$. In this case, we obtain
\begin{eqnarray}
\hat{\sigma}^2 = \frac{1}{\sum_i \frac{1}{\sigma_i^2}}.
\end{eqnarray}
This basic approach leads to a small value of $\hat{\sigma}$ that will be mainly controlled by the measurement with the smallest $\sigma_i$. In principle, this approach does not define a conservative estimate of $\hat{\sigma}$ as we can see clearly that the best-fit points $x_i$ are not stable and strongly depends on the measurement being used. Therefore, in order to improve this basic approach we inflate the error $\hat{\sigma}$ by a reasonable factor which we choose to be five. The result of this approach is depicted in figure \ref{fig:weightedfit:result} where we can see the green shaded area which corresponds to the combined errors passes through most of the best-fit points. Therefore, the new combined result along with the corresponding uncertainty is expected to provide a very good agreement with almost all the measurements within the error bands.  The final result is given by
\begin{eqnarray}
\texttt{StringZ:aExtraDiquark} = 1.03 \pm 0.59.
\label{eq:aDi:weighted}
\end{eqnarray}
To get out a comprehensive uncertainty bands from the variations of the fragmentation function parameters we consider all the possible variations obtained from the errors depicted in equations \ref{eq:tune2018} and \ref{eq:aDi:weighted} and not only the correlated ones: $\{p_1, p_2, p_3, p_4\} = (++++), (----)$. For a parameter space of dimension four, the total number of variations, excluding the nominal value, is given by $N = 4^3 - 1 = 63$ variations. However, given that the sum of \texttt{StringZ:aLund} and \texttt{StringZ:aExtraDiquark} may be thought as a new effective $a_{\rm eff}$ parameter, we can consider their correlated variations, {\it i.e.} they will be changed in the same direction, we end up with $3^3 - 1 = 26$ possible variations. 

\begin{table}[!h]
 \setlength\tabcolsep{4pt}
 \begin{center}
\begin{adjustbox}{max width=\textwidth}
  \centering
    \begin{tabular}{l c c c c}
    \toprule
    \toprule
Tune	& \texttt{StringZ:aLund} &	\texttt{StringZ:avgZLund} & \texttt{StringPT:sigma}	& \texttt{StringZ:aExtraDiquark} \\
\toprule
Central	& $0.601$	& $0.540$ &	$0.307$	& $1.671$ \\
\toprule
\multicolumn{5}{l}{\textit{$1\sigma$ eigentunes}} \\
\toprule
Variation $1^+$	& $0.608$ &  $0.542$ &  $0.307$ &  $1.771$ \\
Variation $1^-$	& $0.592$ & $0.538$ &  $0.307$ &	$1.568$ \\
Variation $2^+$	& $0.498$ &	$0.535$ &  $0.306$ &	$1.679$ \\
Variation $2^-$	& $0.701$ & $0.544$ &  $0.309$ &    $1.662$ \\
Variation $3^+$	& $0.599$ &	$0.575$ &  $0.321$ &	$1.671$ \\
Variation $3^-$	& $0.602$ &	$0.506$ &  $0.295$ &	$1.671$ \\
Variation $4^+$	& $0.601$ & $0.511$ &  $0.384$ &	$1.671$ \\
Variation $4^-$	& $0.600$ & $0.563$	& $0.245$  &    $1.671$ \\
\toprule
\multicolumn{5}{l}{\textit{$2\sigma$ eigentunes}} \\
\toprule
Variation $1^+$	& $0.609$ & $0.542$ &  $0.307$ &    $1.775$ \\
Variation $1^-$	& $0.591$ & $0.538$ &  $0.307$ &	$1.558$ \\
Variation $2^+$	& $0.501$ &	$0.535$ &  $0.306$ &	$1.679$ \\
Variation $2^-$	& $0.700$ & $0.544$ &  $0.308$ &    $1.662$ \\
Variation $3^+$	& $0.597$ &	$0.609$ &  $0.333$ &	$1.670$ \\
Variation $3^-$	& $0.603$ &	$0.474$ &  $0.283$ &	$1.671$ \\
Variation $4^+$	& $0.601$ & $0.478$ &  $0.475$ &	$1.672$ \\
Variation $4^-$	& $0.600$ & $0.581$	&  $0.197$ &    $1.669$ \\
\toprule
\multicolumn{5}{l}{\textit{$3\sigma$ eigentunes}} \\
\toprule
Variation $1^+$	& $0.609$ & $0.542$ &  $0.307$ &    $1.780$ \\
Variation $1^-$	& $0.590$ & $0.538$ &  $0.307$ &	$1.543$ \\
Variation $2^+$	& $0.500$ &	$0.535$ &  $0.306$ &	$1.679$ \\
Variation $2^-$	& $0.700$ & $0.544$ &  $0.309$ &    $1.662$ \\
Variation $3^+$	& $0.595$ &	$0.642$ &  $0.345$ &	$1.669$ \\
Variation $3^-$	& $0.605$ &	$0.447$ &  $0.272$ &	$1.672$ \\
Variation $4^+$	& $0.602$ & $0.445$ &  $0.562$ &	$1.673$ \\
Variation $4^-$	& $0.599$ & $0.595$	& $0.158$  &    $1.669$ \\
\toprule
\bottomrule
\end{tabular}
\hspace{0.2cm}
\end{adjustbox}
\end{center}
    \caption{The Hessian variations (eigentunes) for the nominal tune including all the measurements performed by \textsc{Aleph}. The variations correspond to $\Delta \chi^2 = 1$ ($68\%$ CL), $\Delta \chi^2 = 4$~($95\%$ CL) and $\Delta \chi^2 = 9$~($99\%$ CL) with $\Delta \chi^2$ is defined as $\Delta \chi^2 \equiv \chi^2_{\rm var} - \chi^2_{\rm min}$.}
    \label{tab:eigentunes}
\end{table}

\subsubsection{Hessian errors (eigentunes)} 
The \textsc{Professor} toolkit provides an estimate of the uncertainties on the fitted parameters through the Hessian method (also known as eigentunes). This method consists of a diagonalisation of the $\chi^2$ covariance matrix near the minimum
\begin{eqnarray}
\Delta \chi^2 = \sum_i \sum_j H_{ij}(x_i, x_j) (x_i - x_i^0)(x_j - x_j^0),
\end{eqnarray}
with $H_{ij}$ being the Hessian matrix which consists of second-order derivatives of the covariance matrix with respect to the parameters near the minimum , {\it i.e} $H_{ij} = \partial^2 \chi^2/\partial x_i \partial x_j$. The problem then consists of diagonalising the $H_{ij}$ matrix in the space of the optimised parameters, {\it i.e.} finding the principal directions or eigenvectors and the corresponding eigenvalues. This results in building a set of $2 \cdot N_{\rm params}$ variations. These variations are then obtained as corresponding to a fixed change in the goodness-of-fit measure which is found by imposing a constraint on the maximum variation, defined as a hypersphere with maximum radius of $T$ (defined as the tolerance), i.e. $\Delta \chi^2 \leq T$. Therefore one can define the $\Delta \chi^2$ to match a corresponding confidence level interval; {\it i.e.} one-sigma variations are obtained by requiring that $\Delta \chi^2 \simeq N_{\rm df}$ where $N_{\rm df}$ is the number of degrees-of-freedom defined in equation \ref{eq:NDF}. This approach defines a conservative estimate of the uncertainty if the event generator being used has a good agreement with data (which is usually quantified by $\chi^2_{\rm min}/N_{\rm df} \leq 1$) and the resulting uncertainties provide a good coverage of the errors in the experimental data. To enable for this situation, we have added an extra $5\%$ uncertainty to the MC predictions for all the observables and bins (see section \ref{sec:setup}) which already implied a $\chi^2/N_{\rm df} \leq 1$ in our fits as depicted in Table \ref{tab:experiments}. On the other hand, we enable for large uncertainties by considering not only the one-sigma eigentunes but also the two-sigma eigentunes (correspond to $\Delta \chi^2/N_{\rm df} = 4$) and the three-sigma eigentunes (correspond to $\Delta \chi^2/N_{\rm df} = 9$). The three-sigma eigentunes provide a very good coverage of all the experimental uncertianties in the data for meson and baryon spectra but results in unreasonably large uncertainties that overshot the experimental errors for {\it e.g.} event shapes or jet rates.  The variations corresponding to the one-sigma, two-sigma, and three-sigma eigentunes are shown in Table \ref{tab:eigentunes}. \\

\begin{figure}[!t]
    \centering
    \includegraphics[width=0.49\linewidth]{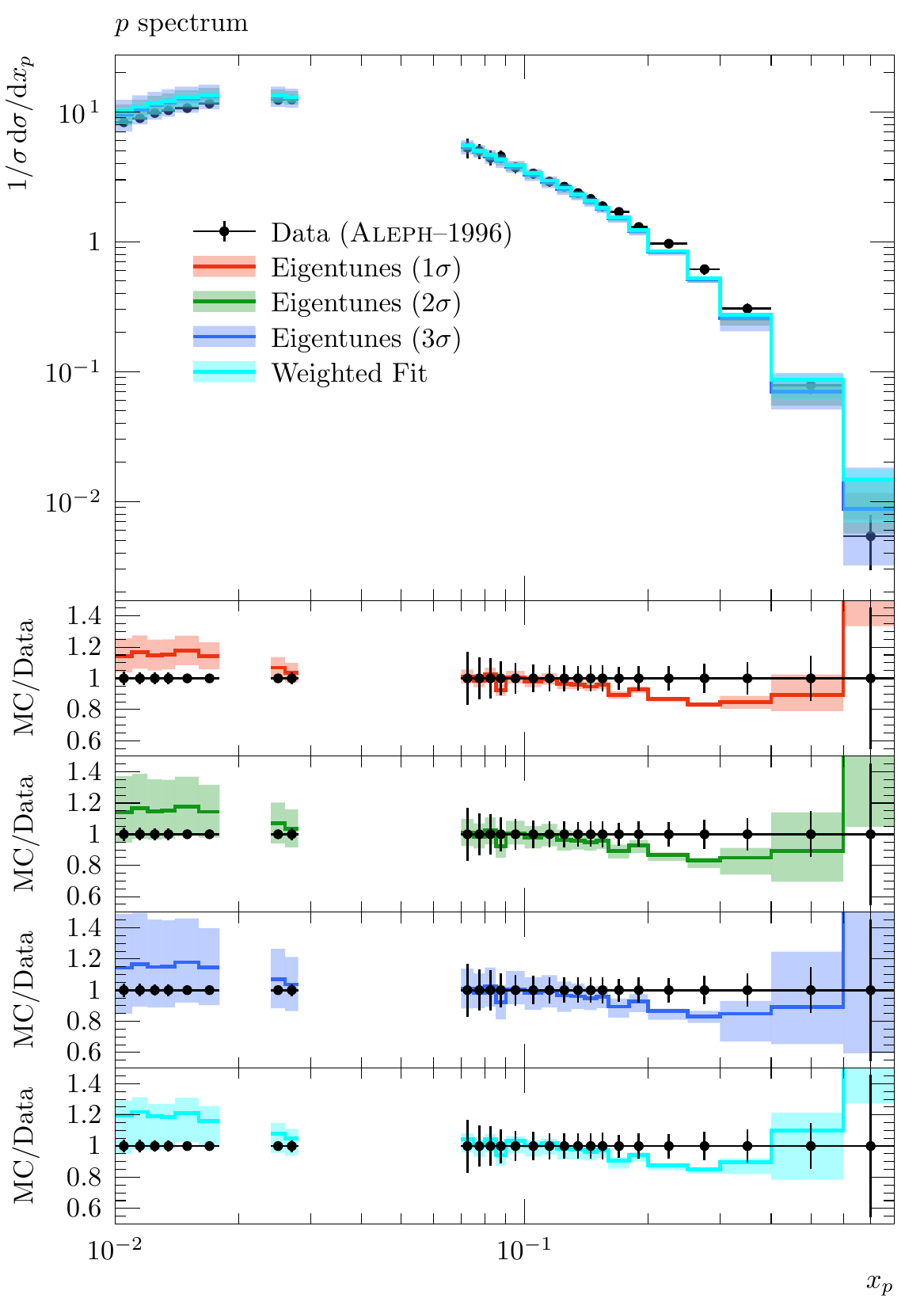}
    \hfill
    \includegraphics[width=0.49\linewidth]{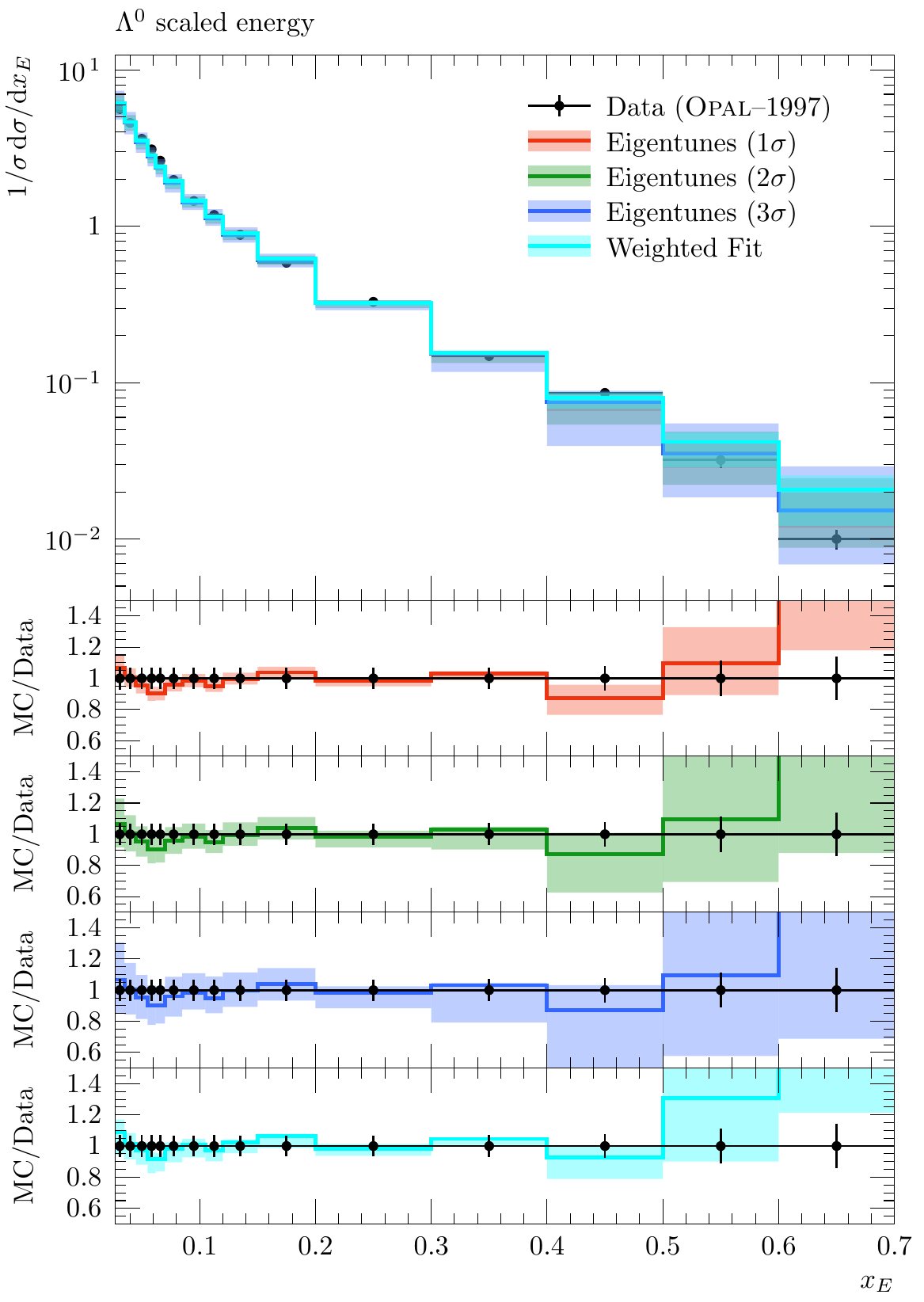}
    \caption{Comparison between the theory predictions and the experimental measurement of $p$ spectrum ({\it left}) and the $\Lambda^0$ scaled energy ({\it right}). The nominal predictions correspond to the fit results shown in Table \ref{tab:experiments} (red, green, and blue) and equations \ref{eq:tune2018}, \ref{eq:aDi:weighted} (cyan). The uncertainty envelopes corresponding to the one-sigma, two-sigma, and three-sigma eigentunes are shown in red, green and blue while the uncertainty envelope from the $26$ variations around the best fit points resulting from the weighted fit is shown as cyan shaded area. Data is taken from \cite{Alexander:1996qj, Barate:1996fi}}.
    \label{fig:comparison}
\end{figure}

\begin{figure}[!t]
    \centering
    \includegraphics[width=0.49\linewidth]{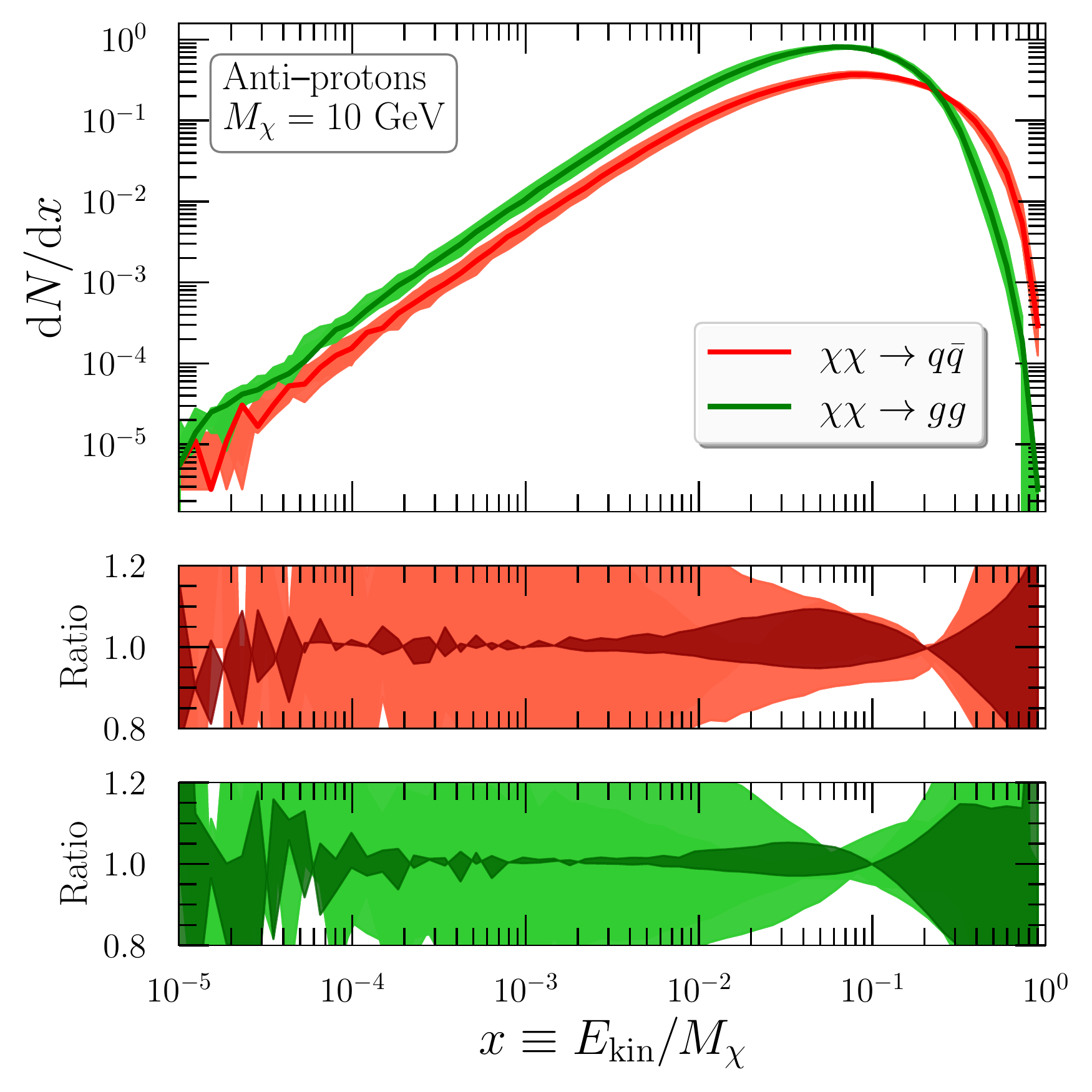}
    \hfill
    \includegraphics[width=0.49\linewidth]{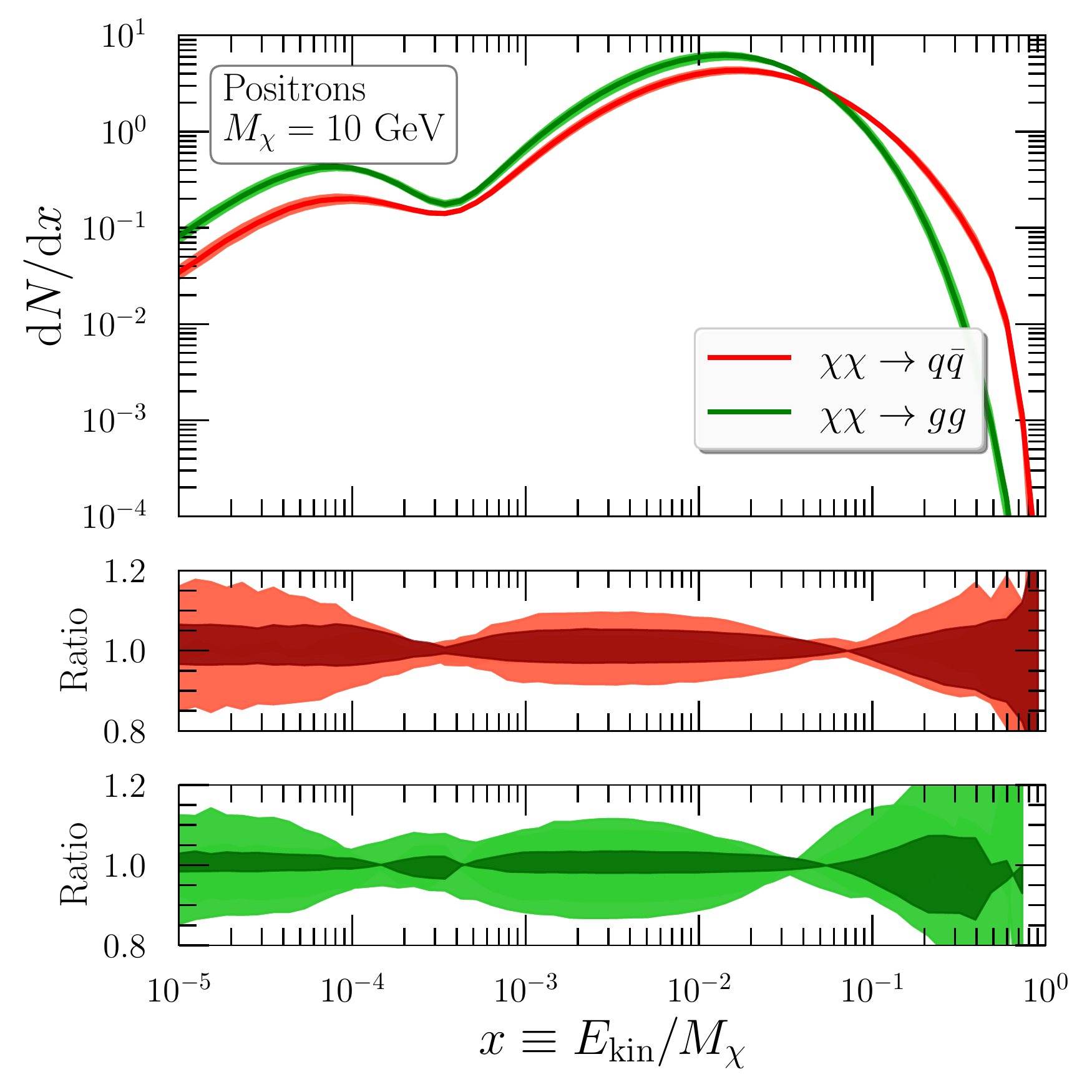}
    \vfill
    \includegraphics[width=0.49\linewidth]{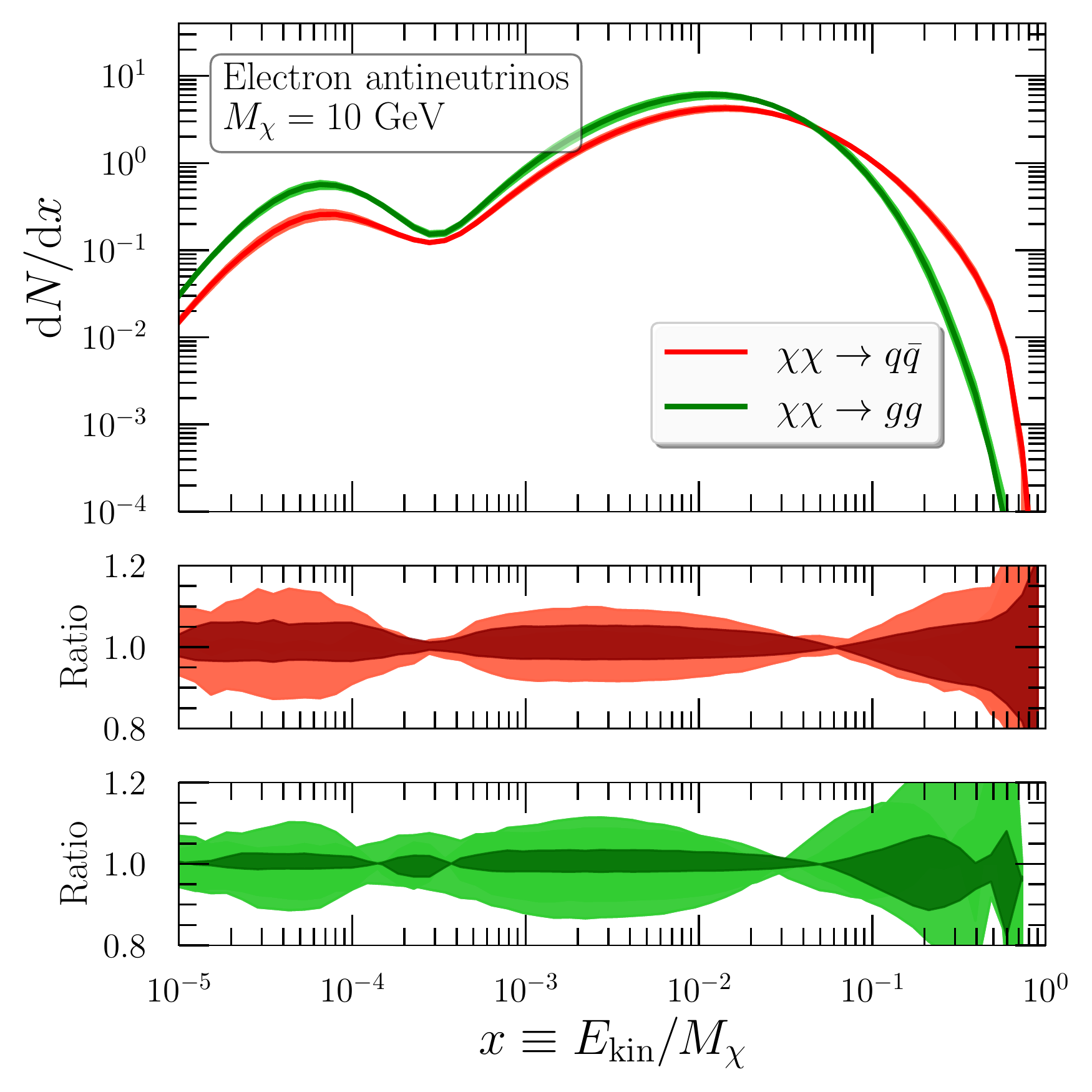}
    \hfill
    \includegraphics[width=0.49\linewidth]{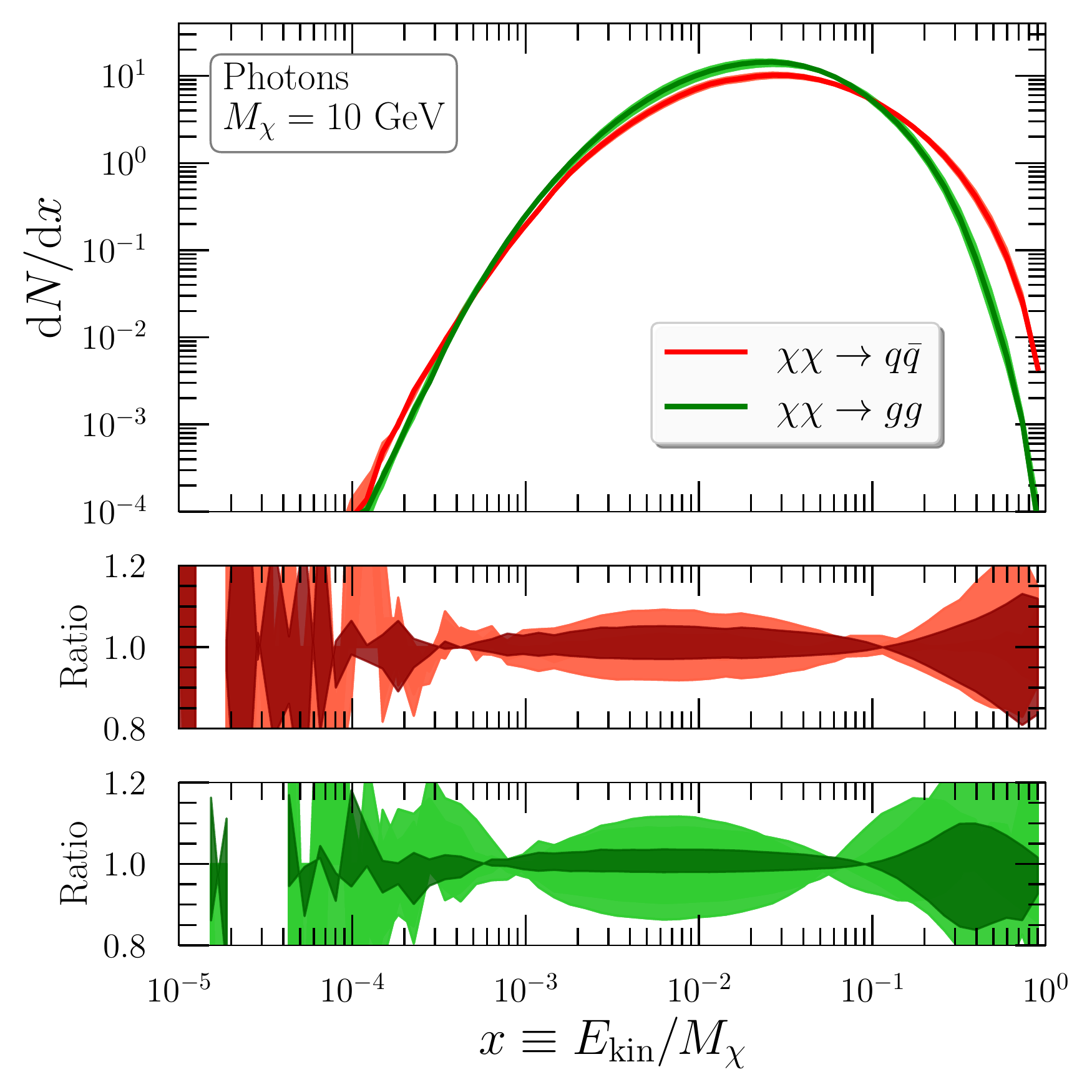}
    \caption{The scaled kinetic energy distribution of anti-protons (left upper panel), positrons (right upper panel), electron antineutrinos (left bottom panel) and photons (right bottom panel) in dark matter annihilation into $q\bar{q}$ (red) and $gg$ (green) and dark matter mass of $10$ GeV. For each pane, the dark shaded band corresponds to the parton-shower uncertainties while the light shaded band corresponds to hadronisation uncertainties.}
    \label{fig:spectra:mDM:10:1}
\end{figure}

\begin{figure}[!t]
    \centering
    \includegraphics[width=0.49\linewidth]{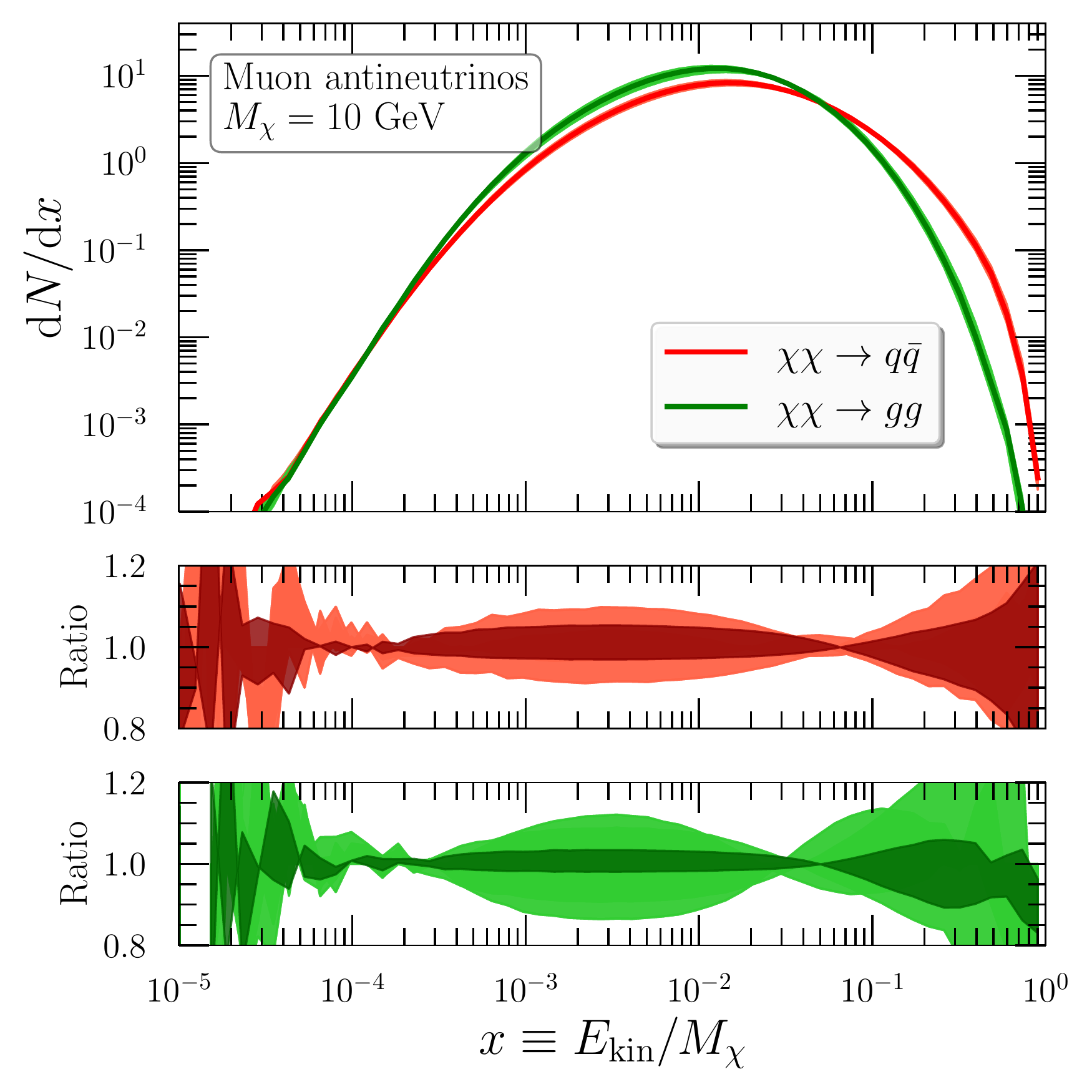}
    \hfill 
    \includegraphics[width=0.49\linewidth]{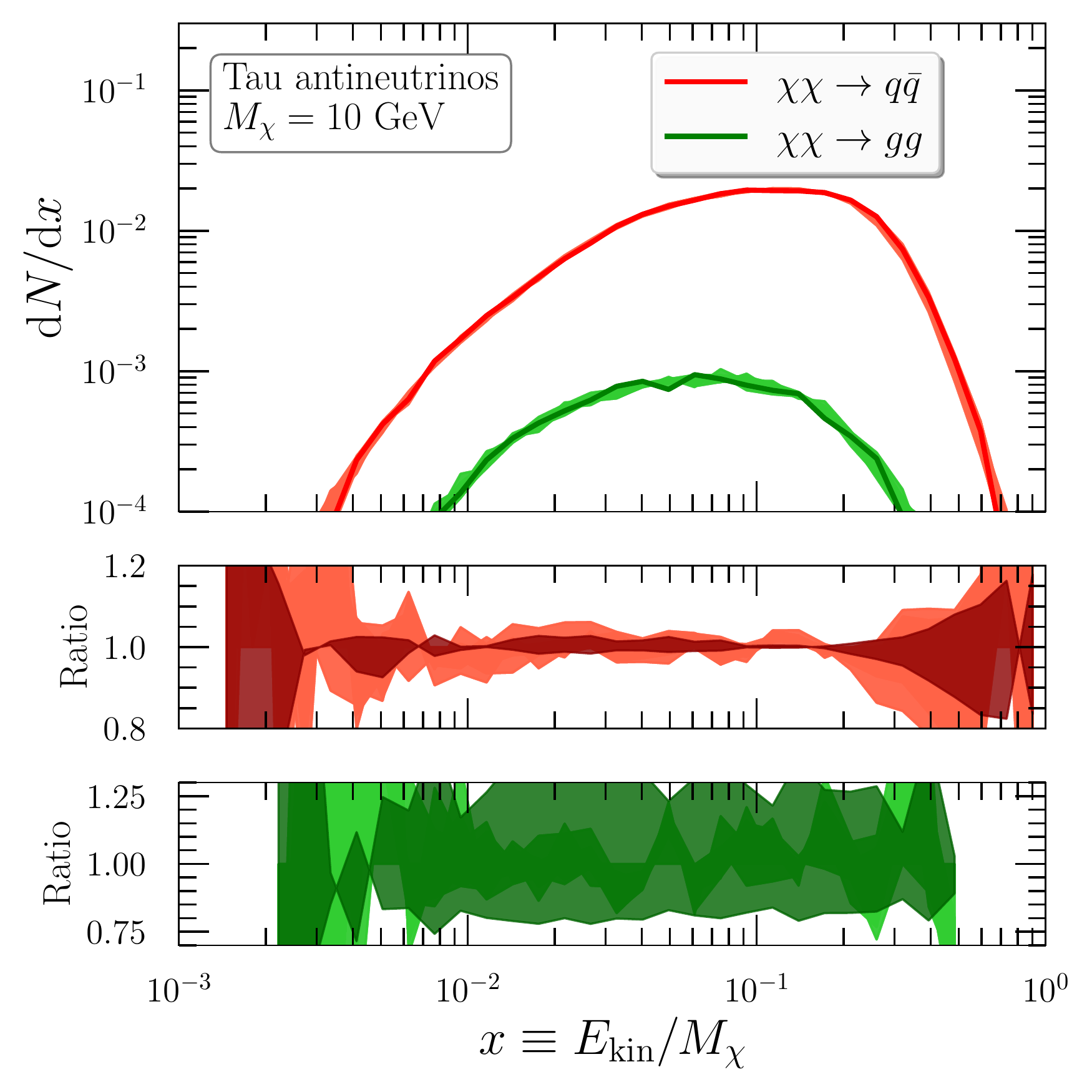}
    \caption{Same as for figure \ref{fig:spectra:mDM:10:1} but for the spectra of muon antineutrinos (left) and tau antineutrinos (right).}
    \label{fig:spectra:mDM:10:2}
\end{figure}
In figure \ref{fig:comparison}, we show the comparison between the theory predictions at the best-fit points (including the envelope from theory uncertainties) and the experimental data for the $p$ spectrum ({\it left}) and $\Lambda^0$ scaled energy ({\it right}). We can see that all the theory predictions agree pretty well with data within the uncertainty envelopes where a large uncertainty coverage clearly visible in the case of two-sigma and three-sigma eigentunes. On the other hand, the envelope spanned from the variations around the best-fit point of the weighted fit (shown in cyan) is somehow located as in between the one-sigma and the two-sigma eigentunes. There are, however, some regions in the $p$ spectrum where all the variations seem to cancel each other (specifically in the few bins between $x_p = 0.1$ and $x_p = 0.3$). The resulting uncertainty can range from about $5\%$ in the low $x_E$ region to about $40\%$ in the high $x_E$ region. In order to be as conservative as possible while in the same do not allow for very large variations we can define either the envelopes from the weighted fit or from the two-sigma eigentunes as our estimate of the uncertainty on the baryon spectra. As the uncertainty estimate from the weighted fits is computationally more expensive, we will use the two-sigma eigentunes in the rest of the paper and also for the new data tables that can be found in this \href{https://github.com/ajueid/qcd-dm.github.io.git}{github} repository\footnote{To obtain the anti-matter spectra along with the hadronisation uncertainties from the weighted fit one needs to generate twenty seven MC samples for each dark matter mass. Therefore, we recommend the user who is interested in producing the spectra himself to use the variations provided in Table \ref{tab:eigentunes} which will require nine runs instead of twenty seven and thus significantly reduce the running time.}.

\subsection{Assessing QCD uncertainties on anti-matter spectra}

In this section, we quantify the impact of QCD uncertainties on particle spectra from dark-matter annihilation for few dark matter masses and annihilation channels. The results will be shown in the $x$ variable defined by 
\begin{eqnarray}
x \equiv \frac{E_{\rm kin}}{M_{\chi}} = \frac{E - m}{M_{\chi}},
\end{eqnarray}
with $E_{\rm kin}$ is the kinetic energy of the particle specie (antiproton, positron, neutrino and photon), $m$ is its mass and $M_{\chi}$ is the dark matter mass. We study the following annihilation channels: 
\begin{table}[!h]
    \centering
    \begin{tabular}{l l l}
    $M_\chi$          \hspace{2cm} &   $\chi \chi \to XX$ \hspace{2cm} & Spectra  \\
    $10~{\rm GeV}$     &   $q\bar{q}$, $gg$ & Figures \ref{fig:spectra:mDM:10:1}--\ref{fig:spectra:mDM:10:2} \\
    $100~{\rm GeV}$   &   $q\bar{q}$, $gg$, $VV$ & Figures \ref{fig:spectra:mDM:100:1}--\ref{fig:spectra:mDM:100:2} \\
    $1000~{\rm GeV}$   &   $q\bar{q}$, $gg$, $VV$, $HH$, $t\bar{t}$ & Figures \ref{fig:spectra:mDM:1000:1}--\ref{fig:spectra:mDM:1000:2} \\
    \end{tabular}
\end{table} \\
For the $q\bar{q}$ annihilation channel, we assume that the dark matter is annihilated to all the quarks except the top quark with ${\rm BR}(\chi \chi \to q\bar{q}) = 0.2, q=u,d,s,c,b$. For the $VV$ channel, we include both the $ZZ$ and $W^+ W^-$ channels with equal probabilities: {\it i.e.} ${\rm BR}(\chi \chi\to W^+ W^-) = {\rm BR}(\chi \chi \to ZZ) = 0.5$. The impact of QCD uncertainties on the particle spectra are shown in figures \ref{fig:spectra:mDM:10:1}--\ref{fig:spectra:mDM:10:2} for $M_\chi = 10$ GeV, figures \ref{fig:spectra:mDM:100:1}--\ref{fig:spectra:mDM:100:2}  for $M_\chi = 100$ GeV and figures \ref{fig:spectra:mDM:1000:1}--\ref{fig:spectra:mDM:1000:2} for $M_\chi = 1000$ GeV for the above mentioned annihilation channels. We can see that the QCD uncertainties resulting from parton-shower variations are subleading for dark matter mass of $10$ GeV and especially in the anti-matter spectra. As far as we go to high dark matter masses, for example $1000$ GeV, these uncertainties become more competitive with the hadronisation uncertainties and reach up to $15\%$ in the peak region. The hadronisation uncertainties on the anti-proton spectra are very important and can reach up to $20\%$ in the low energy region and about $10\%$ in the peak region. In the high energy region, both the perturbative and hadronisation uncertainties are important with the latter are dominant with respect to the former and can reach up to $50\%$. Note that the position of the peak changes for some particle species. There are regions where all the variations result in no uncertainty at all, {\it i.e.} $x \simeq 0.2$ in the anti-proton spectra in the $q\bar{q}$ final state. 
\section{Application to dark-matter indirect detection experiments}
\label{sec:DMfit}
In this section we quantify the effects of QCD uncertainties on two DM observables: the velocity-weighted effective cross section, $\langle \sigma v \rangle$, and the DM mass $M_\chi$. We first discuss the general methodology for determining the DM uncertainties arising from QCD uncertainties. We then discuss the results for an antiproton final state. This is followed by a discussion of electron antineutrinos and photons final states as a proof of principle.

\subsection{Dark Matter uncertainties}
\label{subsec:DMunc}
In DM indirect searches, there are generally two important parameters: the velocity-weighted effective cross section, $\langle \sigma v \rangle$\footnote{In principle, the DM density can also be used instead of the velocity-weighted effective cross section.}, and the DM mass, $M_\chi$. The inclusion of QCD uncertainties will translate into uncertainties in determination of these two DM observables. The effect on $\langle \sigma v \rangle$ is simply an overall shift in the height of the spectrum; varying $\langle \sigma v \rangle$ is identical to changing the normalization of the spectrum. The uncertainty concerning the DM mass is quite straightforward, since a change in the DM mass changes the spectrum including the peak position. To quantify the effects of the QCD uncertainties on the DM mass, we consider two different approaches. First, for a spectrum coming from a DM particle with mass $M_\chi$, which different masses have spectra including QCD uncertainties that can match the spectrum of $M_\chi$, i.e., which masses can look the same as $M_\chi$. Or, the spectra of which masses can match the spectra of $M_\chi$ including QCD uncertainties, i.e., with which masses can $M_\chi$ look similar. We will call these two different approaches to finding the DM mass uncertainty $\Delta M$ and $\Delta S$, respectively. To quantify which spectra agree when QCD uncertainties are included, we require that for the upper and lower bounds of the DM uncertainties $\chi^2 / N_{\text{d.f.}} \approx 1$ holds. Although not proven, we consider it safe to assume that $\chi^2 / N_{\text{d.f.}}$ strictly increases for larger deviating values of DM mass or spectral height.

We have written a public code to both interpolate spectra for a given DM mass, channel, and final state, and to compute the upper and lower bounds for the three aforementioned DM uncertainties: $\Delta \langle \sigma v \rangle$, $\Delta M$, and $\Delta S$. Additionally, the considered fit region can be user supplied. The code can be found in this \href{https://github.com/ajueid/qcd-dm.github.io.git}{github} repository.
 
For both electron antineutrino and photon final states the spectrum does not need to be propagated, as these particles can propagate freely. Thus the DM uncertainties can be determined directly from the annihilation spectrum. However, for the antiproton final state, the spectrum must first be propagated before the upper and lower bounds of the uncertainties can be determined. Our code emulates cosmic-ray propagation by diffusing each bin using tabulated diffusion values, if the bin is not tabulated the diffusion values are linearly interpolated. By performing the diffusion for every bin, the complete propagated spectrum is obtained. We use \textsc{Dragon}~2~\cite{Evoli:2017, Evoli:2018} to compute the tabulated diffusion values by inserting an antiproton spectrum for the DM annihilation spectrum that is peaked at a specific bin. The propagation parameters are determined by fitting the AMS-02~\cite{AMS:2021} proton $p$, antiproton over proton $\bar{p}/p$ and Boron over Carbon ratio $B/C$ spectra using an artificial bee colony as the optimization algorithm~\cite{Dervis:2005, Akay:2012}. The resulting propagation parameters can be found in the README of the code. This fit resulted in a DM mass of $\mathcal{O}(100\text{GeV})$, which may affect results for other DM masses. 

To assess the effects of the different propagation parameters, we cross-compared a set of parameters from~\cite{CR_parameters:Di_Bernardo_2010} and the default settings from \textsc{Dragon} 2. The variation in the DM uncertainties between these sets is about 1-5\%. We consider these discrepancies to be small enough that it is safe to use our fitted parameters for the following results. 

Our default tabulated diffusion values do not account for solar modulation; only a low-energy correction factor for the diffusion coefficient is implemented. To account for this, we have fitted antiproton final-state spectra down to 1 GeV. While solar modulation is relevant at these energies, QCD uncertainties are typically small in this range, so we do not expect the inclusion of solar modulation to significantly change our results. Regardless, the tabulated diffusion values can be input by the user to meet situation-specific requirements. In particular, positron propagation is not implemented, so all results for positron final states were obtained using the annihilation spectrum and should therefore be treated with caution. In addition, the results of our code may be unreliable in certain cases for very small values of differential flux or QCD uncertainties due to numerical problems. We recommend that all results be checked for accuracy.

\subsection{Results for antiprotons}
\label{subsec:antiprot_DMunc}
\begin{figure}[h]
    \centering
    \includegraphics[width=.9\textwidth]{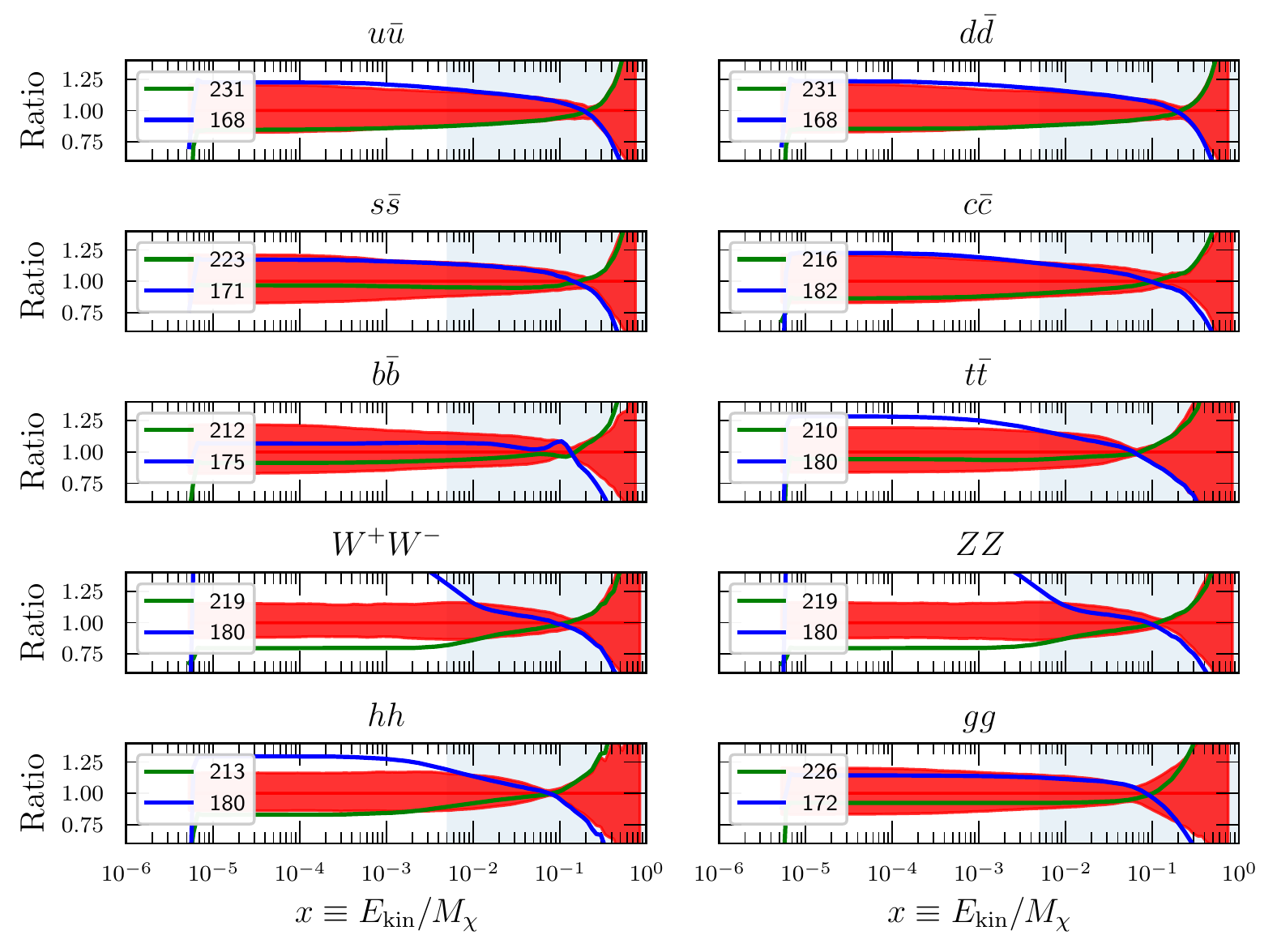}
    \caption{The $\Delta S$ uncertainty of various channels for a nominal DM mass of 200 GeV. The QCD uncertainties and the spectra of the upper and lower bound of $\Delta S$ are shown as the ratio to the spectrum of the 200 GeV DM particle. The shaded region from 1 GeV up to the DM mass of 200 GeV is considered for the fit.}
    \label{fig:antiproton_DM_unc}
\end{figure}

In order to showcase the impact of the QCD uncertainties on the antiproton spectrum and consequently the DM observables, we show $\Delta S$ for all channels in figure \ref{fig:antiproton_DM_unc} for a DM mass of 200 GeV. The QCD uncertainties are shown as the ratio to nominal, and the fitted upper and lower boundaries as a green and blue line respectively. We use a 200 GeV DM particle such that all channels can be produced on shell without having to account for kinematic boundary effects. We use $\Delta S$ as the DM uncertainty simply for visualization purposes; $\Delta S$ only has one set of QCD uncertainties, namely those of the nominal spectrum, as opposed to $\Delta M$ which considers the QCD uncertainties of both the upper and lower limits. The uncertainty on $\langle \sigma v \rangle$ is of course simply a vertical shift of the spectrum. \\
In figure~\ref{fig:antiproton_DM_unc} the shaded region, $x \in [1 / M_\chi, 1]$, is the energy range that is used for the fit, i.e. 1 GeV up to the DM mass of 200 GeV. Due to the requirement on the upper and lower bounds of $\Delta S$ that $\chi^2 / N_{\text{df}}\approx 1$ some averaging over the spectra is to be expected, which indeed occurs mainly in the high-$x$ regions. This is most striking for $b\bar{b}$, where the spectra fits comfortably within the QCD uncertainties, for low $x$, thereby compensating for a poorer fit at high $x$. The relevance of the high-$x$ regions is of course dependent on a case-by-case scenario. In order to make no assumptions about the relevance of the high-$x$ region we fit the entire spectrum up to $x=1$. However, not considering the high-$x$ regions increases the size of the DM uncertainties for most channels, most significantly for $b\bar{b}$. 

The relative DM uncertainties for higher masses DM must be calculated specifically for a given mass and channel, but some general observations can be made from the figure \ref{fig:antiproton_DM_unc}. For different masses, the influence of the low-$x$ regions becomes more important; the relative uncertainties of the spectra do not change significantly for higher DM masses, so the low-$x$ regions can significantly affect the relative magnitude of the upper and lower limits. This is because if one consistently considers the fitting range down to 1 GeV, the fraction of low $x$ regions increases. We have checked the following statements for both $M_\chi = 500$ GeV and $M_\chi = 1000$ GeV. For $u\bar{u}$, $d\bar{d}$, $c\bar{c}$, and $gg$, no significant change occurs in the relative size of $\Delta S$, which is evident when considering figure \ref{fig:antiproton_DM_unc}. The $W^+W^-$, $ZZ$, and $hh$ obtain relatively smaller DM uncertainties, as is also expected since the fitted spectra for these channels diverge in the low-$x$ regions. A relative increase in $\Delta S$ is seen for $b\bar{b}$, the upper limit of $s\bar{s}$, and the upper limit of $t\bar{t}$, as is also expected. 

While these results refer to $\Delta S$, the conclusions can easily be applied to $\Delta M$, since in both cases it is a matter of shifting the mass. The main difference, of course, is that the QCD uncertainties are different for different DM masses, so the numerical values for $\Delta S$ and $\Delta M$ may be different. This will be made explicit in the following section dealing with electron antineutrino and photon final states.

\subsection{Results for electron antineutrinos and photons}
\label{subsec:neutrino_DMunc}
In the following, we consider only electron antineutrinos and photons to show the effects of QCD uncertainties. The results for muon and tau antineutrinos may be significantly different from those for electron antineutrinos because their spectra differ considerably, as can be seen in Figure~\ref{fig:spectra:mDM:100:2}. The DM uncertainties for antineutrinos and photons can be determined directly from their annihilation spectra without the need for CR propagation. We show the three different DM uncertainties, $\Delta \langle \sigma v \rangle$, $\Delta M$, $\Delta S$, for a DM mass of 200 GeV for an electron antineutrino and photon final state in Table~\ref{tab:neutrino_and_photon_DM_unc}. We consider the energy range down to $x=10^{-5}/M_\chi$ for antineutrinos and $x=10^{-3}/M_\chi$ for photons, since for lower values of $x$ the QCD uncertainties become erratic. For both final states, we additionally perform a fit that includes and excludes the high- $x$ region of $x\in(0.5,1]$ to quantify the effects of these regions\\
\begin{table}[!t]
    \centering
    \begin{tabular}{l?cc|cc|cc?cc|cc|cc}
        \toprule
        \toprule
        & \multicolumn{6}{c?}{$\chi \chi \rightarrow XX \rightarrow \bar{\nu}_e$} & \multicolumn{6}{c}{$\chi \chi \rightarrow XX \rightarrow \gamma$} \\
        \midrule
        & \multicolumn{2}{c|}{$\Delta\langle\sigma v\rangle$ [\%]} & \multicolumn{2}{c|}{$\Delta M_\chi$ [GeV]} & \multicolumn{2}{c?}{$\Delta S$ [GeV]} & \multicolumn{2}{c|}{$\Delta\langle\sigma v\rangle$ [\%]} & \multicolumn{2}{c|}{$\Delta M$ [GeV]} & \multicolumn{2}{c}{$\Delta S$ [GeV]}\\
         & $A$ & $B$& $A$ & $B$& $A$ & $B$& $C$ & $D$& $C$ & $D$& $C$ & $D$\\
        \midrule
        $q\bar{q}$ &$^{+7}_{-7}$& $^{+7}_{-7}$ & $^{+33}_{-15}$& $^{+34}_{-28}$ & $^{+17}_{-30}$   & $^{+30}_{-30}$ & $^{+6}_{-6}$& $^{+6}_{-6}$ & $^{+20}_{-3}$ & $^{+23}_{-26}$& $^{+17}_{-25}$   & $^{+36}_{-29}$\\
        \rowcolor{Gray}
        $b\bar{b}$ &$^{+8}_{-7}$& $^{+7}_{-7}$ & $^{+31}_{-14}$& $^{+44}_{-27}$ & $^{+11}_{-29}$  & $^{+29}_{-30}$ & $^{+5}_{-6}$ & $^{+5}_{-6}$& $^{+14}_{-11}$& $^{+14}_{-11}$ & $^{+11}_{-20}$   & $^{+12}_{-20}$\\
        $t\bar{t}$ &$^{+3}_{-3}$& $^{+3}_{-3}$ & $^{+0}_{-0}$ & $^{+0}_{-0}$  & $^{+0}_{-0}$ & $^{+0}_{-0}$ & $^{+1}_{-3}$& $^{+1}_{-3}$ & $^{+10}_{-0}$& $^{+10}_{-0}$ & $^{+0}_{-0}$   & $^{+0}_{-8}$\\
        \rowcolor{Gray}
        $W^+W^-$ &$^{+2}_{-1}$& $^{+4}_{-4}$ & $^{+0}_{-0}$ & $^{+8}_{-9}$ & $^{+0}_{-0}$   & $^{+10}_{-7}$ & $^{+4}_{-2}$& $^{+4}_{-2}$ & $^{+0}_{-0}$& $^{+21}_{-10}$ & $^{+0}_{-0}$   & $^{+10}_{-10}$\\
        $ZZ$ &$^{+3}_{-0}$& $^{+4}_{-0}$ & $^{+0}_{-0}$& $^{+7}_{-5}$ & $^{+0}_{-0}$   & $^{+0}_{-6}$ & $^{+4}_{-4}$& $^{+4}_{-4}$ & $^{+8}_{-8}$ & $^{+10}_{-10}$& $^{+9}_{-7}$   & $^{+11}_{-10}$\\
        \rowcolor{Gray}
        $HH$ & $^{+4}_{-4}$ & $^{+5}_{-5}$& $^{+0}_{-0}$ & $^{+15}_{-12}$ & $^{+0}_{-0}$  & $^{+13}_{-12}$ &$^{+4}_{-2}$ & $^{+4}_{-3}$& $^{+0}_{-0}$ & $^{+11}_{-8}$& $^{+3}_{-0}$   & $^{+9}_{-10}$\\
        $gg$ &$^{+7}_{-7}$ & $^{+7}_{-7}$& $^{+30}_{-20}$ & $^{+43}_{-26}$& $^{+14}_{-28}$   & $^{+29}_{-28}$ & $^{+6}_{-6}$& $^{+6}_{-6}$ & $^{+14}_{-20}$ & $^{+14}_{-20}$& $^{+13}_{-20}$   & $^{+13}_{-29}$\\
        \bottomrule
        \bottomrule
    \end{tabular}
    \caption{The three DM uncertainties, $\Delta \langle \sigma v \rangle$, $\Delta M$, and $\Delta S$, for a DM mass of $M_\chi=200$ GeV for electron antineutrino and photon final states. The uncertainties are fitted for four different ranges: $A = [10^{-5}/M_\chi, 1]$, $B = [10^{-5}/M_\chi, 0.5]$, $C=[10^{-3}/M_\chi,1] $, and $D=[10^{-3}/M_\chi, 0.5]$. The lower bounds of the fit range has been chosen such the QCD uncertainties do not become erratic.}
    \label{tab:neutrino_and_photon_DM_unc}
\end{table}
From the table~\ref{tab:neutrino_and_photon_DM_unc} it can be seen that for the $W^+W^-$, $ZZ$, and $HH$ channels the inclusion of the high-$x$ region is very important, especially for a $\bar{\nu}_e$ final state. Figure~\ref{fig:spectra:mDM:100:1} clearly shows the small QCD uncertainties in the high-$x$ regions for both $W^+W^-$ and $ZZ$, and Figure~\ref{fig:spectra:mDM:1000:2} includes $HH$, where the small QCD uncertainties naturally translate into small DM uncertainties. While for most channels the spectrum falls off smoothly at high $x$ values, this is not the case for $W^+W^-$ or $ZZ$ when the final state is an antineutrino. Thus, while for many channels the inclusion or exclusion of high-$x$ values can generally be justified due to the low differential flux in this region, for both $W^+W^-$ and $ZZ$ this must be judged on a case-by-case basis depending on the specific analysis.

The uncertainties of DM for all non-top quarks and gluons are comparable to an antiproton final state: The upper and lower bounds are approximately at the edges of the QCD uncertainties. Thus, in general, the uncertainties of DM do not change significantly with or without the high-$x$ region. There are some exceptions though, e.g., the upper limit of $\Delta M$ for $\chi\rightarrow gg\rightarrow \bar{\nu}_e$. For a $t\bar{t}$-mediated $\bar{\nu}_e$ or $\gamma$ final state, the QCD uncertainties are more comparable to the $VV$ or $HH$ channels than to the $q\bar{q}$ or $gg$ channels, as can be seen from figure~\ref{fig:spectra:mDM:1000:2}. This is reflected in the DM uncertainties, which are indeed more comparable to $VV$ and $HH$.

In general, the magnitude of the DM uncertainties for both $\bar{\nu}_e$ and $\gamma$ final states depends strongly on the channel and the considered fitting region, even more so than for an antiproton final state. The impact of the QCD uncertainties on the DM observables can be as large as 20\% in certain scenarios and therefore must be considered in future DM indirect detection studies.
\section{Conclusions}
\label{sec:conclusions}

In this work, we have studied the QCD uncertainties on antimatter spectra from dark-matter annihilation and therefore completing the series of analyses related to QCD uncertainties in dark-matter indirect searches where we studied gamma-ray spectra in ref. \cite{Amoroso:2018qga} and antiproton spectra in ref. \cite{Jueid:2022qjg}.  After studying the general features of antimatter production in dark-matter annihilation within \textsc{Pythia}~8, we studied in detail the origin of antimatter for various annihilation channels taking $M_\chi = 1000$ GeV as an example. We found that the spectra of antimatter can be modeled correctly if the spectrum of charged pions, antiprotons and hyperons measured at LEP is modeled properly. We have performed a detailed analysis of baryon production at LEP (especially antiprotons and hyperons). We have found some tensions between the different measurements at LEP and selected the most reliable measurements for our tunes. Then, we have compared between the predictions of the state-of-art MC event generators, {\it i.e.}, \textsc{Herwig}~7, \textsc{Pythia}~8 and \textsc{Sherpa}~2. We found that MC event generators based on the string hadronisation model have a good agreement with data while the MC event generators based on cluster hadronisation model have very poor agreement with the experimental measurements with disagreement reaching up to $50\%$ in some kinematic regions. This comparison also suggests that the envelope spanned between the different MC event generators cannot define a faithful estimate of the QCD uncertainties since for photons and charged pions they agree pretty well in the peak region (see \cite{Amoroso:2018qga}) while for protons and hyperons they have very large differences. Therefore, we studied the alternative scenario where we estimate the QCD uncertainties within \textsc{Pythia}~8. \\

First we performed several returnings of the fragmentation-function parameters in \textsc{Pythia}~8 with the \textsc{Monash} 2013 tune as our baseline and using a set of constraining measurements at LEP totalling 48 measurements and 856 bins. The resulting tune yields very good agreement with the experimental measurements with $\chi^2/N_{\rm df} \simeq 1$ for most of the observables. We then estimated the QCD uncertainties that arise from parton-shower scale evolution variation and from hadronisation modeling using some parametric variations around the best-fit points of the hadronisation model. We studied the impact of these uncertainties on antimatter and photon spectra. We found that the QCD uncertainties are highly dependent on the annihilation channel, the DM mass,  the particle specie and the energy region. A notable example is the antiproton spectrum where the hadronisation uncertainties dominate the particle-physics error budgets and can reach $10\%$--$20\%$ in the bulk and the peak of spectra and up to $50\%$ in the high-$x$ region. The QCD uncertainties on the other antimatter species are highly dependent on the annihilation final state but are around $10\%$--$15\%$ depending on the annihilation channel and DM mass. \\ 

We finally analysed the impact of these QCD uncertainties on the DM indirect detection fits using realistic CR propagation models for antiprotons, electron antineutrinos and photons. We have considered various annihilation channels that lead to hadronic final states -- $u\bar{u}$, $d\bar{d}$, $s\bar{s}$, $c\bar{c}$, $b\bar{b}$, $t\bar{t}$, $W^+ W^-$, $ZZ$, $hh$, and $gg$ --. For antiprotons, we found that the QCD uncertainties impact the DM masses by up to $\Delta M_\chi = 18$--$32$ GeV depending on the annihilation channel. For the electron antineutrinos and photons the effects are much more different and can go anywhere between $0$ GeV to $43$ GeV depending on the annihilation channel and the kinematic region used in the fits. The effects on $\langle \sigma v \rangle$ and $\Delta S$ were also studied where we found important consequences. The size of QCD uncertainties are negligibly dependent on the choice of the diffusion parameters. Further measurements of the antimatter spectra at the Large Hadron Collider can be very important as they will deliver additional information that are necessary to improve the theory predictions and reduce the associated uncertainties. Therefore, we recommend the DM groups to start using these results for their future analyses. For this purpose, we provide the spectra in tabulated form including QCD uncertainties and some code snippets to perform fast DM fits (can be found in this \href{https://github.com/ajueid/qcd-dm.github.io.git}{github} repository). The tables can also be found in the latest releases of \textsc{DarkSusy}~6 \cite{Bringmann:2018lay},  \textsc{MicrOmegas}~5 \cite{Belanger:2018ccd} and \textsc{MadDM} \cite{Ambrogi:2018jqj}.

\section*{Acknowledgements}
The authors would like to thank Sascha Caron for his collaboration at early stages of this work. The work of AJ is supported in part by the Institute for Basic Science (IBS) under the project code, IBS-R018-D1. JK is supported by the NWO Physics Vrij Programme ``The Hidden Universe of Weakly Interacting Particles" with project number 680.92.18.03 (NWO Vrije Programma), which is (partly) financed by the Dutch Research Council (NWO). R. RdA acknowledges the Ministerio de Ciencia e Innovación (PID2020-113644GB-I00). PS is funded by the Australian Research Council via Discovery Project DP170100708 — ``Emergent Phenomena in Quantum Chromodynamics''. This work was also supported in part by the European Union’s Horizon 2020 research and innovation programme under the Marie Sklodowska-Curie grant agreement No 722105 — MCnetITN3.

\appendix
\section{Measurements: Complete list}
\label{sec:appendix1}

\begin{table*}[!h]
\setlength\tabcolsep{18pt}
\begin{center}
\begin{adjustbox}{max width=\textwidth}
\begin{tabular}{l  l  l l}
\toprule
Dataset & {Measurement} & $N_{\rm bins}$ & Reference \\
\toprule
\multicolumn{4}{l}{\textit{$p/\bar{p}$ momentum}} \\
\textsc{Aleph}~\textcolor{red}{(*)} & $x_p = |{\bf{p}}|/|p|_{\rm beam}$ & $26$ & ALEPH\_1996\_S3486095 \cite{Barate:1996fi}  \\
\textsc{Delphi}~\textcolor{red}{(*)} & $|\bf{p}|$  & $8$ & DELPHI\_1995\_I394052 \cite{Abreu:1995cu}  \\
\textsc{Delphi} & $|\bf{p}|$ & $23$ & DELPHI\_1998\_I473409 \cite{Abreu:1998vq}  \\
\textsc{Delphi} & $N_{p/\bar{p}}/N_{\rm charged}$ & $23$ & DELPHI\_1998\_I473409 \cite{Abreu:1998vq}  \\
\textsc{Opal}~\textcolor{red}{(*)} & $|{\bf{p}}|$  & $37$ &  OPAL\_1994\_S2927284 \cite{Akers:1994ez}   \\
\midrule
\multicolumn{4}{l}{\textit{Baryon spectra}} \\
\textsc{Aleph}~\textcolor{red}{(*)} & $\Lambda^0:~x_p$ & $25$ & ALEPH\_1996\_S3486095 \cite{Barate:1996fi}  \\
\textsc{Aleph} & $\Lambda^0:~\xi \equiv \log(1/x_p)$~(all events) & $22$ & ALEPH\_2000\_I507531 \cite{Barate:1999gb} \\
\textsc{Aleph} & $\Lambda^0:~\xi \equiv \log(1/x_p)$~($2$-jet events) & $22$ & ALEPH\_2000\_I507531 \cite{Barate:1999gb} \\
\textsc{Delphi} & $\Lambda^0$ scaled momentum & $11$ & DELPHI\_1993\_I360638 \cite{Abreu:1993mm} \\
\textsc{Opal} & $\Lambda^0$ scaled energy $x_E$ & $15$ & OPAL\_1997\_S3396100 \cite{Alexander:1996qj} \\ 
\bottomrule
\end{tabular}
\hspace{0.2cm}
\end{adjustbox}
\end{center}
\caption{\label{tab:measurements:protons} 
Measurements used in the optimisation process, and the corresponding number of bins for $p/\bar{p}$ and $\Lambda/\bar{\Lambda}$ momenta. The data is taken from \textsc{Aleph} \cite{Barate:1996fi, Barate:1999gb}, \textsc{Delphi} \cite{Abreu:1993mm, Abreu:1995cu, Abreu:1998vq}, and \textsc{Opal} \cite{Akers:1994ez, Alexander:1996qj}. The measurements marked by \textcolor{red}{(*)} are not used in the four-dimensional tunes (see the text for more details).}
\end{table*}

\begin{table*}[!h]
\setlength\tabcolsep{20pt}
\begin{center}
\begin{adjustbox}{max width=\textwidth}
\begin{tabular}{l  l  l l}
\toprule
Dataset & {Measurement} & $N_{\rm bins}$ & Reference \\
\toprule
\multicolumn{4}{l}{\textit{Charged multiplicity}} \\
\textsc{Delphi} & $N_{\rm ch}$,~2-jets, $y_{\text{cut}}=0.01$ & $19$ & DELPHI\_1992\_I334948 \cite{Abreu:1992gp}  \\
\textsc{Delphi} & $N_{\rm ch}$,~2-jets, $y_{\text{cut}}=0.02$ & $19$ & DELPHI\_1992\_I334948 \cite{Abreu:1992gp}  \\
\textsc{L3} & $N_{\rm ch}$ & $28$ & L3\_2004\_I652683 \cite{Achard:2004sv}  \\
\toprule
\multicolumn{4}{l}{\textit{Charged Particle Momentum}} \\
\textsc{Delphi} & $\log(1/x_p)$  & $27$ & DELPHI\_1996\_S3430090 \cite{Abreu:1996na}  \\
\textsc{Delphi} & $|\bf{p}|$ (all events)  & $27$ & DELPHI\_1998\_I473409 \cite{Abreu:1998vq}  \\
\textsc{Aleph} & $\log(1/x_p)$ (charged)  & $42$ & ALEPH\_1996\_S3486095 \cite{Barate:1996fi}  \\
\textsc{L3} & $\log(1/x_p)$ & $40$ & L3\_2004\_I652683 \cite{Achard:2004sv}  \\
\textsc{L3} & $\log(1/x_p)$, $udsc$ events & $40$ & L3\_2004\_I652683 \cite{Achard:2004sv}  \\
\textsc{Opal} & All events $\log(1/x_p)$  & $29$ & OPAL\_1998\_S3780481 \cite{Ackerstaff:1998hz}  \\
\bottomrule
\end{tabular}
\hspace{0.2cm}
\end{adjustbox}
\end{center}
\caption{\label{tab:measurements:charged} 
Same as for Table \ref{tab:measurements:protons} but for the charged multiplicity distributions. Data is taken from \cite{Abreu:1992gp, Achard:2004sv}.}
\end{table*}

\begin{table*}[!t]
\setlength\tabcolsep{9pt}
\begin{center}
\begin{adjustbox}{max width=\textwidth}
\begin{tabular}{l  l  l l}
\toprule
Dataset & {Measurement} & $N_{\rm bins}$ & Reference \\
\toprule
\multicolumn{4}{l}{\textit{Mean charged multiplicity}} \\
\textsc{Aleph} & $\langle N_{\rm ch} \rangle$ & $1$ & ALEPH\_1996\_S3486095 \cite{Barate:1996fi}  \\
\textsc{Aleph} & $\langle N_{\rm ch} \rangle$~for $|Y| < 0.5$ & $1$ & ALEPH\_1996\_S3486095 \cite{Barate:1996fi}  \\
\textsc{Aleph} & $\langle N_{\rm ch} \rangle$~for $|Y| < 1.0$ & $1$ & ALEPH\_1996\_S3486095 \cite{Barate:1996fi}  \\
\textsc{Aleph} & $\langle N_{\rm ch} \rangle$~for $|Y| < 1.5$ & $1$ & ALEPH\_1996\_S3486095 \cite{Barate:1996fi}  \\
\textsc{Aleph} & $\langle N_{\rm ch} \rangle$~for $|Y| < 2.0$ & $1$ & ALEPH\_1996\_S3486095 \cite{Barate:1996fi}  \\
\textsc{Delphi} & $\langle N_{\rm ch} \rangle$  & $1$ &  DELPHI\_1996\_S3430090 \cite{Abreu:1996na}   \\
\textsc{Delphi} & $\langle N_{\rm ch} \rangle$~(all events)  & $1$ &  DELPHI\_1998\_I473409 \cite{Abreu:1998vq}   \\
\textsc{Opal} & Mean charged multiplicity  & $1$ &  OPAL\_1992\_I321190 \cite{Acton:1991aa}   \\
\textsc{Opal} & All events mean charged multiplicity  & $1$ &  OPAL\_1998\_S3780481 \cite{Ackerstaff:1998hz}   \\
\toprule
\multicolumn{4}{l}{\textit{Identified Particle multiplicities}} \\
\textsc{Delphi} & Mean $\Lambda^0, \bar{\Lambda}^0$ multiplicity  & $1$ &  DELPHI\_1993\_I360638 \cite{Abreu:1993mm}   \\
\textsc{Delphi} & Mean $\pi^+/\pi^-$ multiplicity  & $1$ &  DELPHI\_1996\_S3430090 \cite{Abreu:1996na}   \\
\textsc{Delphi} & Mean $\pi^0$ multiplicity  & $1$ &  DELPHI\_1996\_S3430090 \cite{Abreu:1996na}   \\
\textsc{Delphi} & Mean $\rho$ multiplicity  & $1$ &  DELPHI\_1996\_S3430090 \cite{Abreu:1996na}   \\
\textsc{Delphi} & Mean $\Lambda^0$ multiplicity  & $1$ &  DELPHI\_1996\_S3430090 \cite{Abreu:1996na}   \\
\textsc{Delphi} & $\langle N_{\pi^\pm} \rangle$~(all events)  & $1$ &  DELPHI\_1998\_I473409 \cite{Abreu:1998vq}   \\
\textsc{Delphi} & $\langle N_{p/\bar{p}} \rangle$~(all events)  & $1$ &  DELPHI\_1998\_I473409 \cite{Abreu:1998vq}   \\
\bottomrule
\end{tabular}
\hspace{0.2cm}
\end{adjustbox}
\end{center}
\caption{\label{tab:measurements:mean} 
Same as for Table \ref{tab:measurements:protons} but for the mean multiplicity of charged particles $\langle N_{\rm ch} \rangle$ and of identified mesons and baryons. Data is taken from \cite{Barate:1996fi, Decamp:1991uz, Abreu:1990cc}. }
\end{table*}

\begin{table*}[!t]
\setlength\tabcolsep{13pt}
\begin{center}
\begin{adjustbox}{max width=\textwidth}
\begin{tabular}{l  l  l l}
\toprule
Dataset & {Measurement} & $N_{\rm bins}$ & Reference \\
\toprule
\multicolumn{4}{l}{\textit{Identified particle spectra}} \\
\textsc{Delphi} & $\pi^\pm$ momentum (all events)  & $23$ & DELPHI\_1998\_I473409 \cite{Abreu:1998vq}  \\
\textsc{Aleph} & $\pi^\pm$ momentum (charged)  & $39$ & ALEPH\_1995\_I382179 \cite{Buskulic:1994ft}  \\
\textsc{Opal} & $\pi^\pm$ momentum & $51$ & OPAL\_1994\_S2927284 \cite{Akers:1994ez}  \\
\textsc{Aleph} & $\pi^\pm$ spectrum  & $8$ & ALEPH\_1996\_S3486095 \cite{Barate:1996fi}  \\
\textsc{Delphi} & $\pi^0$ scaled momentum, all events  & $24$ & DELPHI\_1996\_I401100 \cite{Adam:1995rf}  \\
\textsc{Aleph} & $\pi^0$ spectrum  & $23$ & ALEPH\_1996\_S3486095 \cite{Barate:1996fi}  \\
\textsc{Opal} & $\pi^0$ scaled momentum, $\log(1/x_p)$  & $20$ & OPAL\_1998\_S3749908 \cite{Ackerstaff:1998ap}  \\
\bottomrule
\end{tabular}
\hspace{0.2cm}
\end{adjustbox}
\end{center}
\caption{\label{tab:measurements:mesons} 
Same as for table \ref{tab:measurements:protons} but for the spectrum of charged and neutral pions. Data is taken from \cite{Abreu:1998vq, Buskulic:1994ft, Akers:1994ez, Barate:1996fi, Adam:1995rf, Ackerstaff:1998ap}.}
\end{table*}

\begin{table*}[!t]
\setlength\tabcolsep{11pt}
\begin{center}
\begin{adjustbox}{max width=1.1\textwidth}
\begin{tabular}{l  l  l l}
\toprule
Dataset & {Measurement} & $N_{\rm bins}$ & Reference \\
\toprule
\multicolumn{4}{l}{\textit{$C$-parameter}} \\
\textsc{Aleph} & $C$ parameter (charged) & $24$ & ALEPH\_1996\_S3486095 \cite{Barate:1996fi}  \\
\textsc{Aleph} & $C$-parameter~($E_{\rm CMS} = 91.2~{\rm GeV}$) & $50$ & ALEPH\_2004\_S5765862 \cite{Heister:2003aj}   \\
\textsc{Delphi} & $C$ parameter & $23$ & DELPHI\_1996\_S3430090 \cite{Abreu:1996na}  \\
\textsc{L3} & $C$-parameter, $udsc$ events & $20$ & L3\_2004\_I652683 \cite{Achard:2004sv}  \\
\textsc{Opal} & $C$-parameter at $91$~GeV & $12$ & OPAL\_2004\_S6132243 \cite{Abbiendi:2004qz}  \\
\toprule
\multicolumn{4}{l}{\textit{Thrust}} \\
\textsc{Aleph} & $1-T$ (charged) & $21$ & ALEPH\_1996\_S3486095 \cite{Barate:1996fi}  \\
\textsc{Aleph} & Thrust (charged) & $42$ & ALEPH\_2004\_S5765862 \cite{Heister:2003aj}  \\
\textsc{Delphi} & Thrust, $1-T$ & $20$ & DELPHI\_1996\_S3430090 \cite{Abreu:1996na}  \\
\textsc{L3} & Thrust, $udsc$ events ($91.2~{\rm GeV}$) & $17$ & L3\_2004\_I652683 \cite{Achard:2004sv}  \\
\textsc{Opal} & Thrust, $1-T$, at $91~{\rm GeV}$ & $11$ & OPAL\_2004\_S6132243 \cite{Abbiendi:2004qz}  \\
\bottomrule
\end{tabular}
\hspace{0.2cm}
\end{adjustbox}
\end{center}
\caption{\label{tab:measurements:eventshapes} 
Same as in table \ref{tab:measurements:protons} but for the Thrust and the $C$--parameter. Data is taken from \cite{Barate:1996fi, Heister:2003aj, Abreu:1996na, Achard:2004sv, Abbiendi:2004qz}.}
\end{table*}

\section{QCD uncertainties in anti-matter spectra: Additional plots}

\begin{figure}
    \centering
    \includegraphics[width=0.49\linewidth]{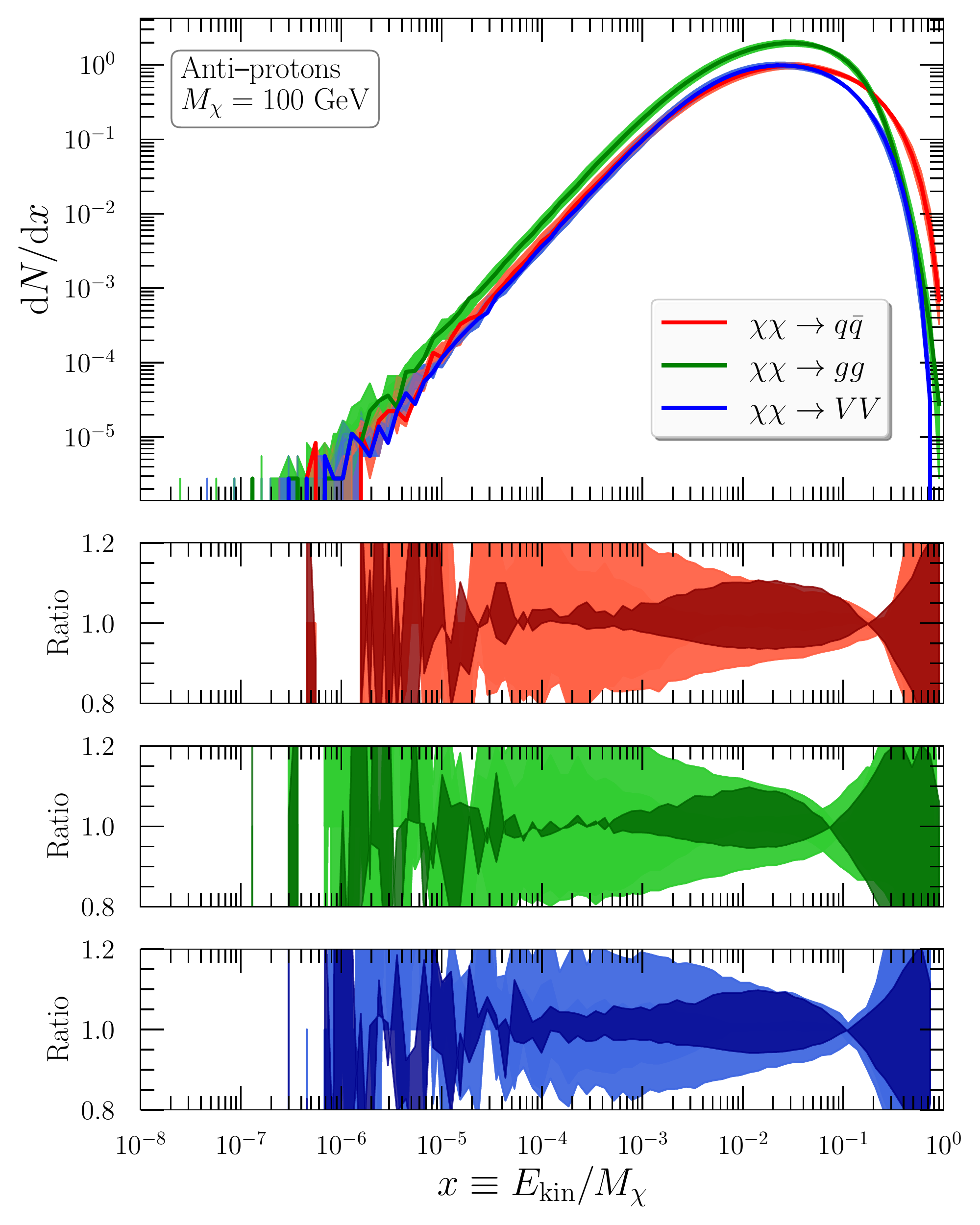}
    \hfill
    \includegraphics[width=0.49\linewidth]{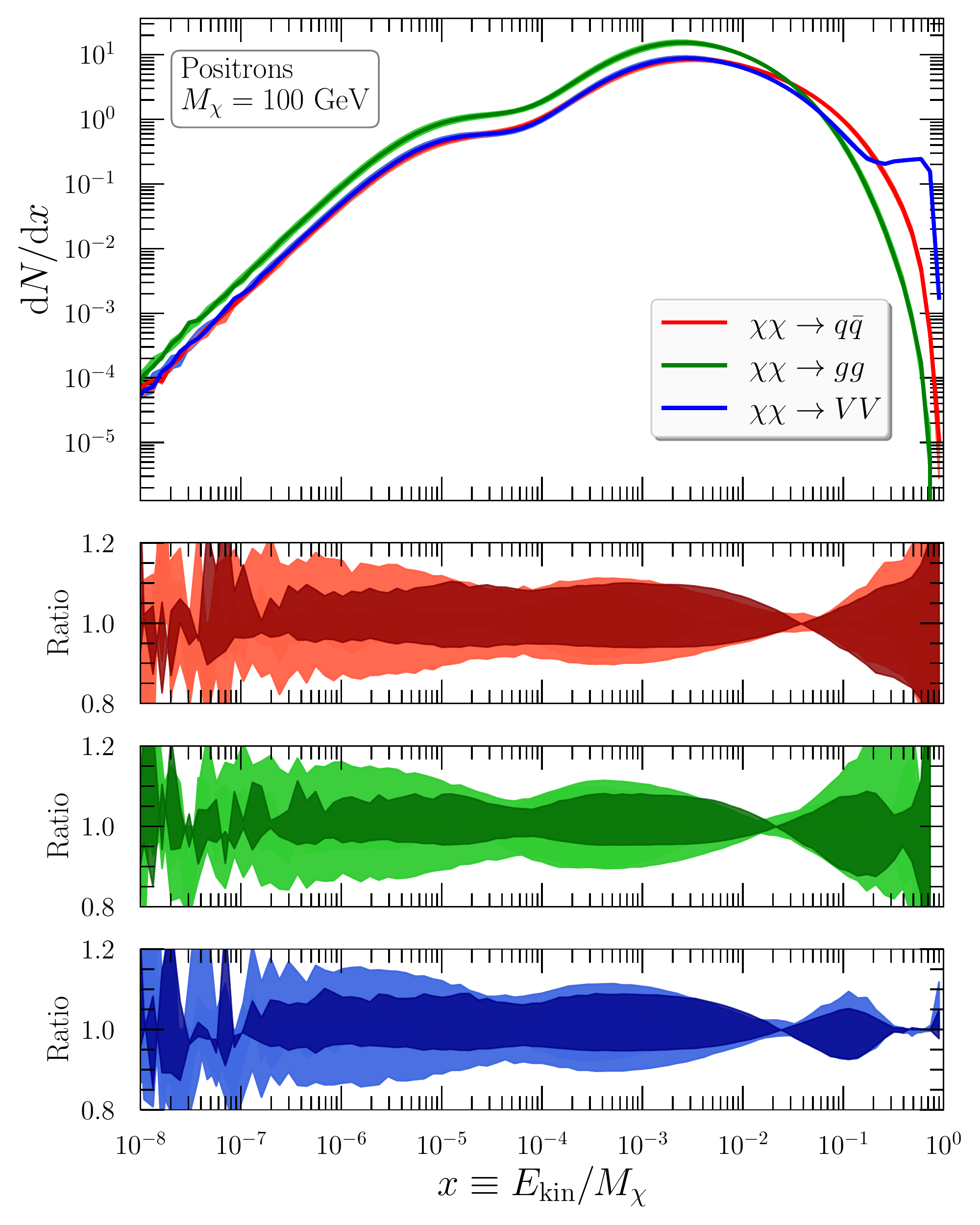}
    \vfill
    \includegraphics[width=0.49\linewidth]{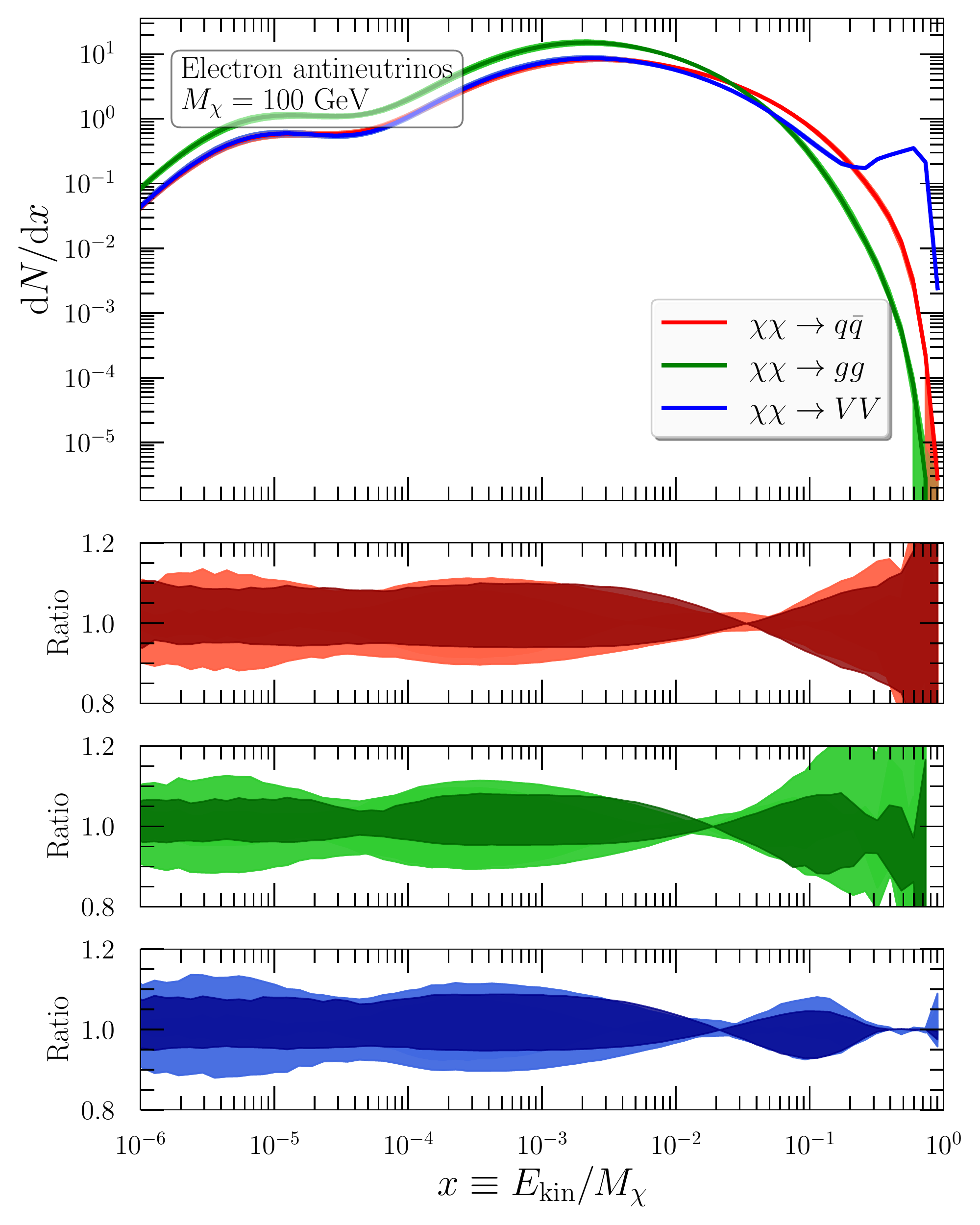}
    \hfill
    \includegraphics[width=0.49\linewidth]{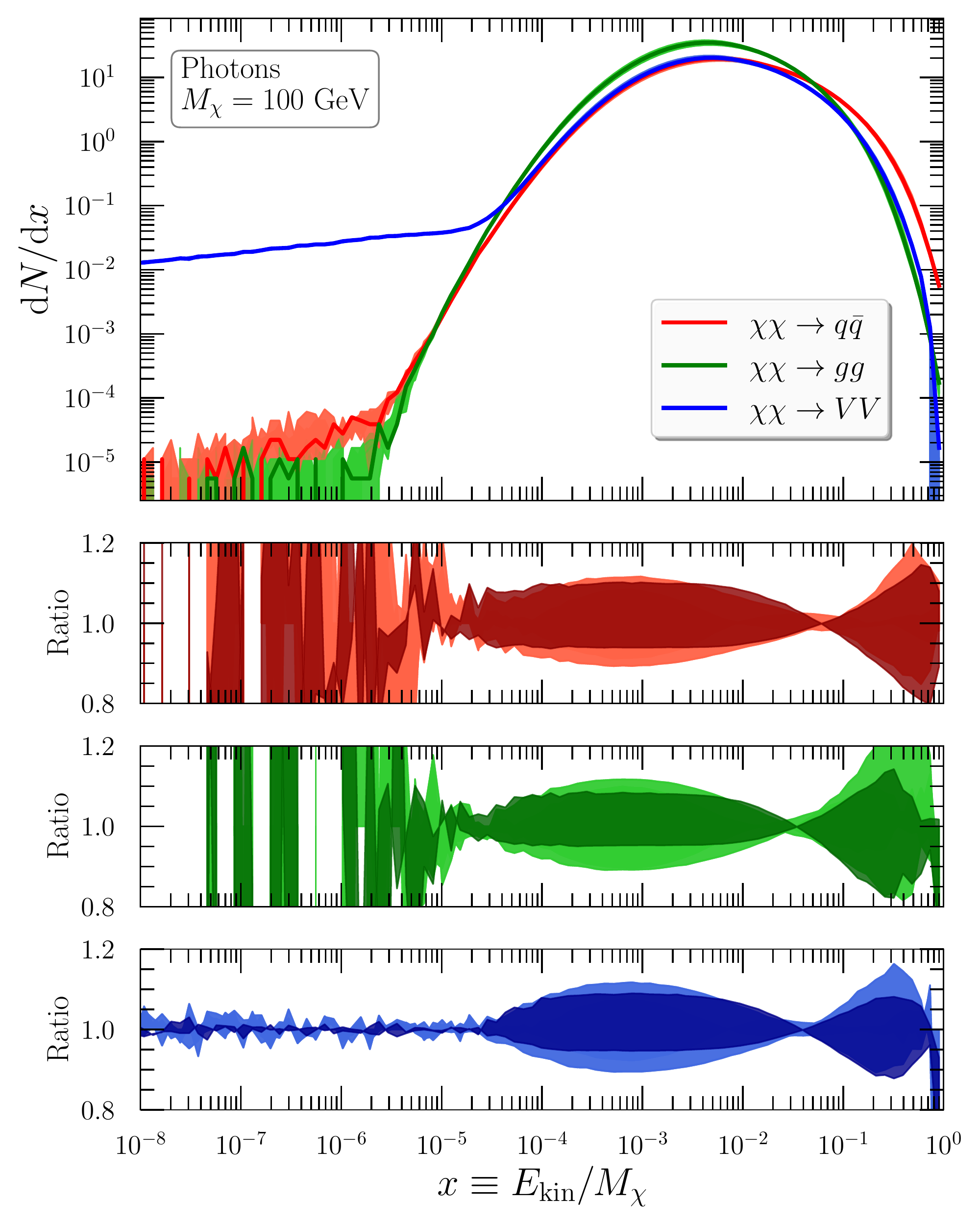}
    \caption{The scaled kinetic energy distribution of anti-protons (left upper panel), positrons (right upper panel), electron antineutrinos (left bottom panel) and photons (right bottom panel) in dark matter annihilation into $q\bar{q}$ (red), $gg$ (green) and $VV$ (blue). Here, the dark matter mass is chosen to be $100$ GeV. For each pane, the dark shaded band corresponds to the parton-shower uncertainties while the light shaded band corresponds to hadronisation uncertainties.}
    \label{fig:spectra:mDM:100:1}
\end{figure}

\begin{figure}[!t]
\centering 
\includegraphics[width=0.49\linewidth]{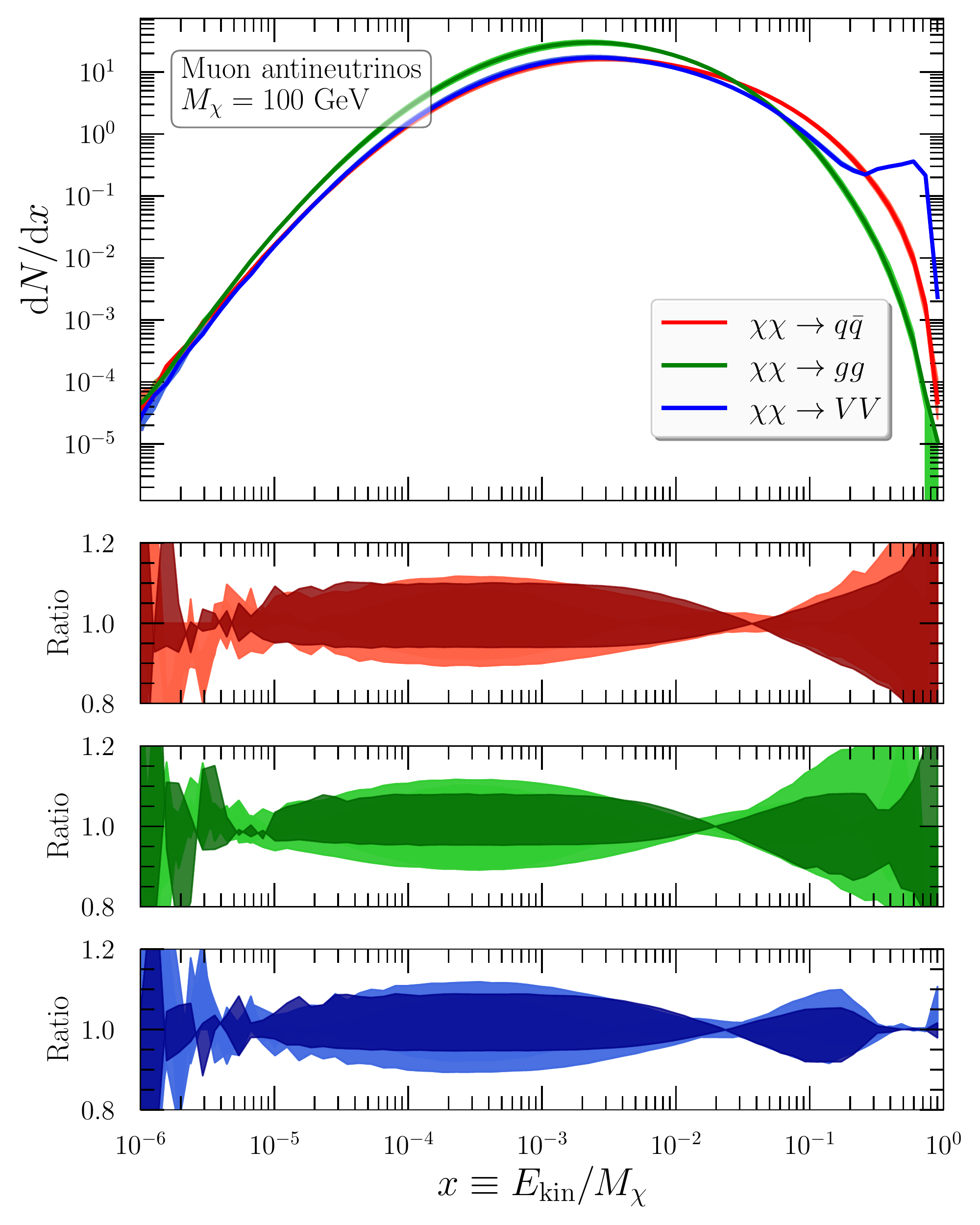}
\hfill 
\includegraphics[width=0.49\linewidth]{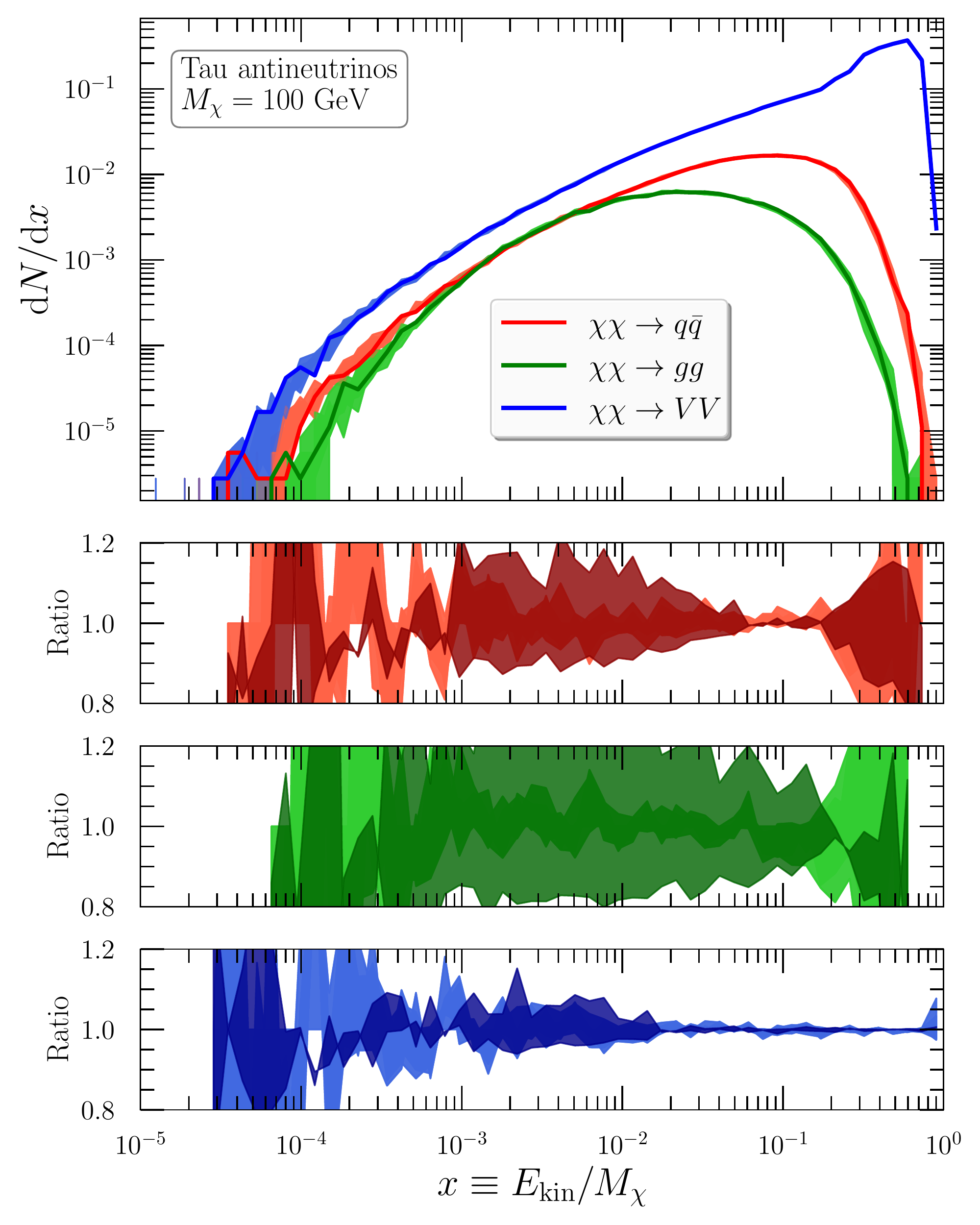}
\caption{Same as for figure \ref{fig:spectra:mDM:100:1} but for muon antineutrinos (left) and tau antineutrinos (right).}
\label{fig:spectra:mDM:100:2}
\end{figure}

\begin{figure}[!h]
    \centering
    \includegraphics[width=0.49\linewidth, height=10.5cm]{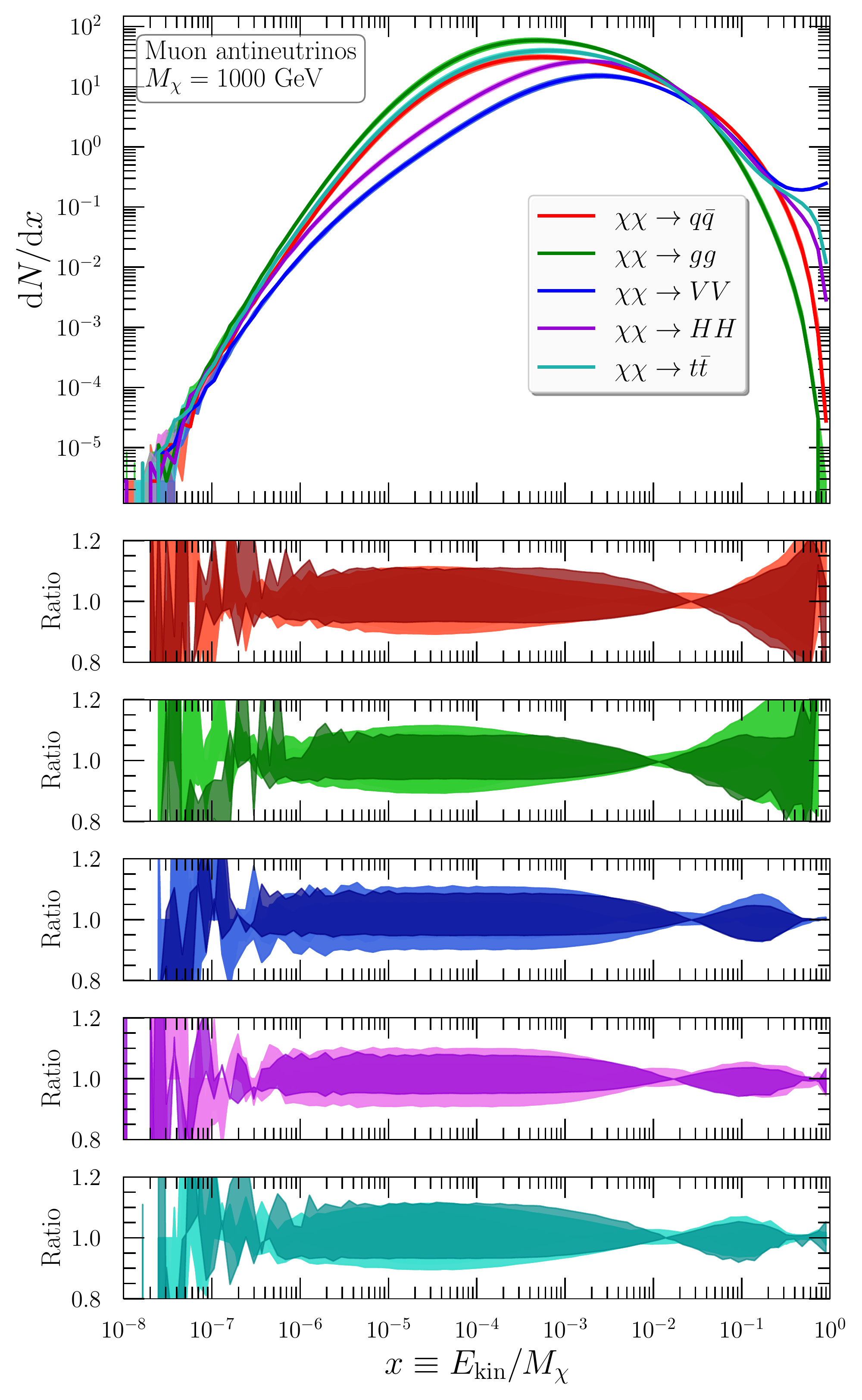}
    \hfill 
    \includegraphics[width=0.49\linewidth, height=10.5cm]{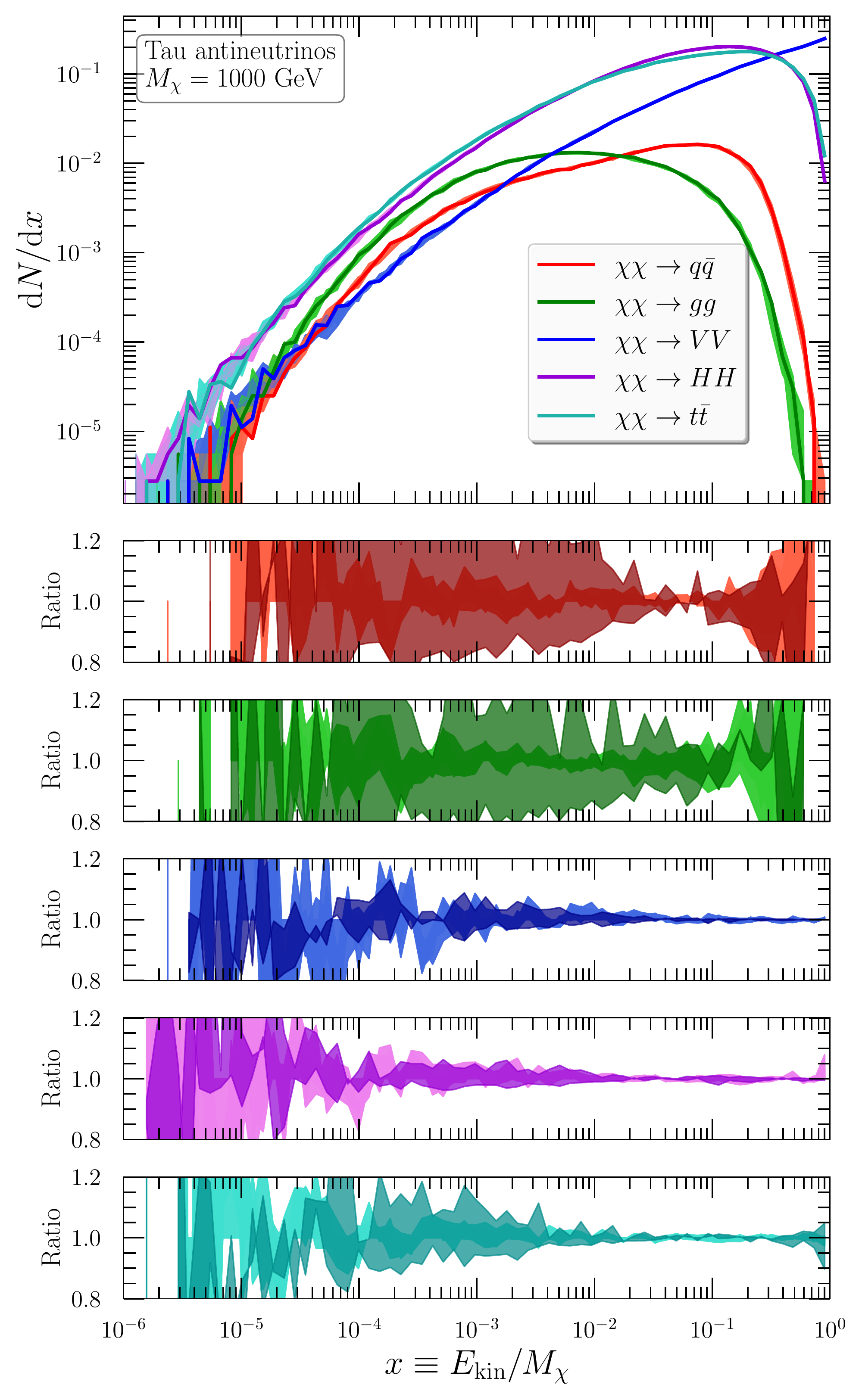}
    \caption{Same as for figure \ref{fig:spectra:mDM:100:2} but for $M_\chi = 1000$ GeV.}
    \label{fig:spectra:mDM:1000:1}
\end{figure}

\begin{figure}[!h]
    \centering
    \includegraphics[width=0.49\linewidth, height=10.5cm]{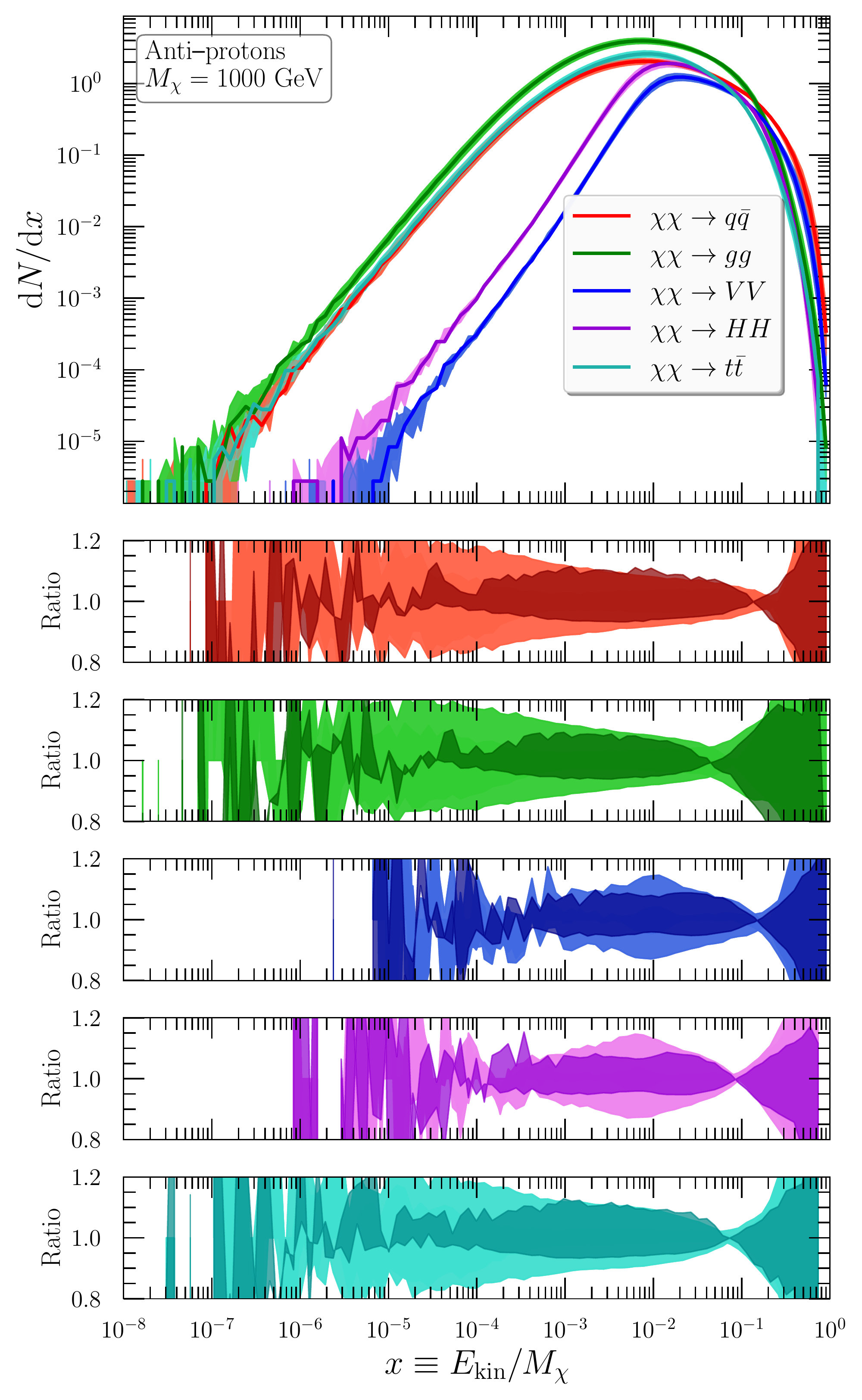}
    \hfill
    \includegraphics[width=0.49\linewidth, height=10.5cm]{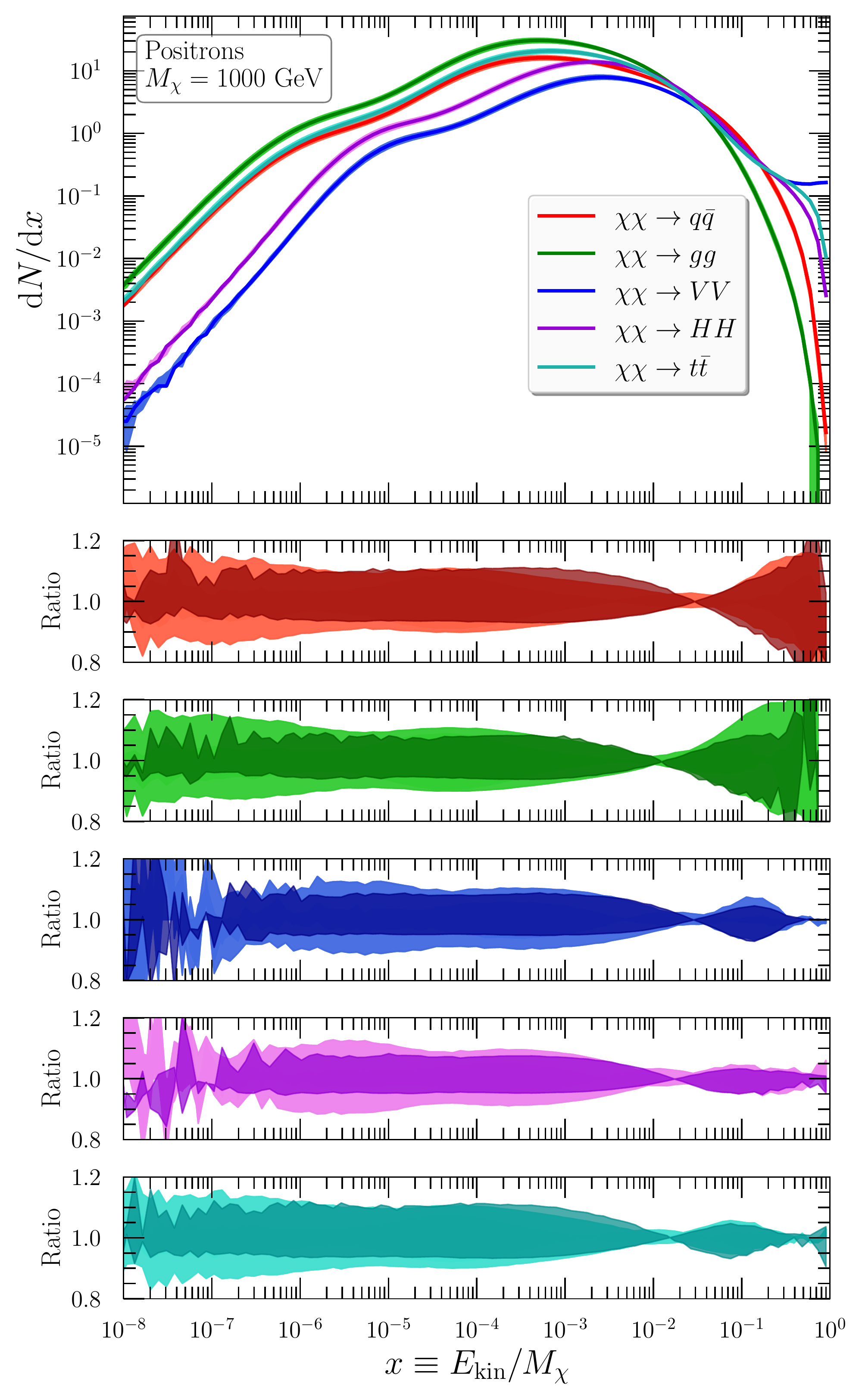}
    \vfill
    \includegraphics[width=0.49\linewidth, height=10.5cm]{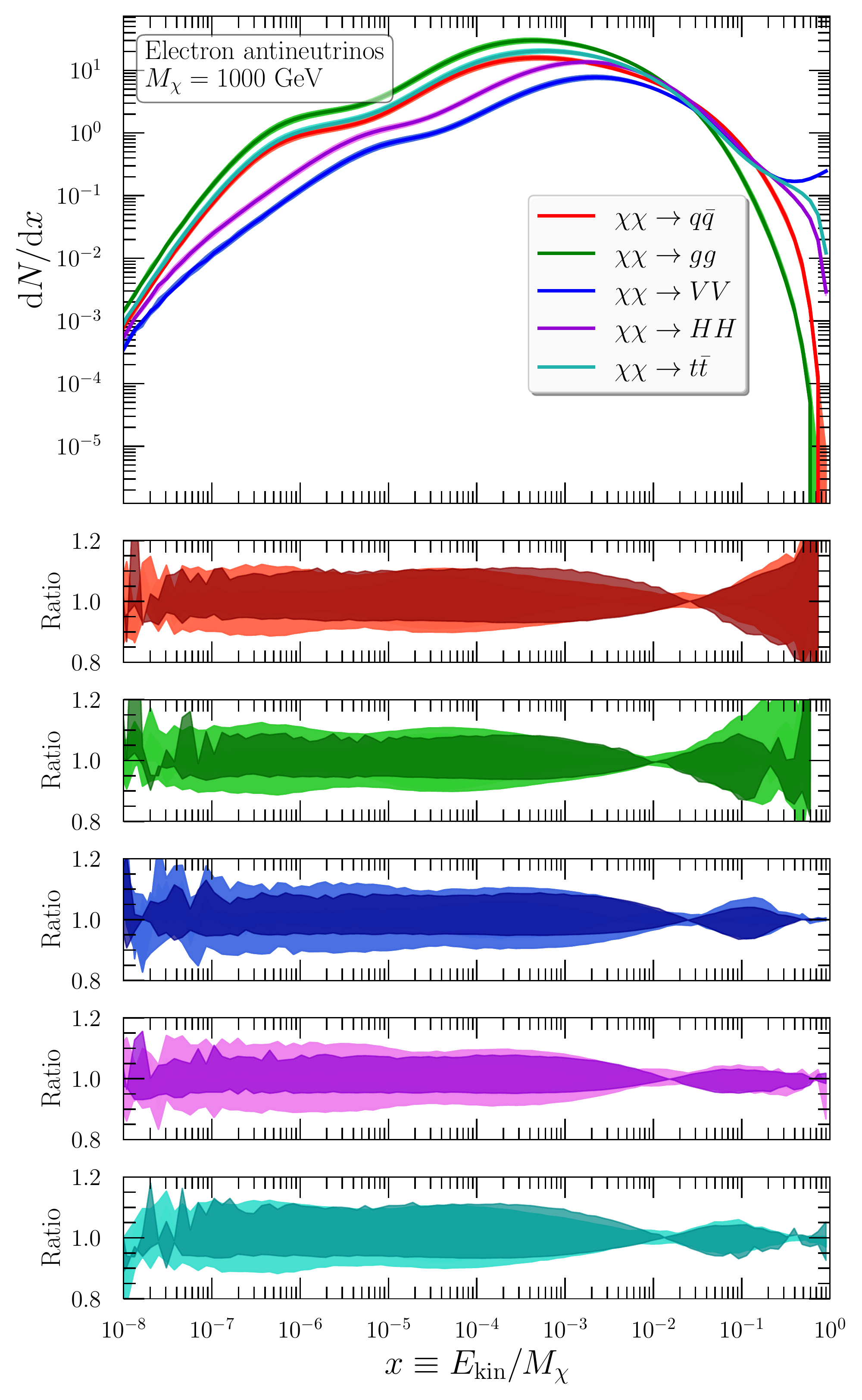}
    \hfill
    \includegraphics[width=0.49\linewidth, height=10.5cm]{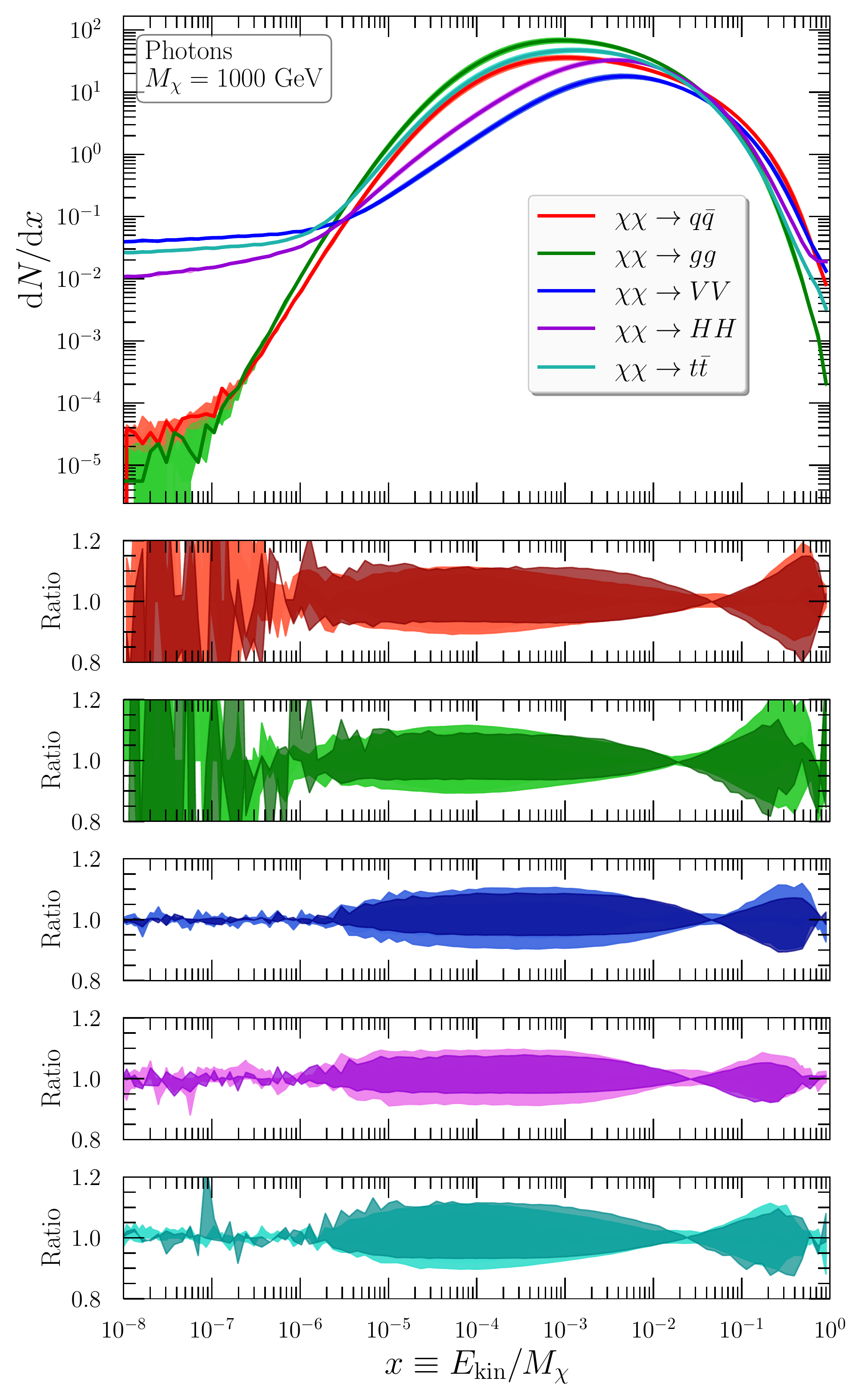}
    \caption{The scaled kinetic energy distribution of anti-protons (left upper panel), positrons (right upper panel), electron antineutrinos (left bottom panel) and photons (right bottom panel) in dark matter annihilation into $q\bar{q}$ (red), $gg$ (green), $VV$ (blue), $HH$ (purple) and $t\bar{t}$ (turquoise). Here, the dark matter mass is chosen to be $1000$ GeV. For each pane, the dark shaded band corresponds to the parton-shower uncertainties while the light shaded band corresponds to hadronisation uncertainties.}
    \label{fig:spectra:mDM:1000:2}
\end{figure}

In this section, we should the spectra of antimatter and photons in dark-matter annihilation for $M_\chi = 100$ GeV (figures \ref{fig:spectra:mDM:100:1}--\ref{fig:spectra:mDM:100:2}) and for $M_\chi = 1000$ GeV (figures \ref{fig:spectra:mDM:1000:1}--\ref{fig:spectra:mDM:1000:2}).

\clearpage

\bibliographystyle{JHEP}
\bibliography{bibliography.bib}


\end{document}